\documentclass[12pt]{article}

\usepackage{scicite}
\usepackage{times}
\usepackage{titling} % two title pages

\def\NN{{\mathbb N}}

\def\RR{{\mathbb R}}

\usepackage[mode=buildnew]{standalone}
\usepackage{amsmath,amssymb,amsfonts,amsthm}
\usepackage[utf8]{inputenc}
\usepackage{csquotes}
\usepackage{xspace}
\usepackage{textcomp} 
\usepackage{tikz}
\usepackage{caption}
\usepackage{subcaption}
\usepackage{placeins}
\usetikzlibrary{calc,shapes} 
\usepackage{tikz-cd}
\usepackage{stmaryrd}
\usepackage{stackengine,graphicx}
\usepackage{ifoddpage}
\usepackage{faktor}
\usepackage{enumitem}
\usepackage{booktabs}
\usepackage{algorithm2e} 
\usepackage{standalone}
\usepackage{pgfplots}
\usepackage{xcolor}
\usepackage{mdframed}

\usepgfplotslibrary{external} 
\pgfplotsset{compat=1.14}
\usepackage{color} 
\usepackage{hyperref}
\hypersetup{
    colorlinks,
    citecolor=black,
    filecolor=black,
    linkcolor=black,
    urlcolor=black
}

\newcommand\cS{\mathcal{S}}

\newcommand\subdivision{\cS}

\newcommand\SetOf[2]{\left\{#1\,\vphantom{#2}\right|\left.\vphantom{#1}\,#2\right\}}

\DeclareMathOperator{\conv}{conv}
\DeclareMathOperator{\supp}{supp}
\DeclareMathOperator{\nvol}{nvol}

\theoremstyle{plain}

\theoremstyle{definition}

\newtheorem*{remark}{Remark} 
\newtheorem{boxm}{Box}

\mdfdefinestyle{makebox}{
    backgroundcolor=gray!20
}
\surroundwithmdframed[style=makebox]{boxm}
\newcommand\polymake{{\tt poly\-make}\xspace}
\newcommand\jupyter{{\tt jupyter}\xspace}
\newcommand*{\inlineequation}[2][]{%
  \begingroup
    \refstepcounter{equation}%
    \ifx\\#1\\%
    \else
      \label{#1}%
    \fi
    \relpenalty=10000 %
    \binoppenalty=10000 %
    \ensuremath{%
      #2%
    }%
    ~\@eqnnum
  \endgroup
}

\newenvironment{sciabstract}{%
\begin{quote} \bf}
{\end{quote}}

\title{Master regulators of evolution and the microbiome in higher dimensions}
\author
{Holger Eble,$^{1}$ Michael Joswig,$^{1,2\ast}$ Lisa Lamberti$^{3,4}$,William B. Ludington$^{5,6\ast}$\\
\\
\normalsize{$^{1}$Chair of Discrete Mathematics/Geometry, TU Berlin, Germany}\\
\normalsize{$^{2}$MPI MiS Leipzig, Germany}\\
\normalsize{$^{3}$Department of Biosystems Science and Engineering, ETH Z\"urich, Basel, Switzerland}\\
\normalsize{$^{4}$
SIB Swiss Institute of Bioinformatics, Basel, Switzerland}\\
\normalsize{$^{5}$
Department of Embryology, Carnegie Institution for Science, USA}\\
\normalsize{$^{6}$
Department of Biology, Johns Hopkins University, Baltimore, MD, USA}\\
\\
\normalsize{$^\ast$To whom correspondence should be addressed;} \\
\normalsize{E-mail: joswig@math.tu-berlin.de, ludington@carnegiescience.edu.}
}

\date{}

\begin{document} 


\baselineskip24pt


\maketitle

\begin{sciabstract}
%Life is an emergent property of interacting parts, including genes, molecules, and organisms, in many dimensions. Frameworks to explore this complexity are limiting. Here we develop a new mathematical formalism for high dimensional fitness landscapes based on triangulations, which we use to decompose fitness landscapes into their simplest parts. that overcomes this challenge, allowing us to compare interactions across multiple dimensions. We examined four combinatorically-complete datasets, two each for genetics and microbiomes, revealing frequent distortions of the landscape by specific mutations or microbiome species. High dimensional interactions with significant effects on organismal fitness were sparsely-distributed, often arose as projections from lower-dimensions, but sometimes arose uniquely in higher dimensions. The decreased prevalence of interactions above three dimensions, suggests a limit to biological complexity, albeit a large one. 

A longstanding goal of biology is to identify the key genes and species that critically impact evolution, ecology, and health. Network analysis has revealed keystone species that regulate ecosystems \cite{Paine1969} and master regulators that regulate cellular genetic networks \cite{Schuldiner2005,Venkatesan2009,Costanzo2010}. 
Yet these studies have focused on pairwise biological interactions, which can be affected by the context of genetic background \cite{Weinreich2018,Kuzmin2018} and other species present \cite{Gould232959,Case1981,Billick1994,Grilli2017} generating higher-order interactions. 
The important regulators of higher-order interactions are unstudied. 
To address this, we applied a new high-dimensional geometry approach that quantifies epistasis in a fitness landscape \cite{Eble2019} to ask how individual genes and species influence the interactions in the rest of the biological network. We then generated and also reanalyzed 5-dimensional datasets (two genetic, two microbiome). We identified key genes (e.g. the \emph{rbs} locus and \emph{pykF}) and species (e.g. \emph{Lactobacilli}) that control the interactions of many other genes and species. These higher-order master regulators can induce or suppress evolutionary and ecological diversification \cite{Bajic2018} by controlling the topography of the fitness landscape. Thus, we provide mathematical intuition and justification for exploration of biological networks in higher dimensions.
\end{sciabstract}

%\keywords{Fitness Landscapes; Epistasis; Biological Networks; Cluster Filtrations; Microbiome; Higher-Order Interactions; Sparsity}

%\tableofcontents

\section{Introduction}

% introduce master regulators
	Master regulators are nodes in a network that control the rest of the network. They are often identified as highly connected nodes. For example, in eukarotic cells, the protein, target of rapamycin (TOR), interacts with many other proteins and pathways to control cellular metabolism \cite{Wullschleger2006}. Identifying TOR unified studies in many areas of cell biology, including regulation of transcription, translation, and the cytoskeleton around a central signaling pathway, with druggable targets for therapeutics of cancer, autoimmunity, metabolic disorders, and aging \cite{Wullschleger2006}. Ecological master regulators are called keystone species, a classical example being the starfish, \emph{Pisaster}, which regulates the biodiversity of intertidal zone by eating many other species \cite{Paine1969}. Identifying these key nodes in biological networks provides control points that can be used for instance in cancer therapy (through TOR) or ecological restoration (through starfish). 
	
	Epistasis is a framework to quantify biological networks, specifically gene networks, in terms of which genes (the nodes) interact and are thus connected by an edge. 
	Constructing a gene network using epistasis works by iteratively mutating a set of individual genes and pairs of these genes, and then using the phenotypes of the mutants to construct the network. For instance, if genes $A$ and $B$ both affect a phenotype, ${\bf C}$, we make the single mutants $a$ and $b$ and the double mutant $ab$. 
	By measuring the effects on the output phenotype, e.g. fitness, it can be determined if $A$ and $B$ operate in parallel to affect ${\bf C}$  ($A \rightarrow {\bf C}$ and $B \rightarrow {\bf C}$) or in serial ($A \rightarrow B \rightarrow {\bf C}$). 
	These two possibilities are differentiated based on the degree of non-additivity: if the phenotypes of $a$ and $b$ add up to the phenotype of $ab$, the genes do not interact and thus operate in parallel. 
	If they are non-additive, the genes interact and thus operate in serial. More specifically, if $A \rightarrow B \rightarrow {\bf C}$, then mutants $a$, $b$, and $ab$ will each produce the same phenotype, thus, $a+b \neq ab$, indicating non-additivity or epistasis. The concept has been applied to map pairwise connections for protein structure \cite{Wu2016}, genetics \cite{Carlborg2004,Weinreich2018,Kuzmin2018,Costanzo2010,Collins2007}, microbiomes \cite{Gould232959}, and ecology \cite{Case1981,Billick1994,Grilli2017}. 
	
	Epistatic interactions are important in nature \cite{Weinreich2005}, for instance when mutations occur \cite{Smith1970,Kauffman1987,McCandlish2018} or when sex, recombination, and horizontal gene transfer bring groups of genes together \cite{Schumer2018,WEINREICH2013700,Weinreich2018,Crona2020,SailerPLOS,McDonald2016}, making multiple loci interact. 
	Applying epistasis to genome-wide measurement of pairwise genetic interactions has revealed biochemical pathways composed of discrete sets of genes \cite{Costanzo2010,Collins2007} as well as complex traits, such as human height, that are affected by almost every gene in the genome \cite{Liu2019,Boyle2017}. 
	New innovations have applied epistasis to broader data types \cite{McCandlish,Kuzmin2018} and at different scales, making epistasis a widely valuable tool. 
	For instance, epistasis between bacteria in the microbiome has functional consequences \cite{Ratzke2020,Friedman2017,Gould232959,Sundarraman2020,Piccardi2019,Sanchez-Gorostiaga2019} when community assembly combines groups of species in a fecal transplant. 
	In this case, the nodes in the network are bacterial species. 
	The master regulators of biological networks are identified by their position in the network, often as nodes with a higher degree of edges than average \cite{Padi2015}. 

% introduce genotopes, hypercubes.
	A known challenge of biological networks is that they are high-dimensional, meaning the interactions can change depending on the biological context or the genetic background \cite{Weinreich:2007signepistasis}, cf. \cite{Krug:2021EpistasisandEvolution} and references therein. 
	This is important because such networks cannot be fully captured by pairwise interactions.
	Higher-order epistatic interactions are interactions that require three or more interacting parts, for instance genetic loci. 
	From a network standpoint, loci that affect the interactions of many other loci play a key role in regulation of network structure.
	
% introduce fitness landscapes
	Identifying such regulators requires a high-dimensional formulation of network structure. 
	We recently developed such a formulation based on epistasis of fitness landscapes \cite{Eble2019}. 
	Fitness landscapes depict biological fitness as a function of genotype space \cite{Wright1932,Smith1970,Kauffman1987}. 
	Sewall Wright defined the genotype space as a hypercube with each genetic locus represented as an independent dimension \cite{Wright1932}. 
	Previous work formalized the fitness landscape of this genotype space and quantified epistasis on the fitness landscape \cite{BPS:2007,Crona2017,Crona2020,Eble2019}. 
	We developed the {\bf epistatic filtration} technique, which segments the high-dimensional fitness landscape into local subregions and quantifies their epistasis in higher dimensions, allowing a researcher to hone in on important subregions of the landscape. 
	
	Here we develop that framework further in order to apply it to identify regulators of high-dimensional interactions. 
	% Note what is new
	Rather than the traditional approach of assigning significance to a gene or species based on its pairwise interactions \cite{Paine1969,Paine1992,Schuldiner2005,Venkatesan2009,Costanzo2010}, we assign significance based on how the presence of that gene or species influences the structure and magnitude of interactions in the rest of the network. 
	In order to compare interaction magnitudes across different dimensions, we develop a dimensionally-normalized definition of epistasis. 
	We also develop a graphical approach to determine whether high-dimensional epistasis has lower-dimensional roots and what they are. 
	We then analyze four data sets for 5-dimensional genotypes. 
	Two are genetic datasets for (i) mutations that arose in \emph{E. coli} evolution \cite{Khan1193} and 
	(ii) $\beta$-galactosidase antibiotic resistance \cite{PhysRevLett.106.198102}. 
	Two are microbiome datasets measuring the impact of bacterial interactions on \emph{Drosophila} lifespan, with one previously published \cite{Gould232959} and another generated here.
	Our framework identifies regulators of higher-dimensional network structure in both the genetics and microbiome datasets. 
	We find that specific genes and bacterial species suppress interactions in the rest of the network, meaning they regulate the higher-order network structure. 

\goodbreak

\section{Results}

\subsection{Epistatic filtrations describe higher-dimensional biological networks} 

% describe Box 1 overview and link to network
	Our goal is to identify master regulators of biological interactions in higher dimensions. We use epistasis as a measure of interactions, and in higher dimensions, these occur on a fitness landscape. Our approach is to first measure epistasis on the high dimensional fitness landscape and then ask how individual loci, e.g. genes, change the shape of the landscape. We use the epistatic filtration technique to quantify epistasis on the fitness landscape. We use parallel epistatic filtrations to quantify the changes in the landscape due to each locus. 
	
	First, we describe epistatic filtrations. Epistatic filtrations are analogous to analyzing the drainage sectors within a watershed (see~\ref{box:conceptual+introduction}), which is a real physical landscape with altitude as a function of latitude and longitude. The topography sets where water will flow. 
	Boundaries of a watershed are set by ridges, which enclose sectors within the watershed. These sectors feed tributary creeks, which join with other tributaries to form larger sectors within the watershed.  
	 We can think of a fitness landscape as having sectors as well. In a fitness landscape, the topography is set not by altitude but by measurements of organismal fitness as a function of genotype. The longitude and latitude of a watershed correspond to genotypes in the fitness landscape.
	Because the biological entities are discrete (i.e., a gene is either wildtype or mutant), our framework is discrete too. We represent each gene with a separate dimension as proposed by Wright \cite{Wright1932}. The space of all genotypes has many dimensions, one per mutated gene \cite{Wright1932,Krug:2021EpistasisandEvolution}. This high-dimensional space is a \emph{genotype hypercube} \cite{Wright1932,Smith1970,Kauffman1987}.
	We next quantify the epistasis of the fitness landscape. This requires that we define sets of genotypes to compare. We do so by segmenting the genotype cube into sectors (see Box \ref{box:conceptual+introduction}). This approach is different from previous approaches that defined sets of genotypes called circuits that traverse paths across the landscape \cite{BPS:2007}. 
	An advantage of our approach is that there are orders of magnitude fewer sectors in a landscape than circuits (c.f. Table~\ref{tab:complexity:gen-micro} versus Table~\ref{tab:bipyramids}), reducing the search space and the associated statistical constraints from multiple testing comparisons. 
These sectors are sets of adjacent genotypes in the hypercube. Geometrically speaking, these sectors are simplices, meaning each vertex (genotype) is directly connected to every other vertex in the set. For instance in $2D$, each vertex in a triangle is connected to the other two.
	To perform the segmentation, we use a triangulation. In Box \ref{box:definition+epistatic+filtration}, we illustrate how a two dimensional fitness landscape is triangulated using the phenotypes of the genotypes, which form a third dimension that we depict on the vertical axis. We use the topography provided by the phenotype data to uniquely determine the ridges of the landscape. Projecting these ridges back to the $2D$ genotype plane forms a triangulation of the genotypes into sectors (see Box \ref{box:definition+epistatic+filtration}). This diagram is similar to previous illustrations of epistasis on a two-dimensional landscape (c.f. \cite{Weinreich:2007signepistasis,Krug:2021EpistasisandEvolution}), but our approach is unique in that we use the triangulation to sector the fitness landscape.
	Next, we construct a network representation of the sectored genotype space to depict the pairwise adjacency of neighboring simplices (nodes) \cite{Eble2019}. An edge in this network indicates that two simplices are adjacent, meaning they share a face. 
	Next, we locate the epistasis on this network topology. Our definition of epistasis is unique yet consistent with previous ones in lower dimensions (see Box \ref{box:definition+epistatic+filtration}). We assess the magnitude of epistasis of each pair of adjacent sectors in the triangulation by calculating the volume spanned by the fitness phenotypes corresponding to the genotypes of the vertices if the adjacent sectors. This definition makes the framework consistent when applying it to higher dimensions. 
	We next rank the magnitudes of the adjacent sectors from smallest to largest. Plotting these merges gives an epistatic filtration (see Box \ref{box:conceptual+introduction} \& \ref{box:example+epistatic+filtration}). 
	
	To determine how an individual locus, e.g. gene or species, affects the interactions in the rest of the network, we compare the epistasis for each pair of adjacent sectors with the locus of interest added or removed. 
This {\bf parallel filtration} quantifies how adding or removing a locus affects the epistasis of the individual sectors of the high-dimensional network (see Box \ref{box:example+parallelfiltration}). 
Discovering loci that have outsized effects on their network allows a new approach to identify master regulators that operate in higher dimensions.

%\pagebreak

\begin{boxm}{Conceptual introduction to epistatic filtrations.}\label{box:conceptual+introduction}

  \noindent
  
  \includegraphics[height=10cm]{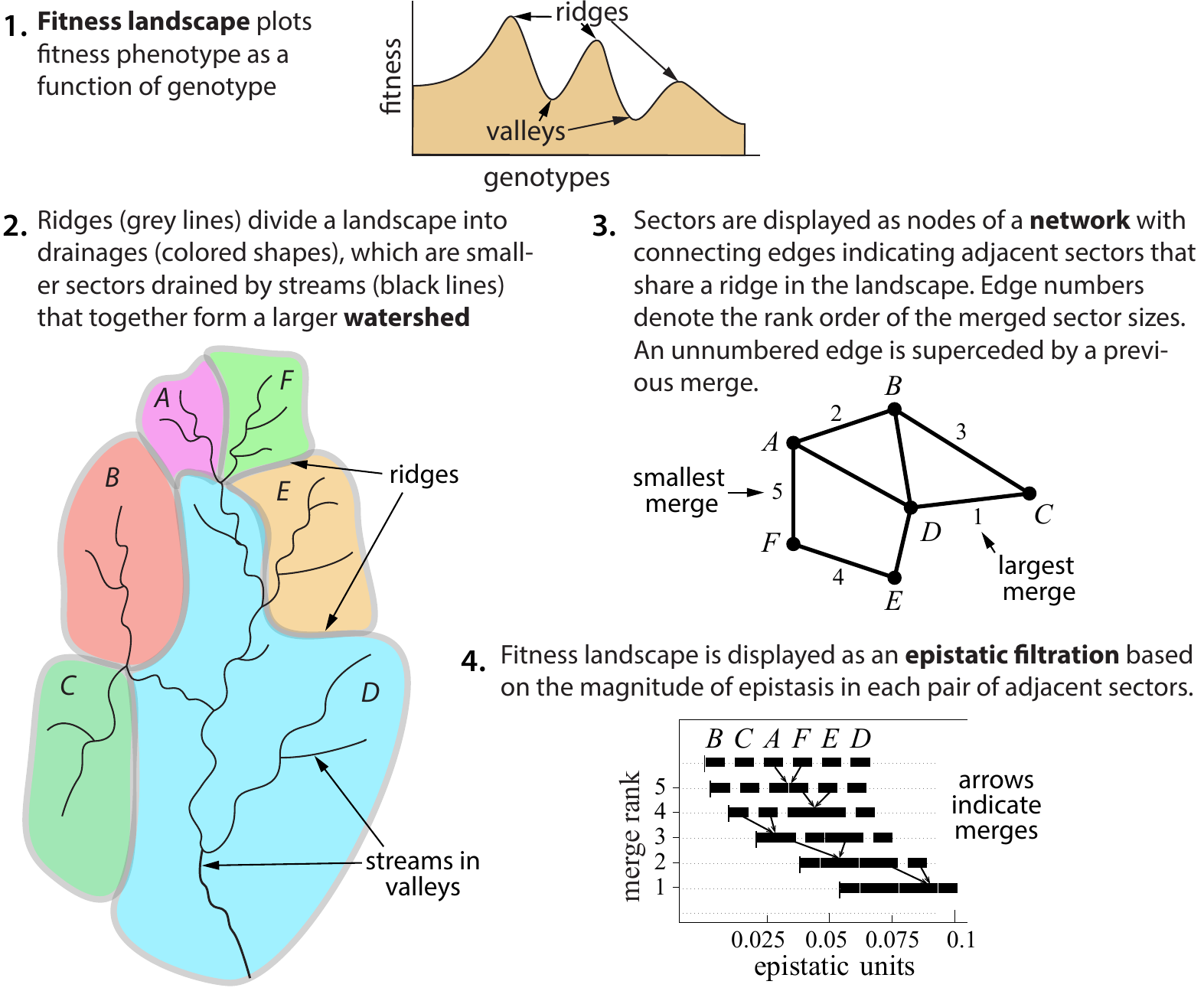}\hspace{0pt}
  \small
 
  \noindent
  An epistatic filtration depicts the epistasis of a fitness landscape. By analogy with a watershed, producing the filtration can be conceptualized in four steps: ({\bf a}) the fitness landscape defines topography; ({\bf b}) the landscape is segmented into sectors based on the topography; ({\bf c}) epistasis is calculated as the shared area of adjacent sectors and displayed on a graph that depicts the adjacency relationships of sectors; ({\bf d}) the epistatic filtration depicts the rank order of epistasis magnitude in the adjacent sectors as a set of merges. Formal definitions follow in Box \ref{box:definition+epistatic+filtration}, Box \ref{box:example+epistatic+filtration}, and text.
  \normalsize
\end{boxm}

\subsection{A volume-based definition of epistasis is valid across many dimensions} 

In this section, we explain the definition of epistasis that we employ throughout. We start by explaining the $2D$ genotype case. With two loci and two alleles (0 or 1) at each locus, we plot the genotypes as a unit square in the x-y plane and the measured phenotypes of each genotype on the z-axis (Box \ref{box:definition+epistatic+filtration}a). The phenotypes thus \emph{lift} the genotypes into one higher dimension, here going from $2D$ to $3D$.
Connecting the four phenotypes gives a simplex, shown as the green polytope in Box \ref{box:definition+epistatic+filtration}a. 
Depending on the relative magnitudes of the phenotypes, the green polytope can be larger or smaller, with the perfectly additive (no epistasis) case giving zero volume (Box \ref{box:definition+epistatic+filtration}a inset). 
We define epistasis as the euclidean volume of the green polytope, which in $2D$ is proportional to the absolute value of the established formula for epistasis, $\epsilon=h(00)+h(11)-(h(10)+h(01))$ \cite{BPS:2007}. 
We call our definition the \emph{epistatic volume} and note that it is of one dimension higher than the genotype space due to the measured phenotype (Box \ref{box:definition+epistatic+filtration}). 
This definition of epistasis based on volume is important because it applies equally well in higher dimensions (Box \ref{box:definition+epistatic+filtration}a,b; \ref{primer+epistatic+filtrations}), as we discuss in the next section. 

\pagebreak

\begin{boxm}{Definition of epistatic filtrations for a genotype space with two loci.}\label{box:definition+epistatic+filtration}

\noindent
\includegraphics[width=15cm]{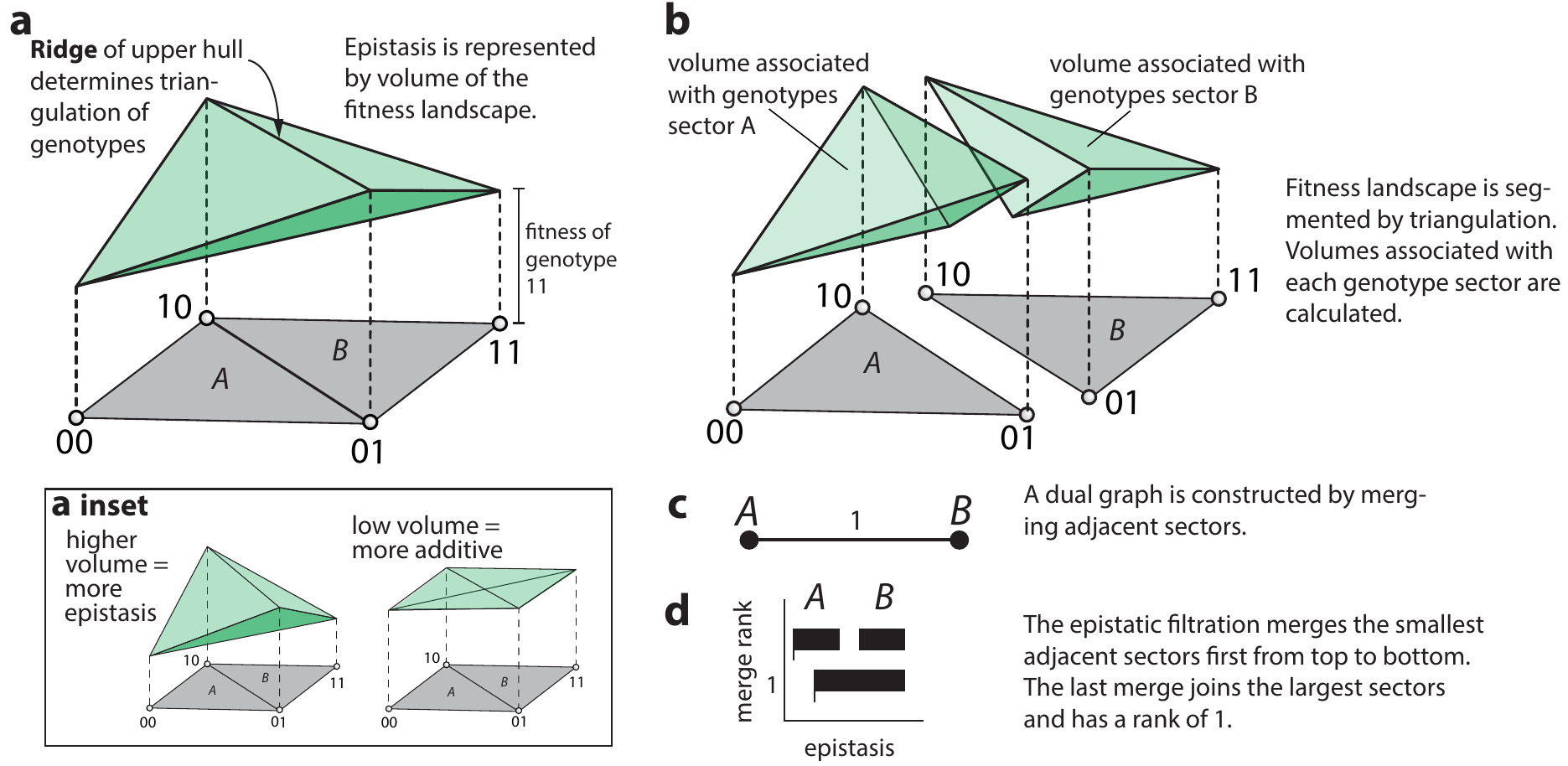}\hspace{0pt}

\small
\noindent

({\bf a}) The biallelic, $2$D genotype set has two loci, each of which can be 0 or 1: $\{00,01,10,11\}$. Each genotype gets \emph{lifted} into $3$D space by appending the phenotype $h(v)$ to each genotype coordinate in the set, $v\in\{(00),(01),(10),(11)\}\subset\RR^2$. Connecting these lifted phenotype points forms a \emph{convex hull}, depicted as the green $3$D body $G^{(3)}$ above the grey genotype set. The upper surface of the green body is two green triangles, which are divided by the {\bf ridge}. The euclidean  volume of the $3$D body $G^{(3)}$ yields a measure for epistasis (c.f. \cite{Khan1193}). \emph{Inset:} A higher degree of epistasis produces a larger volume, and lower epistasis produces a lower volume of the green body. ({\bf b})~The ridge sets a triangulation of the genotype space in grey (a.k.a.\ \emph{genotope} \cite{BPS:2007}). This is done by removing the phenotype dimension from the ridge vertices, which projects it back to the $2D$ genotype space. The ridge thus splits the space into sectors, which are two adjacent triangles,  $\{00,01,10\}$ and $\{01,10,11\}$, denoted as \emph{A} and \emph{B}. We note that the euclidean volume of $G^{(3)}$ equals the absolute value of the established formula $\epsilon=h(00)+h(11)-(h(10)+h(01))$ for epistasis in the two-dimensional case, scaled by a dimension related constant factor. ({\bf c}) The dual graph connecting the adjacent triangles \emph{A} and \emph{B} is trivial in $2D$ as is the ({\bf d}) epistatic filtration. Generalizing to higher dimensions, the triangles become simplices. These are explained further in Box~\ref{box:example+epistatic+filtration} for the $3D$ case. 

\normalsize
\end{boxm}

%\subsection{Epistatic filtrations: The three-loci case} 

\pagebreak

% Box 3
\begin{boxm}{Example epistatic filtration for three loci.}\label{box:example+epistatic+filtration}

\noindent
\includegraphics[height=10cm]{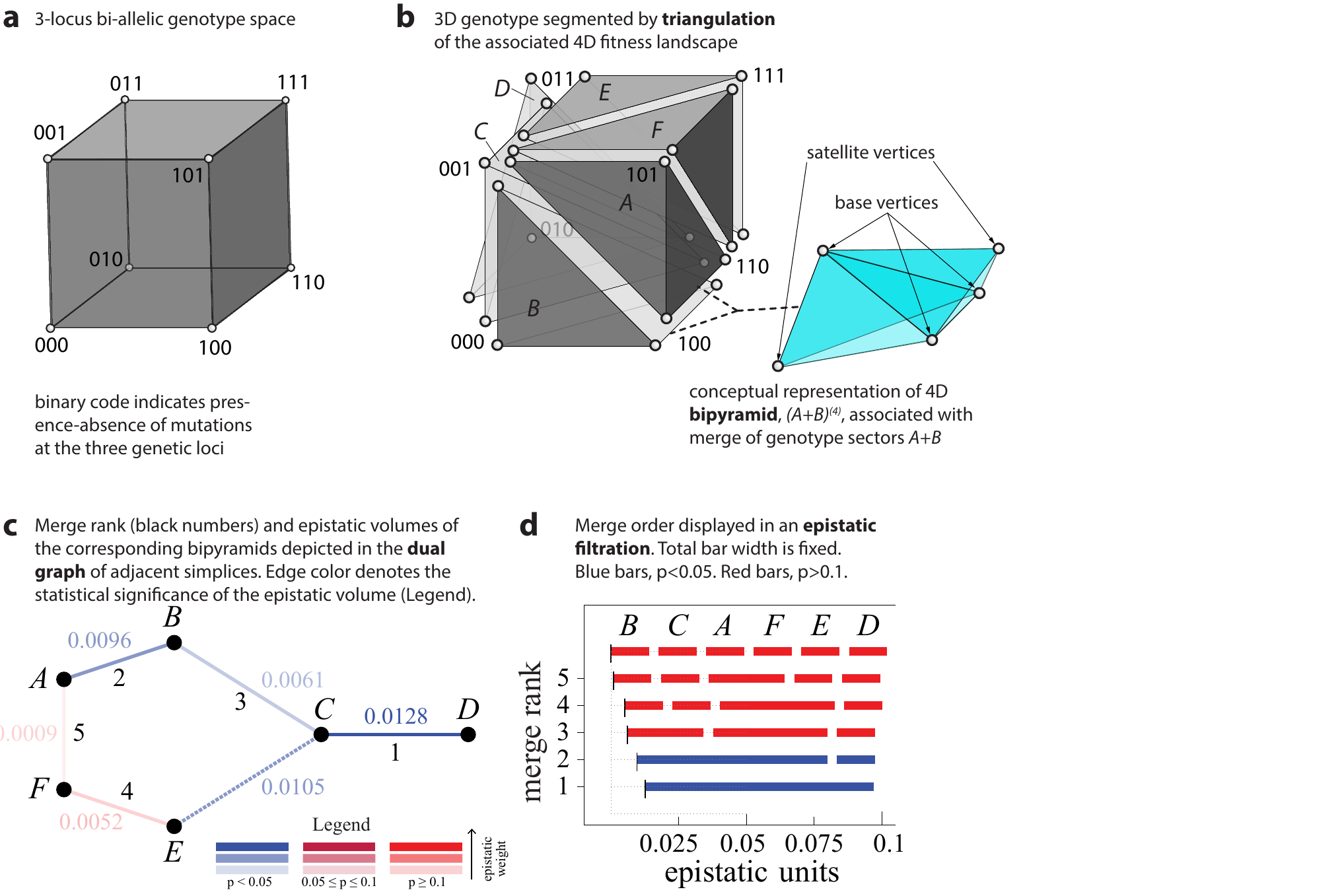}\hspace{0pt}

\small
\noindent
({\bf a}) The $3$D genotype set forms a cube, and, as before, mapping the phenotypes onto the genotypes, $h(v)$, adds an extra dimension. The convex hull of the phenotypes, $h(v)$, forms a convex body $G^{(4)}$ in dimension $4$, which yields ridges (see Box~\ref{box:definition+epistatic+filtration}). 
({\bf b}) The ridges produce a {\bf regular triangulation}, $\subdivision$, which consists of the six tetrahedra, A, B, C, D, E and F. 
%These tetrahedra represent fittest populations as explained in \cite{Eble2019}. 
Epistasis is calculated from the union of adjacent tetrahedra, which form a convex body in $4$D, cartooned in blue. The blue is called a {\bf bipyramid} because it is comprised of two neighboring tetrahedra that share a face. The vertices of the \emph{shared} face are called {\bf base} vertices. The unshared vertices of the two tetrahedra are called {\bf satellites}.
({\bf c}) The adjacency relations of the tetrahedra give rise to a network, which is the {\bf dual graph} of $\subdivision$. 
In this graph, for instance, the edge $(A,F)$ refers to the {\bf bipyramid} comprised of $A$ and $F$ with vertices
\inlineequation[eq:bipyramid_AF3d]{\{010\} + \{011\, ,\, 110\, ,\, 001\} + \{111\}}.
The set $\{011,110,001\}$ is the base where $A$ and $F$ meet, and it separates the two satellites $010$ and $111$. 
Analogous to the two-loci case, appending the $h(v)$ phenotypes to the genotypes in (\ref{eq:bipyramid_AF3d}) yields a $4$D simplex $(A,F)^{(4)}$. The volume of $(A,F)^{(4)}$ is the {\bf epistatic weight} $e_h(A,F)$ (see Appendix \ref{primer+epistatic+filtrations}. 
Color of edges indicates statistical significance (Legend; see Appendix for method; \cite{Eble2019}).
({\bf d}) The {\bf epistatic filtration} of the genotype-phenotype map depicts the iterative process of glueing bipyramids in a non-redundant manner, going from lowest to highest epistatic weight. For example, rank 5 is the merge between $A$ and $F$ and has the lowest epistasis, rank 4 is the merge between $E$ and $F$, and so forth. The black vertical tick mark at the left end of each row of blocks gives the epistasis added to the filtration at that rank.
({\bf e}) The epistatic filtration is analogous to merging drainage sectors in a watershed.
%In this way, a concise description of global epistasis is obtained. 
\normalsize
\end{boxm}

\subsection{Epistatic filtrations: The $n$-loci case} 
In the $n$-loci case, the genotype set is given by $\{0,1\}^n$, i.e. every genotype is encoded as a bitstring of length $n$, and the genotype-phenotype assignment $h$ is a map $h\colon \{0,1\}^n\to \RR$, meaning each vertex $v$ in the hypercube of genotype space has an associated phenotype $h(v)$.
This is shown in Box~\ref{box:definition+epistatic+filtration} and Box~\ref{box:example+epistatic+filtration} which visualize the two smallest cases $n=2$ and $n=3$, respectively.
As in these lower dimensional cases, the lifted convex body $G^{(n+1)}\subset \RR^{n+1}$ is given by the convex hull of the lifted points $(v,h(v))$ for genotypes $v\in\{0,1\}^n$.  The  upper hull of $G^{(n+1)}$ consists of many facets and, as before, removing the phenotype coordinate, $h(v)$, from the vertices of the ridges (see Box \ref{box:definition+epistatic+filtration}a) yields the regular triangulation $\subdivision(h)$ of the genotype space.
%This {\bf fitness}, which is characterized mathematically in terms of a linear program, is maximized at the upper convex hull, cf.\ \cite[\S2.2]{Eble2019}.
Every sector $s$ of $\subdivision(h)$ is an $n$-dimensional simplex and, as such, it is spanned by $n+1$ vertices $v^{(1)},\ldots, v^{(n+1)}\in\{0,1\}^n$, cf. Box \ref{box:example+epistatic+filtration}b). Given another simplex $t$ of $\subdivision(h)$, the pair $(s,t)$ describes a bipyramid if the two are adjacent, which is true when $t$ is spanned by vertices $v^{(2)},\ldots, v^{(n+2)}\in\{0,1\}^n$. We use the notation 
\begin{equation}
  \{v^{(1)}\} + \{v^{(2)},\dots,v^{(n+1)}\} + \{v^{(n+2)}\} 
\end{equation}
for the bipyramid $(s,t)$ in order to emphasize its satellite vertices $v^{(1)}$ and $v^{(n+2)}$. As before, the lifted bipyramid $(s,t)^{(n+1)}\subset\RR^{n+1}$ is the convex hull of the points $(v^{(i)}, h(v^{(i)}))$ for $1\leq i\leq n+2$ and the epistatic weight $e_h(s,t)$ of the bipyramid $(s,t)$, defined in equation (\ref{eq:epistatic-weight}) of Appendix \ref{primer+epistatic+filtrations}, can be seen as a variant of the euclidean volume of the lifted bipyramid $(s,t)^{(n+1)}$. 
Since that volume is non-negative, there are only two cases.
Either $e_h(s,t)=0$, which signals perfect additivity.
Or we have $e_h(s,t)>0$, which means that $G^{(n+1)}$ breaks at the ridge $\{v^{(2)},\ldots, v^{(n+1)}\}$.
In that case the phenotype of the satellite $v^{(1)}$ lies below the expected value, assuming that $h$ extends additively from the simplex $t$ to the whole bipyramid $(s,t)=\{v^{(1)}\} + t$.
A similar statement applies for the other satellite $v^{(n+2)}$.
In this case, the $n+2$ genotypes of the bipyramid $(s,t)$ form an {\bf epistatic interaction}, and the value $e_h(s,t)$ measures its strength.

Visualizing an $n$-dimensional polytope can be non-intuitive, but as for the $3$-dimensional case, we can visualize the topography of the {\bf epistatic landscape} by forming the {\bf dual graph} of the triangulation $\subdivision(h)$, where the nodes are $n$-dimensional simplices and the edged are bipyramids formed by adjacent simplices. We then calculate the volume of each bipyramid to determine the epistasis. 
We rank the bipyramids by their epistasis and depict the order with what we call an epistatic filtration.  

As in lower dimensions, this visualization of a fitness landscape, ranked by epistasis, can be thought of intuitively like a watershed. Ridges enclose sectors that are iteratively merged with progressively larger sectors to form the entire landscape. Epistatic filtrations break apart a high-dimensional fitness landscape into sectors using a triangulation to define the ridges. In higher dimensions, the sectors are $n$-dimensional simplices. The dimensionality of the simplices is the dimensionality of the fitness landscape. Epistasis within these sectors is calculated using the full dimensionality. A statistical test determines significance of each epistatic interaction. The epistatic filtration of the fitness landscape depicts the path from smallest to largest epistasis by merging adjacent simplices to form connected clusters. Therefore, this is not a dimensional reduction but rather an approach that allows a global view of epistasis on a fitness landscape in higher dimensions.  
	This process rests on the mathematical theory of linear optimization, convex polyhedra, and regular subdivisions \cite{Triangulations,Eble2019}.

It is often useful to restrict the analysis to subsystems which are characterized by assuming the presence or absence of specific genes.
These subsystems correspond to {\bf faces} of the fitness cube $[0,1]^n$, which are cubes of lower dimensions.
We denote these faces as a string of zeros, ones and stars.
For instance, ${0}{*}{*}{*}{*}$ in Fig.~\ref{fig:Eble-0ssss_khanraw} is the $4$-loci subsystem where the first gene is wildtype, and only mutations among the remaining four loci are studied.
The analysis applies to such subsystems by restricting the genotype-phenotype map, which is important in our approach for identifying master regulators, as discussed later.

%This method has many advantages over parameter fitting, including that it does not depend on the statistical constraints of determining a best fit.  Filtrations are also not constrained by the sign of epistasis, which depends on which genotype is considered \emph{wildtype}, a somewhat arbitrary decision given varied ancestries (see Appendix B1). 

%\subsection{Epistatic weights and bipyramids}\label{Section:bipyramids}
%This is the face 0**** of the khan raw data
\subsection{Epistatic filtrations reveal higher-order structure in \emph{E. coli} evolution} 

% STOPPED HERE 3 April 2022

\begin{figure}[!h]\centering
	\hspace*{\fill}
  \begin{subfigure}[t]{1\textwidth}		\vspace{0pt}		
    \centering
    \includegraphics[width=1\textwidth]{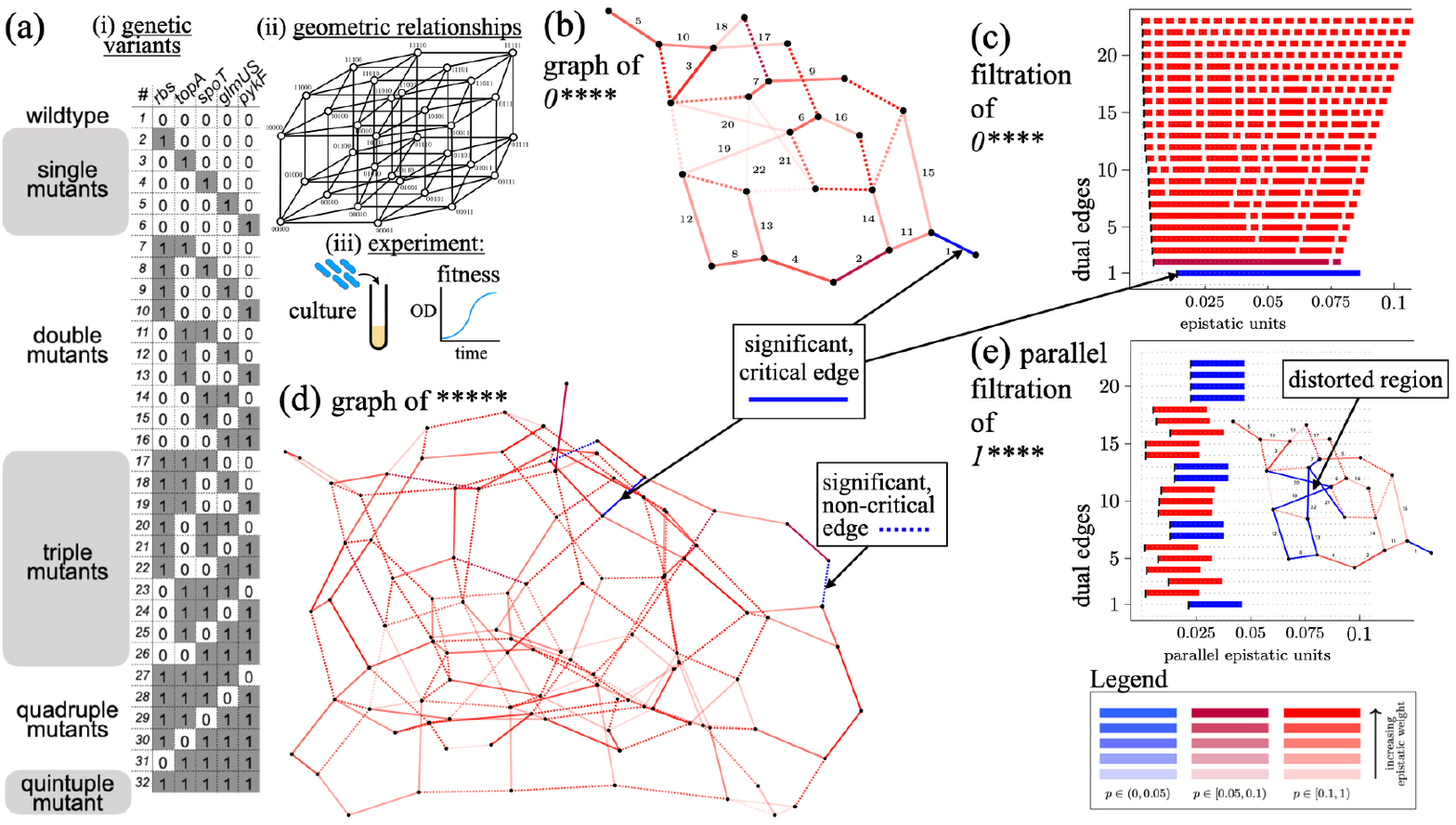}\hspace{0pt}
  \end{subfigure}

  \caption[\emph{E. coli} evolution is guided by epistatic landscape distortions.]{{\bf \emph{E. coli} evolution is guided by epistatic landscape distortions.}
    (a) (\emph{i}) \emph{E. coli} mutants examined \cite{Khan1193}, (\emph{ii}) their geometric relationships, and (\emph{iii}) experimental approach to measure fitness. (b) Edge labeled dual graph and (c) epistatic filtration restricted to $n=4$ mutations in \emph{topA} (locus 2), \emph{spoT} (locus 3), \emph{glmUS} (locus 4) and \emph{pykF} (locus 5). Locus 1, \emph{rbs}, is fixed 0 (\emph{wildtype}).
    Note that the left edge of the bars in (c) indicates there is very little epistatic weight added to the filtration except for the final merge, where the single genotype $00001$ gives weight to the entire filtration.
    This final interaction corresponds to the vertices $\{00001\} + \{00000, 01001, 00101, 00011\} + \{00010\}$.
    (d) Dual graph for the complete Khan data set. 
    %The edge degree distribution in the format (degree, number of maximal cells incident with that number of dual edges) is $((1,1),(2,11),(3,28), (4,35),(5,21),(6,5))$.
    Black indices in (b) label the critical dual edges of ~$\subdivision(h)$. (e) In the parallel filtration, for ${1}{*}{*}{*}{*}$, where the \emph{rbs} mutation is present, the landscape is disorted by a concentrated area of higher epistasis. Inset: graph in (b) recolored with weights from (e). The lengths of the bars in the parallel transport figure (e) have no meaning. Only the horizontal position of the black marks, the vertical position of the bars and its coloring encode information. The horizontal shift represents the value of the epistatic weight, the vertical position of the bar indicates which dual edge is transported and the color expresses if the epistatic weight is significant after parallel transport.}
  \label{fig:Eble-0ssss_khanraw}
\end{figure}

To illustrate our approach, we examined an existing data set from Lenski's~\cite{Lenski_NatGenetReview} classic experimental evolution of \emph{Esherichia coli}, in a set of strains with each combination of five beneficial mutations \cite{Khan1193} (Fig.~\ref{fig:Eble-0ssss_khanraw}a). 
We first examine $n=3$ loci, corresponding to biallelic mutations in \emph{topA}, \emph{spoT}, and \emph{pykF}. 
Epistasis was generally low in magnitude \cite{Khan1193,Sailer2017}, and occurs in two ways: (\emph{i}) either from merging groups of groups of simplices (c.f.\ BC + AFE in line \#2 of Box \ref{box:example+epistatic+filtration}e, or (\emph{ii}) from merging a single simplex, c.f.\ D, with the aggregated rest of the simplices (c.f. line \#1 of Box \ref{box:example+epistatic+filtration}e, much like a dominant effect in the NK model \cite{Kauffman1987}. 
This second way is consistent with a fitness landscape distortion, which occurs when certain mutations influence the interactions of many other genes \cite{Ito2020}. 
Geometrically, such a distortion constitutes a vertex split \cite{HerrmannJoswig:2008}.
We next add a fourth biallelic mutation, in the \emph{glmUS} locus (Fig.~\ref{fig:Eble-0ssss_khanraw}b,c), encoding peptidoglycan availability, which is an essential component of the cell wall.

The filtration reveals a smooth, additive landscape with one dominant cell where epistasis arises only in the final merge of the filtration (Fig.~\ref{fig:Eble-0ssss_khanraw}c),
meaning the epistatic topography of the entire landscape (Fig.~\ref{fig:Eble-0ssss_khanraw}d) rests upon the single vertex, $00001$, \emph{pykF}. 
While the previous analysis detected a significant, marginal effect of \emph{pykF} \cite{Khan1193}, filtrations reveal the geometric structure in terms of which specific combinations of loci are responsible for the effect (Fig.~\ref{fig:Eble-0ssss_khanraw}e): we establish an interaction between  \emph{glmUS}, $\{00001\}$, and  \emph{pykF}, $\{00010\}$. The interaction depends on the genotypes  $\{00000, 01001, 00101, 00011\}$ in the bipyramid base. 
%is needed as only the genotypes $\{00000, 01001, 00101, 00011\},\{00001\},\{00010\}$ together span a bipyramid with significant non-zero epistatic weight. 
%Thus, this $4$D interaction in the 4-locus system $0****$ involves all the genotypes of $\{00001\},\{00010\},\{00000, 01001, 00101, 00011\}$ and that this interaction cannot be seen by considering expressions involving fewer genotypes (Fig.~\ref{fig:Eble-0ssss_khanraw}c), or a locus system with fewer mutable loci.
Interestingly, the four loci context involves genotypes with the wild type and only up to double mutants. 
%This shows an instance of how higher order epistasis can arise from lower dimensional interactions. 
%Indeed, the order of the epistatic interaction is at least four, as it can first be seen in the 4-locus system ${0}{*}{*}{*}{*}$. 
But these double mutants must be present together to yield a higher dimensional interaction. 
This conclusion is consistent with recent genome-wide work on trans-gene interactions~\cite{Liu2019}, suggesting that complex traits may arise from genome-wide epistasis, where each mutation's contribution to the trait depends on the presence of other mutations. 
Additionally, we observe that the interaction of $\{00001\},\{00000, 01001, 00101, 00011\},\{00010\}$
in the $4D$ case (with the first locus wildtype) remains significant in the full 5-locus setting, ${*}{*}{*}{*}{*}$, see the blue critical edge in the dual graph of Fig.~\ref{fig:Eble-0ssss_khanraw}d), indicating an interaction in lower dimensions that is unaffected when a mutation is introduced in the first locus. 
%This yields another interpretation of how higher order epistasis can originate from low order epistasis. To see this, notice that the first epistasis has order 5 since it appears in a 5 locus system, while the lower order epistasis has order 4 since it originates in the 4 locus subsystem ${0}{*}{*}{*}{*}$.

\subsection{Parallel epistatic filtrations reveal master regulators in \emph{E. coli} evolution}

To discern the role of each locus on the $4$D network structure, we applied {\bf parallel filtrations} \cite[\S6.6]{Eble2019}. This technique measures context-dependence in the fitness landscape by assessing changes in the epistasis of sectors that occur when a particular locus is mutated versus wildtype. For example, the epistatic filtration can be calculated for ${0}{*}{*}{*}{*}$, where the first locus is fixed as wildtype and the filtration is performed for the remaining $4$ loci. This yields a set of bipyramids for which the epistasis is calculated. In the parallel filtration, we compare the epistasis for ${0}{*}{*}{*}{*}$ with the epistasis for ${1}{*}{*}{*}{*}$ using the triangulation set by ${0}{*}{*}{*}{*}$ as well as the rank order. In this way, two parallel faces of the $5-$cube are compared (see Box 4 and Fig.~\ref{fig:khan:0ss0s-1ss0s}).
Parallel filtrations extend the concepts of conditional, marginal, and sign epistasis \cite{Gill1965,Weinreich2005} into the epistatic filtrations context.

\newpage
% Box 4
\begin{boxm}{Parallel epistatic filtration for three loci when a 4th locus is modified.}\label{box:example+parallelfiltration}

\noindent
\includegraphics[height=10cm]{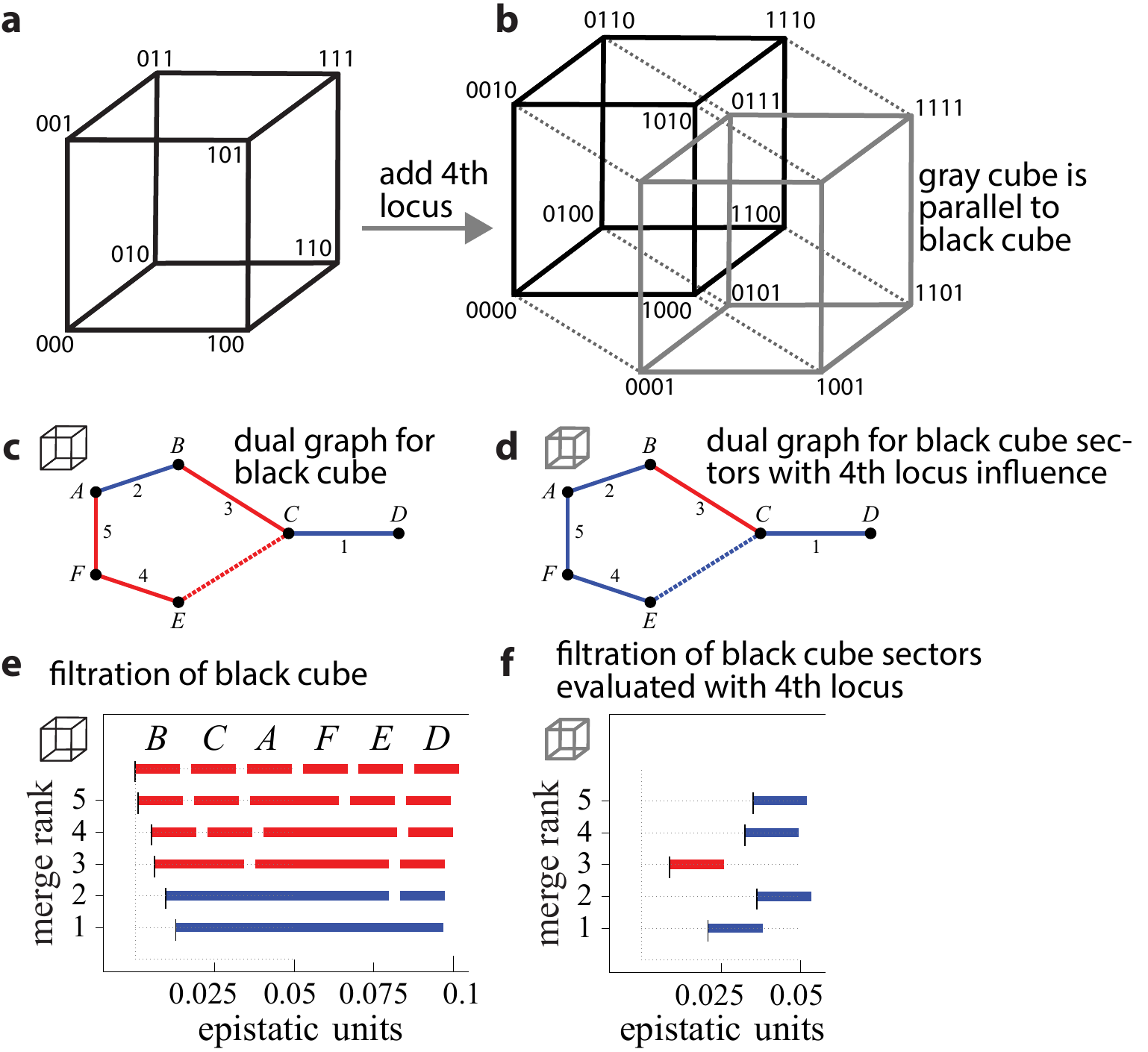}\hspace{0pt}

\small
\noindent
({\bf a}) The $3$D genotype space. 
({\bf b}) Adding a locus produces a $4$D genotype space that can be visualized as two parallel $3$D genotype spaces, depicted in black and grey, where the grey genotype space has a mutation in the 4th locus and the black is wildtype at the 4th locus. 
({\bf c}) The dual graph of $\subdivision$ for the black genotype space. 
({\bf d}) The parallel dual graph for the grey genotype space. Note several edges in {\bf c} (black cube) shift to significant in {\bf d} (grey cube), indicating the context of the 4th locus influences the interactions.
({\bf e}) The {\bf epistatic filtration} of the black genotype space.
({\bf f}) The {\bf parallel filtration} calculates epistasis of the black genotype sectors with the phenotypes of the parallel cube (i.e. when the 4th locus is present). This approach measures the influence of the 4th locus on the rest of the epistatic interactions in the network. Specifically, note the shift in the x-values of the black vertical tick marks on the left sides of the left-most colored bars in {\bf e} versus the corresponding tick mark and bar in {\bf f}. 
\normalsize
\end{boxm}

    Examining the Khan data with and without the \emph{pykF} mutation \cite{Khan1193} (Fig.~\ref{fig:khan_parallel_ssss0_ssss1}) showed increased significance in 8 out of 22 of the dual edges, when \emph{pykF} was mutated. Each bipyramid in Fig.~\ref{fig:khan_parallel_ssss0_ssss1}e) matches a bipyramid in Fig.~\ref{fig:khan_parallel_ssss0_ssss1}c) via the parallel transport operation \cite{Eble2019}. In particular, both filtrations have 22 dual edges. 
%    The very top of Fig.~\ref{fig:khan_parallel_ssss0_ssss1}c) is not counted as a dual edge as it is a mere list of the simplices involved in the merging operation, but does not represent a dual edge.
  
    The biological interpretation of the parallel transport operation is simple. It changes the context in which the epistatic weights associated to the dual edges are measured. For Fig.~\ref{fig:Eble-0ssss_khanraw}e) this means that epistatic weights in the genotype system with wildtype \emph{rbs} are different when \emph{rbs} is mutated. 
    Since this locus is fixed in the parallel transport operation, comparing the wildtype and mutant, we call this locus the bystander.
    Here, changing the bystander state modifies the magnitude and significance status of the epistatic weights (Fig.~\ref{fig:Eble-0ssss_khanraw}c,e), with epistatic weights generally higher when \emph{rbs} is mutated. 
    %Additionally, epistatic weights tend to become significant after changing the status of the bystander \emph{rbs} from wildtype to mutated: there are 9 blue dual edges in Fig.~\ref{fig:Eble-0ssss_khanraw}e), as opposed to    Fig.~\ref{fig:Eble-0ssss_khanraw}c) where there is only 1 blue dual edge. 
    Thus mutating the \emph{rbs} locus distorts the fitness landscape. We note that the precise locations of the distortions are concentrated as a set of adjacent blue edges in the dual graph (Fig. \ref{fig:Eble-0ssss_khanraw}e Inset). 
    Examining the restoration of \emph{pykF} to wildtype (Fig.~\ref{fig:khan_parallel_ssss1_ssss0}), only 3 of 22 edges changed significance and just one critical edge lost significance, emphasizing the importance of context in the fitness landscape.
%	Applying our methods to previous data on the $\beta$-lactamase enzyme \cite{PhysRevLett.106.198102}, we found similar context-dependence and directionality (Appendix B7).
	Filtrations thus provide a new perspective on how genes regulate biological network structure in higher dimensions. 
%	More data sets are discussed below and more examples of possible biological interpretations are provided. 

%This is Eble_norm_0ssss.
\subsection{Lactobacilli produce microbiome distortions}
 
Up to this point, we have focused on genetic epistasis, but our framework is equally valid for interactions of environmental parameters, including bacterial species in the gut microbiome. 
Like the genome, which is composed of many genes that interact to determine organismal fitness, the microbiome is also composed of many smaller units, i.e. bacterial species, that affect host fitness. 
Hosts are known to select and maintain a certain core set of microbes \cite{Ley2008,Risely2020}; the interactions of these bacteria can affect host fitness~\cite{Gould232959}; and it is debated to what extent these interactions are of higher order, cf. \cite{Friedman2017}. See also
\cite{Krug:2021EpistasisandEvolution} for a broad overview on papers elaborating on possible meanings and instances of higher-order epistasis.
While vertebrates have a gut taxonomic diversity of $\approx 1000$ species, precluding study of all possible combinations, the laboratory fruit fly, \emph{Drosophila melanogaster}, has naturally low diversity of $\approx 5$ stably associated species \cite{5}. 

\begin{figure}[!h]
  \includegraphics[width=1\textwidth]{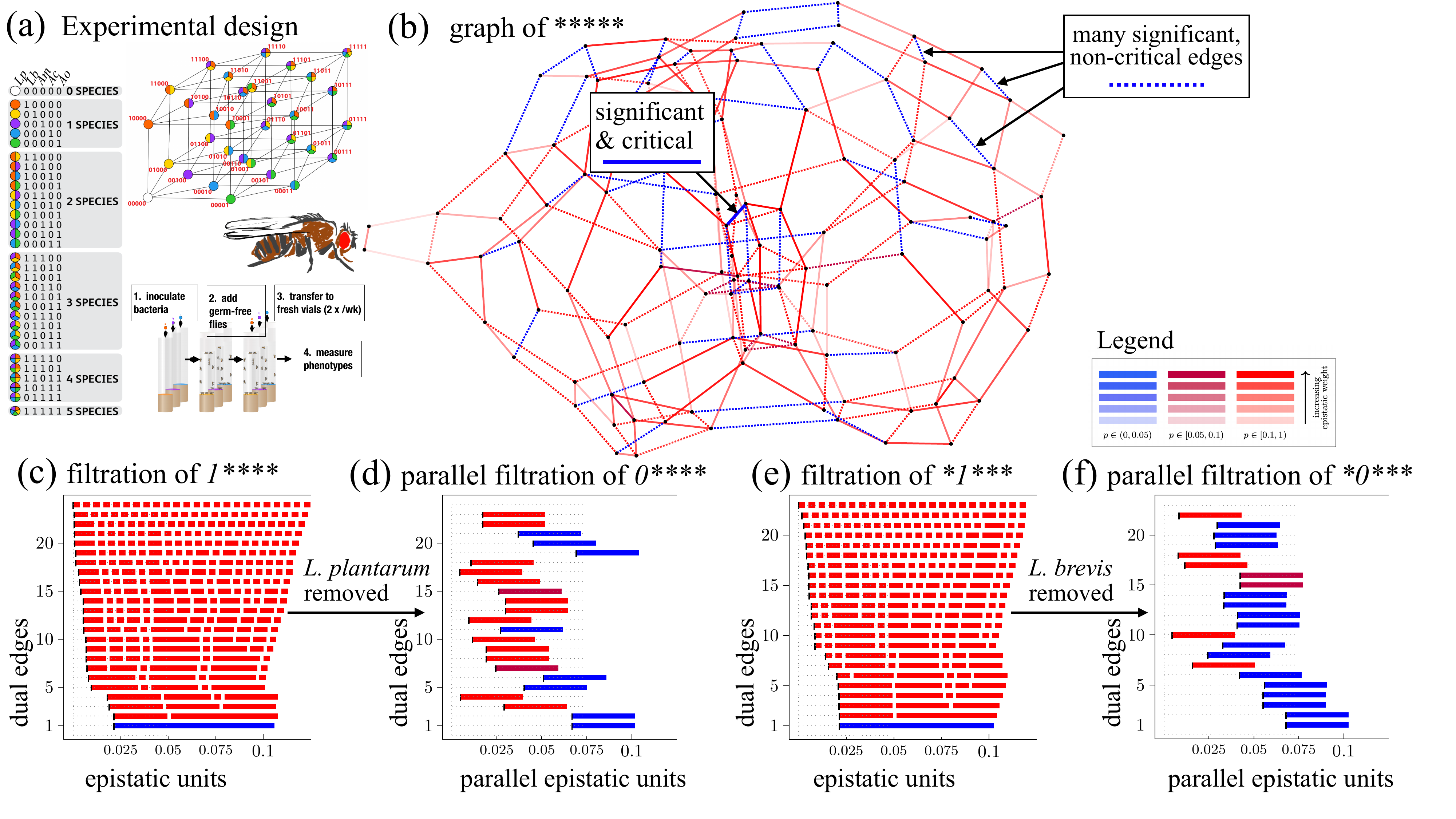}\hspace{0pt}

 \caption[Loss of lactobacilli causes global distortion of the microbiome epistastic landscape.]{{\bf Loss of lactobacilli causes global distortion of the microbiome epistastic landscape.} 
(a) Experimental design for Eble and Gould \cite{Gould232959} microbiome manipulations in flies. 
(b) Full graph of ${\ast}{\ast}{\ast}{\ast}{\ast}$ for the Eble data. 
 %The dual edge degree distribution in the format (degree, number of maximal cells incident with that number of dual edges) is $((1,0),(2,5),(3,30),(4,45),(5,36),(6,0))$. 
(c) Filtration of $\subdivision(h)$ for the 4-face, ${1}{\ast}{\ast}{\ast}{\ast}$, of Eble data, where \emph{L.~plantarum} is present, indicates epistasis where two clusters of maximal cells merge. 
(d) Parallel filtration with \emph{L.~plantarum} removed shows a landscape distortion. 
(e) Filtration for ${\ast}{1}{\ast}{\ast}{\ast}$, where \emph{L.~brevis} is present has similar structure to ${1}{\ast}{\ast}{\ast}{\ast}$. 
(f) Parallel filtration with \emph{L.~brevis} removed shows a landscape distortion. }

  \label{fig:Eble-0ssss}
\end{figure}

We made gnotobiotic flies inoculated with each combination of a set of $n=5$ bacteria ($2^5=32$ combinations) that were isolated from a single wild-caught \emph{D. melanogaster}, consisting of two members of the \emph{Lactobacillus} genus (\emph{L.~plantarum} and \emph{L.~brevis}) and three members of the \emph{Acetobacter} genus (Fig.~\ref{fig:Eble-0ssss}a). 
We measured fly lifespan, which we previously identified as a reproducible phenotype that is changed by the microbiome~\cite{Gould232959}.
Overall a reduction of microbial diversity (number of species) led to an increase in fly lifespan as with a taxonomically similar set of bacteria we examined previously, which came from multiple hosts \cite{Gould232959}. 

\goodbreak

The dual graph for the $5$-loci genotype space revealed a single significant and critical epistatic interaction (Fig. \ref{fig:Eble-0ssss}b). Abundant non-critical edges were distributed throughout the graph (Fig. \ref{fig:Eble-0ssss}c) indicating prevalent interactions that weakly affect the fitness landscape. We note that such interactions were absent from the \emph{E. coli} fitness landscape (compare the number of blue edges in \ref{fig:Eble-0ssss}b versus Fig.~\ref{fig:Eble-0ssss_khanraw}d).
%contrasting starkly with the Khan data (Fig.~\ref{fig:Khan-5cube}). 
%Examining the filtration (Fig.~\ref{fig:Eble-0ssss}d), the epistatic weight (i.e.\ magnitude) for the microbiome data generated $\approx 5\%$ effect, roughly three times the weight in the Khan data and half that in the Tan $\beta$-lactamase landscapes \cite{PhysRevLett.106.198102} (cf. $x$-axis between Fig.~\ref{fig:Eble-0ssss}, \ref{fig:Khan-5cube}, \ref{fig:weinreich:0ssss-1ssss}), indicating that the rugosity of microbiome interactions is comparable to genetic ones.
Using parallel filtrations to measure the role of individual bacterial species on the overall network, we found that the \emph{Lactobacilli} drive changes in the global structure (Fig~\ref{fig:Eble-0ssss}d,e). In 46 out of 128 (36\%) interactions, significance changed due to adding or removing a \emph{Lactobacillus} (Fig~\ref{fig:Eble-0ssss}c-f,~\ref{fig:Eble:parallel:0to1****},~\ref{fig:Eble:parallel:*0to1***}). These changes in significance primarily derive from non-significant interactions when \emph{L.~brevis} is present that become significant when it is removed and vice versa, indicating \emph{L.~brevis} suppresses epistatic interactions that affect fly lifespan.    
    
Microbiome abundances could drive the effects on host lifespan, however, comparing the epistatic landscapes for CFUs and lifespan, we found that only 2 of 99 dual edges were significant for both the bacterial abundance and fly lifespan data sets (Fig.~\ref{fig:Gould:parallel:CFUtoTTD:0****},~\ref{fig:Gould:parallel:CFUtoTTD:1****},~\ref{fig:Gould:parallel:CFUtoTTD:*0***},~\ref{fig:Gould:parallel:CFUtoTTD:*1***}, Tables~\ref{tab:Gould:parallel:CFUtoTTD:0****},~\ref{tab:Gould:parallel:CFUtoTTD:1****},~\ref{tab:Gould:parallel:CFUtoTTD:*0***},~\ref{tab:Gould:parallel:CFUtoTTD:*1***}), and there was a lack of correlation between the epistatic weights of the bipyramids (Spearman rank correlations: $p=0.7$, $p=0.5$, $p=0.3$, and $p=0.3$ respectively). This discord between the epistatic landscapes for microbiome fitness and host fitness could e.g. diminish the rate of co-evolution.

\FloatBarrier

%\subsection{Context-dependence generates higher-order epistasis.}  
%    
%    
%    moved to appendix

\subsection{The epistatic landscape within a single enzyme is rugged}

As a point of comparison with the Khan data set, we re-analyzed data from a fully factorial $5$-mutation data set in the $\beta$-lactamase gene, where each mutation is in a separate residue of the same enzyme \cite{PhysRevLett.106.198102,Weinreich2006}. 
We note that the data are discrete (growth/no growth for a given set of antibiotic concentrations), and this type of microbiology experiment does not show variation in general. Thus, we can generally treat the calculated interaction magnitudes as accurate. We therefore discuss the meanings of the magnitudes. Due to a lack of the raw replicate data, our computations are based on the reported mean values, and $p$-values are not calculated. 

The filtration holds a high magnitude of epistasis (Fig.~\ref{fig:weinreich:0ssss-1ssss},~\ref{fig:weinreich:ss0ss-ss1ss}) compared with the Khan data set (Fig.~\ref{fig:Khan-5cube},~\ref{fig:khan_parallel_ssss0_ssss1}). Note that we can directly compare magnitudes ($x$-axis) due to the normalization procedure (see section \ref{Sec:Terminology}.3). 
The epistasis arises in many steps (note slope of filtration adds magnitude in each step; (Fig.~\ref{fig:weinreich:0ssss-1ssss},~\ref{fig:weinreich:ss0ss-ss1ss})), consistent with the low number of possible evolutionary paths observed by Weinreich \cite{Weinreich2006}, and distortions are apparent in the shifted magnitude of epistasis by parallel transport (Fig.~\ref{fig:weinreich:0ssss-1ssss},~\ref{fig:weinreich:ss0ss-ss1ss}). 
The filtration also reveals a tiered structure to the epistasis, cf.\ the largest weight merges two clusters of simplices (Fig.~\ref{fig:weinreich:0ssss-1ssss},~\ref{fig:weinreich:ss0ss-ss1ss}) in contrast to the Khan data set, where epistasis came from one individual simplex on the periphery of the dual graph, indicating a more complex epistatic landscape in the $\beta$-lactamase. 

Comparing the filtrations between the different datasets (Fig.~\ref{fig:Eble-0ssss}d), the epistatic weight (i.e.\ magnitude) for the microbiome data generated $\approx 5\%$ effect, roughly three times the weight in the Khan data and half that in the Tan $\beta$-lactamase landscapes \cite{PhysRevLett.106.198102} (cf. $x$-axis between Fig.~\ref{fig:Eble-0ssss}, \ref{fig:Khan-5cube}, \ref{fig:weinreich:0ssss-1ssss}), indicating comparable interactions.

%To further compare the global effect of context across different datasets, we developed a method to compute epistasis, based on the triangulation of dual landscapes, which we call the epistatic product [Appendix~\emph{\nameref{se:product-model}}] (Fig.~\ref{fig:khan:product:TssTs}, \ref{fig:Weinreich-ss0ss-ss1ss-productmodel}, \ref{fig:Weinreich-0ssss-1ssss-productmodel}, \ref{fig:Khan-sss0s-sss1s-productmodel}, \ref{fig:Eble-0ssss-1ssss-productmodel}, \ref{fig:Eble-s0sss-s0sss-productmodel}). The total epistasis was highest for the $\beta$-lactamase experiment \cite{PhysRevLett.106.198102}, which carries much higher context-dependence than either the microbiome \cite{Gould232959} or \emph{E.~coli} evolution data sets \cite{Khan1193}, indicative of overall high epistasis at the smallest, within enzyme, scale.

\subsection{Interactions are sparse in higher dimensions}
We used epistatic filtrations to systematically evaluate the prevalence of higher-order interactions as a function of the number of dimensions. 
Critical, significant, higher-order interactions were less frequent than pairwise interactions ($p<10^{-6}$, \emph{Z-test}) for each of the Khan, Eble, and Gould data sets, with a decreasing probability as a function of the face dimension (Table ~\ref{tab:complexity:gen-micro}). 
This occurs for three primary reasons. 
First, the degrees of freedom increase fast in higher dimensions. 
Second, the probability of selecting a significant interaction from the set of all possible interactions decreases because the total number of interactions increases with increasing dimensions. 
Finally, the absolute number of significant interactions decreases in higher dimensions (Table~\ref{tab:complexity:gen-micro}), meaning they are biologically less prevalent. 
Overall, $\approx 10\%$ of possible dual edges were significant at higher order, with $\approx 1\%$ significant for $n=5$ dimensions (Table~\ref{tab:complexity:gen-micro}), suggesting limits to the dimensions of biological complexity.

% Tab S5
\begin{table}[h!]
  \caption[Prevalence of interactions at different levels of complexity in genetics and microbiome data.]{Prevalence of interactions at different levels of complexity in genetics and microbiome data sets.
    Significant versus all critical dual edges ($p<0.05$).}
\label{tab:complexity:gen-micro}                
\begin{tabular*}{\linewidth}{@{\extracolsep{\fill}}llll@{}}
  \toprule 
  \multicolumn{1}{l}{} & \multicolumn{1}{l}{Dataset:} & \multicolumn{1}{l}{Dataset:} & \multicolumn{1}{l}{Dataset:}\\
  \multicolumn{1}{l}{Interaction dimension} & \multicolumn{1}{l}{Khan} & \multicolumn{1}{l}{Eble} & \multicolumn{1}{l}{Gould}\\
  \midrule
  \multicolumn{1}{l}{2:}& \multicolumn{1}{l}{20/80 (25\%)}& \multicolumn{1}{l}{24/80 (30\%)} & \multicolumn{1}{l}{22/80 (28\%)}\\
  \multicolumn{1}{l}{all higher order:}& \multicolumn{1}{l}{29/508 (5.7\%)}& \multicolumn{1}{l}{58/540 (10\%)}& \multicolumn{1}{l}{21/520 (4.0\%)}\\
  \multicolumn{1}{l}{\strut\quad 3:}& \multicolumn{1}{l}{21/194 (11\%)}& \multicolumn{1}{l}{35/199 (17\%)}& \multicolumn{1}{l}{14/194 (7.2\%)}\\
  \multicolumn{1}{l}{\strut\quad 4:}& \multicolumn{1}{l}{7/214 (3.2\%)}& \multicolumn{1}{l}{22/226 (10\%)}& \multicolumn{1}{l}{6/216 (2.7\%)}\\
  \multicolumn{1}{l}{\strut\quad 5:}& \multicolumn{1}{l}{1/100 (1.0\%)}& \multicolumn{1}{l}{1/115 (0.8\%)}& \multicolumn{1}{l}{1/110 (0.9\%)}\\
  \multicolumn{1}{l}{total:}& \multicolumn{1}{l}{49/588 (8.3\%)}& \multicolumn{1}{l}{82/620 (13\%)} & \multicolumn{1}{l}{43/600 (7.1\%)}\\
  \bottomrule
\end{tabular*}
\end{table}

\FloatBarrier

The epistatic filtration of the Eble microbiome data in (Fig.~\ref{fig:Eble-0ssss}) has a much richer texture than the epistatic filtration of the Khan data set.

For instance, in the Eble microbiome data there are two top 4-dimensional epistatic weights which greatly impact the topograpy of the fitness landscape, in the following sense. 
The two epistatic weights are
\[
\begin{array}{clr}
\{01001\} + \{00000, 01000, 01101, 01111\} + \{01100\}  &  0.0451 & \#2\\ % 0.04507199929
\{01001\} + \{00000, 01000, 01011, 01111\} + \{01110\} &  0.0485 & \#1   % 0.04845372022
\end{array}
\]
here given with their spanning genotypes, magnitude of the interaction, and edge ID number. The edge ID matches the position of the dual edge in the filtration of the left panel in Fig.~\ref{fig:Eble:parallel:0to1****} when counting from down up. The magnitudes of these two
interactions combined have a
$9\%$ effect on fitness (sum of the magnitudes of the epistatic weights) with the largest accounting for
$\simeq 5\%$, indicating a region of the landscape where epistasis is concentrated. 
Proximal to these genotypes are two additional cells with nearly significant epistatic weight:
\[
\begin{array}{cr}
\{01011\} + \{00000, 01001, 00111, 01111\} + \{01101\} & \#8 \\
\{01011\} + \{00000, 01000, 01001, 01111\} + \{01101\}  & \#7
\end{array}
\]
The corresponding dual edges are purple in the left panel in Fig.~\ref{fig:Eble:parallel:0to1****}. 

	The genotypes in the interactions form a cluster relating the interactions between \emph{L.~brevis} and increasing numbers of \emph{Acetobacters}. Because the interaction is detected based on the phenotype of fly lifespan, it suggests there may be interesting cellular and molecular mechanisms to investigate. For instance, the interactions could derive from metabolic crossfeeding between the \emph{Acetobacters}, which produce many co-factors, and \emph{L.~brevis}, which produces lactate, stimulating \emph{Acetobacter} growth \cite{Consuegra2020,Henriques2020,Aranda-Diaz2020}.
Note that the support sets of all four interactions above contain both the wild type $00000$ and $01111$, which are the genotypes with maximum and minimum fitness respectively, indicating that all loci contribute to the higher-dimensional epistatic effect, even ones with low fitness.

\subsection{Higher-order interactions can arise from lower-order interactions}\label{sec:high from low} 

%The notion of a higher-order interaction suggests a system property that emerges when all components are present. 
Lower-order interactions can produce interactions in higher dimensions \cite{Sailer2017}. 
%and our method can detect these as non-linearities of hyperplanes across lifted points and projections into higher dimensions. 
In examining the higher-order epistasis present in our data sets, we noted that the clusters where significant epistatic weights occur are often preceded by clusters with nearly significant epistatic weights in lower dimensions (Fig.~\ref{fig:Khan-5cube}).
These lower dimensional interactions involve fewer genotypes than the higher-order interactions that they set up, meaning that the addition of genotypes pushes nearly significant interactions to significance.

We developed a graphical approach to distinguish these interactions from those that arise \emph{de novo} (Fig.~\ref{fig4}b,c; Appendix B11). More specifically, these graphics are intended to answer
the question of to what extent higher-order epistatic effects are induced by lower dimensional ones or, put in other terms, which lower dimensional epistatic effects maintain significance when embedded into higher dimensions?

In (Fig.~\ref{fig4}b) we exhibit an example for the Eble data set, with 5 loci, where we take the three 4-dimensional faces $0{\ast}{\ast}{\ast}{\ast}$, ${\ast}0{\ast}{\ast}{\ast}$ and ${\ast}{\ast}0{\ast}{\ast}$ into consideration.
For each such face, we computed the corresponding filtration of epistatic weights.
We then repeat this procedure, and display the filtrations for relevant 3-dimensional subspaces  ((Fig.~\ref{fig4}b) second row), and finally filtrations for 2-dimensional subspaces (Fig.~\ref{fig4}b) last row).
The reasoning behind this is similar to what happens in regression-based epistasis calculations, where one can extract a certain portion of a higher dimensional space into lower dimensional spaces.
%Some steps in these filtrations are linked to steps in the filtrations over smaller dimensional spaces.
%These links highlight sources for higher dimensional interaction.  

Performing the same operations on the Gould data, there are over all fewer significant epistatic weights. In this data set, we also observe examples of lower order interactions inducing higher order ones, as explained above, but for which the statistical significance status changes - here, from not significant (red bars) to significant (blue bars) (Fig.~\ref{fig4}c). 
Linking the observed higher-order interactions to their lower-dimensional sources can help design biological experiments into the molecular mechanisms, for instance by designating two interacting bacteria to focus on from a larger community where the higher-order interactions emerge. 

We also observe that several higher order interactions in the Eble, Gould and Khan data could not be attributed to lower-order effects (see (Fig.~\ref{fig4}b,c) as well as Table~\ref{Tab:special+bipyramids}). By this we mean that
the interactions could not be linked to
subsets with four, three, or two loci inside the 5-locus system, regardless of their significance (cf. Fig.~\ref{fig4}c). Thus, some interactions arise only in the higher dimensional context and cannot be discovered or predicted by studying lower-order interactions.

As we noted, the 4-dimensional interaction in the \emph{E. coli} evolution experiment involved loci with two genes (Fig.~\ref{fig:Eble-0ssss_khanraw}), whereas in the microbiome, interactions involved loci with four species, suggesting there may be different types of underlying geometries for the interactions between genes in evolution versus between species in the microbiome (Table~\ref{Tab:special+bipyramids}).

\section{Discussion and Conclusions}

\subsection{New biological findings}
From an evolutionary perspective, the Red Queen hypothesis emphasizes how conflicts with other organisms can drive continuous genetic innovation \cite{VanValen1974}. 
In our analysis of the shapes of fitness landscapes, we find that epistasis in higher dimensions reshapes the fitness landscape. 
Thus, the continuous diversification observed in long term evolution experiments \cite{Good2017} could be generated by the continuously changing fitness landscape as new mutations occur. 
In particular, we identify master regulators that operate in higher dimensions by significantly enhancing or suppressing interactions in the rest of the biological network. In the microbiome these are lactobacilli, and in \emph{E. coli} evolution we identified \emph{rbs} and \emph{pykF}. While it would require future experiments, it might be expected that such higher-order master regulators may also regulate the onset and progression of cancers.  
%Epistatic interactions may constrain evolutionary paths or ecological community assembly. 
%In higher dimensions, we lack simple terminology to describe the many types of interactions that may occur, whether between quadruples and singles, pairs and triples, or different genetic backgrounds. 

The prevalence and importance of higher-order interactions is debated, with some studies 
suggesting pairwise interactions predict the vast majority of interactions in complex communities \cite{Friedman2017}, and others suggesting a large influence of context-dependent effects \cite{Gould232959} \cite{Sundarraman}, 
which would make higher-order interactions unpredictable. Ample evidence that higher-order epistasis has at least some evolutionary impact was established in recent publications, see \cite{Krug:2021EpistasisandEvolution} and its references. 
Our analyses suggest limitations on the existence of epistasis in higher dimensions. 
This could arise due to e.g.\ limited phenotypic dimensions where interactions can be detected or to a lower dimensional manifold that absorbs the majority of the effects \cite{Husain2020} (e.g.\ lifespan and fecundity are anti-correlated, making fitness robust to changes in one or the other). 

In Section \ref{sec:high from low},
we analyzed how higher-order interactions in three data sets can arise from lower order ones. We found that in the majority of cases, the full biological information can only be obtained by analyzing epistatic weights in the full dimensional genotope space and that lower-order interactions are not sufficient to describe \emph{all} interactions. In a few cases, however, the source of the higher dimensional interaction
is rooted in a lower dimensional space and no additional biological information is obtained by increasing the dimension.

Our analysis also shows that significant epistatic interactions are increasingly sparse as the number of dimensions for interaction increase, indicating some limits to biological complexity.

\subsection{Relation between epistatic filtrations and other measures of epistasis}

From a methodological point of view, the present work lays the geometric groundwork 
for detecting epistasis via interactions of higher-order as well as other geometric properties 
of large fitness landscapes. Our work relies on polytope theory, following the shape approach of
\cite{BPS:2007,BMC:2007}, as this is the only framework allowing a 
mathematical definition of epistasis in a fine grained manner for a 
general $n$-locus system. By this we mean, that our interactions 
involve a minimal number of genotypes in the sense of a minimal set of dependent points \cite{Triangulations}. 
The motivation for this is that these sets generalize the notion of adjacent triangles 
in a 2-locus system to an $n$-locus system. Additionally, in this way interactions have
a geometric meaning, which makes them comparable across data sets.
Although our method has similarities with \cite{BPS:2007,BMC:2007}, 
it also has significant theoretical and computational differences and improvements. 
For example, our analyses heavily rely on studying the dual graph of the induced 
triangulation together with colored filtrations. 
This is a novelty in the theory and provides a number of new biological findings. 
For example, we localize regions of epistasis in four fitness landscapes, we quantify the sparsity of these regions, we compare portions of fitness landscapes via the parallel transport operation or by changing bystander species. We also further develop \cite{Eble2019} by 
providing a new framework to detect and interpret how higher-order epistasis 
arises from lower order epistasis via meta-epistatic charts. 
%Finally, we introduce a new mathematical technique designed to analyze tuples of fitness landscapes associated to different height functions. {\bf I think we should remove this method and put it in a separate arxiv posting.}

More specifically, epistatic weights capture new properties 
of fitness landscapes even in the 3-locus case. 
In this case, there are between four and six epistatic weights, as these are the number of adjacent pairs of simplices in the subdivision of the $3$D cube, which appear as edges in the dual graph \cite[Fig.1]{10.230740234576}. In contrast, there 
are 20 circuit interactions \cite[Ex.3.9]{BPS:2007} and many more possible 
and potentially relevant interactions that must be checked in a randomized, exhaustive search.
In addition to reducing the search space, epistatic weights can be localized in the fitness 
landscape, allowing the occurrence of mutations to be linked to changes in the topography of the epistatic landscape. 
Furthermore, we can link these changes across dimensions, tracking the source of the interactions. 
%Thus, there is a clear meaning of what it means to consider epistatic weights jointly, 
%and what it means that significant epistatic weights in the three-loci case originate from significant epistatic weights in two-loci cases. 
%{\bf I'm not sure I see this. Do you mean meta-epistatic charts? I added some text in the preceding sentences.}
%This all provides new biological information that is not accessible with all other methods. 

Our method relates to other measures of epistasis, for example to linear regression 
approaches, as we explain in Section \ref{srm}, see also the recent work \cite{McCandlish}. 
It also relates to methods originating from harmonic analysis, cf.~\cite{Weinberger,Sailer2017,Weinreich:2018aa}; and to correlations between the effects of pairwise mutations, as we pointed out in \cite{Eble2019}. 
More concretely, in a 2-locus, biallelic system, all these methods can easily be recovered from one another; some of them even agree. This is also true for some ecological approaches, including the generalized Lotka-Voleterra equations, which yield a mathematically equivalent form to epistasis for certain situations cf.\ see equation 9 of \cite{Case1981}. In higher dimensional systems, these methods remain conceptually closely related but they generally yield different insights about the problem, such as which interactions are considered, whether the interactions are significant, what their magnitude is, and what their sign is. Because these previous methods make specific, \emph{a priori} assumptions about the forms of interactions, they are limited by these assumptions. Epistatic filtrations add a global perspective, determining the structure of interactions from the shape of the fitness landscape in a parameter-free approach.

Finally, rank orders play an important 
role in the recent fitness landscape theory \cite{Crona2017a,Crona2017,Lienkaemper2018}. For an overview and for references to relevant work in the theory, see the review article \cite{Krug:2021EpistasisandEvolution}. 
It is straightforward to recast the fitness landscapes presented here into a rank-order fitness graph and then count the number of peaks, i.e.\ the number of sinks in a fitness graph. The technical details are beyond the scope of the present paper.

\subsection{Interactions in higher dimensions}
We found that biologically-significant epistatic interactions in four and 
five dimensions are sparse and often rooted in lower order, meaning that a limited number 
of regions of epistasis and hence of distortion exist in these fitness landscapes. 
This extends to higher dimensions the trend that 3-way interactions 
are often predicted from 2-way interactions \cite{Kuzmin2018,Friedman2017,Gould232959}.
However, our finding that key genes and species cause distortions emphasizes the need to identify the significant higher-order interactions from the vast number of possible ones, a task that epistatic filtrations enable. 

In a five-loci case, we also found that the fitness landscape 
in the Eble data set is much more distorted, i.e. non-linear, 
than the Khan fitness landscape. We also found the precise
locations of distortions inside the corresponding fitness landscapes and 
contextualize them in terms of distortions visible in lower dimensional sub-fitness landscapes.
These findings are new and cannot be established with the old methods. 
%In particular, these results could not be found with the work of \cite{BPS:2007,BMC:2007}. 

\subsection{Strength and limitations of epistatic filtrations}
A major advance of this work is that we provide a way to discover high dimensional regulators of biological networks. Rather than identifying key nodes as having a high number of low dimensional edges, we developed a method to identify nodes that regulate the higher-dimensional interactions in the rest of the network. This operation is performed by the parallel transport function, and we provide a web-based tool to perform the analysis (see Appendix~\ref{tab:datasets}). The implications of these findings are that certain genes and species modulate the interactions in the rest of the network, and perturbing these loci can destabilize the network. 
Destabilizing an unhealthy biological network could be crucial to restoring a degraded ecosystem, a sick  microbiome, or curing a cancer, while destabilization of a healthy biological network could have the opposite consequences. 

Methodologically, we also improve the framework in which higher-order epistasis can be mathematically formalized and analyzed geometrically. 
We provide concrete tools to find epistatic interactions in the fitness landscape and to distinguish if the landscape is locally flat, i.e. a hyperplane of a certain dimension. 
Our work additionally allows us to localize and contextualize regions inside the fitness landscape which are not flat and hence distorted.
  
Our approach does not provide a distinction between positive and negative epistasis, but only between presence and absence of epistasis. 
However, this limitation is shared with other methods including the circuit, linear regression, and Fourier expansion approaches.
To give an example, the circuit interactions in \cite{BPS:2007} can produce positive or negative values, but the sign
depends on the choice of a basis for the interaction space, without a real biological motivation.
The biallelic case provides an elementary case.
In traditional terms, the epistasis in the Example from Box~\ref{box:definition+epistatic+filtration} is negative since the lifted genotype $11$ lies \emph{below} the plane spanned by the lifted genotypes $00$, $10$ and $01$.
Picking that particular plane for choosing the sign rests on the basis where the wild type is $00$.
If instead we use the genotype $10$ as a basis, then the lifts of that genotype and its two neighbors $00$ and $11$, span a plane such that the lifted fourth genotype $10$ lies \emph{above} that plane of reference. 
However, while circuit interactions use signs to locate epistatic effects, in our approach this is not necessary, as the location information is concisely encoded in the regular triangulation induced by the phenotypes as described (c.f. Box~\ref{box:definition+epistatic+filtration}).
In this sense, the lack of sign is not a limitation of epistatic filtrations but a consequence of the high-dimensional approach.

A second limitation is a computational one which arises when one considers a multi-allelic system.
In that setting our method still applies in theory, but the computational bottlenecks are reached rather quickly (at around $n=10$ alleles without large hardware).
However, it should be pointed out that the number of circuits of the cube $[0,1]^n$ grows even faster with~$n$; cf.\ Table~\ref{tab:bipyramids}.
So methods based on these also suffer from combinatorial explosion.

\subsection{Outlook} 
This geometric approach could be extended, e.g.\ to GWAS \cite{Fang2019,Liu2019,Carlborg2004}, ecosystems \cite{Case1981,Billick1994}, or neuronal networks \cite{Reimann2017}, to discover non-additive higher-order structures at different scales. 
It should be noted that the polyhedral geometry methods for analyzing epistasis deserve to be developed further from the mathematical point of view. We believe that more concepts related to curvature for piecewise linear manifolds will be useful \cite{Sullivan:2008}.

Taken together, our approach offers a number of new insights on higher-dimensional properties of fitness landscapes and their biological implications, and we think these will be useful as higher throughput experiments enable more combinatorial approaches.

\section{Acknowledgements}
The authors acknowledge L.J. Holt, O. Brandman, and J. Derrick for insightful comments on the manuscript. Research by M.J. is carried out in the framework of Matheon supported by Einstein Foundation Berlin. Further partial support by Deutsche Forschungsgemeinschaft (SFB-TRR 109: “Discretization in Geometry and Dynamics” and SFB-TRR 195: “Symbolic Tools in Mathematics and their Application”. W.B.L. acknowledges NIH grant DP5OD017851, NSF IOS award 2032985,  and the Carnegie Institution for Science Endowment. 

\section{Competing interests}
The authors declare no competing interests.

\section{Supplementary Materials}
Materials and Methods \\
Appendices\\
Fig S1 – S22\\
Tables S1 – S9 \\

%\bibliography{2020biologyPaper}{}
%\bibliographystyle{unsrt}
%%%%\bibliographystyle{Science.bst}

% restart the numbering of the figures/tables as S1, S2, ...

\setcounter{figure}{0}
\setcounter{table}{0}
\makeatletter 
\renewcommand{\thefigure}{S\@arabic\c@figure}
\renewcommand{\thetable}{S\@arabic\c@table}
\makeatother

\title{Supplementary Materials for: \\
Fitness landscapes distortions alter evolution and microbiomes in higher dimensions}

\author
{Holger Eble,$^{1}$ Michael Joswig,$^{1,2\ast}$ Lisa Lamberti$^{3,4}$,William B. Ludington$^{5,6\ast}$\\
\\
\normalsize{$^{1}$Chair of Discrete Mathematics/Geometry, TU Berlin, Germany}\\
\normalsize{$^{2}$MPI MiS Leipzig, Germany}\\
\normalsize{$^{3}$Department of Biosystems Science and Engineering, ETH Z\"urich, Basel, Switzerland}\\
\normalsize{$^{4}$
SIB Swiss Institute of Bioinformatics, Basel, Switzerland}\\
\normalsize{$^{5}$
Department of Embryology, Carnegie Institution for Science, USA}\\
\normalsize{$^{6}$
Department of Biology, Johns Hopkins University, Baltimore, MD, USA}\\
\\
\normalsize{$^\ast$To whom correspondence should be addressed;} \\
\normalsize{E-mail: joswig@math.tu-berlin.de, ludington@carnegiescience.edu.}
}

\date{}
\maketitle

\newpage

\FloatBarrier

\appendix

\section{Materials and Methods}
\subsection{Fly husbandry} Flies were reared germ-free and inoculated with one combination of bacteria on day 5 after eclosion. $N{\geq}100$ flies were assayed for lifespan in $n{\geq}5$ independent vials per bacterial combination for a total of $3200$ individual flies. Food was 10\% autoclaved fresh yeast, 5\% filter-sterilized glucose, 1.2\% agar, and 0.42\% propionic acid, pH 4.5. Complete methods are described in Gould \emph{et al} \cite{Gould232959}. 

\subsection{Bacterial cultures} Bacteria were cultured on MRS or MYPL, washed in PBS, standardized to a density of $10^7$ CFU/mL and 50 µL was inoculated onto the fly food. Strains are indicated in Table~\ref{tab:datasets}. See Gould \emph{et al} \cite{Gould232959} for complete methods.

\subsection{Genetics data} Existing genetics data sets were gotten from Sailer and Harms 2017~\cite{Sailer2017} github repository (\url{https://github.com/harmslab/epistasis}) or from Tan \emph{et al} \cite{PhysRevLett.106.198102}. 

For the Khan data in Fig. \ref{fig:Eble-0ssss_khanraw}, the fitness function $h$ is defined for (b) by assigning the following normalized values to the 16 genotypes:
\begin{center}
    $
   \protect \begin{array}{llll}
0 0 0 0 0 \mapsto 0.1524\; &
0 1 0 0 0 \mapsto 0.1745\; &
0 0 1 0 0 \mapsto 0.1689\; &
0 0 0 1 0 \mapsto 0.1569\; \\
0 0 0 0 1 \mapsto 0.1528\; &
0 1 1 0 0 \mapsto 0.1842\; &
0 1 0 1 0 \mapsto 0.1756\; &
0 1 0 0 1 \mapsto 0.1823\; \\
0 0 1 1 0 \mapsto 0.1718\; &
0 0 1 0 1 \mapsto 0.1810\; &
0 0 0 1 1 \mapsto 0.1642\; &
0 1 1 1 0 \mapsto 0.1836\; \\
0 1 1 0 1 \mapsto 0.1956\; &
0 1 0 1 1 \mapsto 0.1858\; &
0 0 1 1 1 \mapsto 0.1813\; &
0 1 1 1 1 \mapsto 0.1987\ \enspace.  
  \protect  \end{array}
    $
\end{center}

The Tan data set is different from the other fitness values in that only median and mean values are given, meaning we cannot compute $p$-values to assess the statistical significance. The fitness values are minimum inhibitory concentrations of antibiotics from a well-standardized assay with little experimental variation. Thus, the measurements and our analysis are believed to be robust.
We note that the regular subdivision resulting from the corresponding height function of $[0,1]^5$ is degenerate in the sense that it is not a triangulation. 
This degeneracy arises because the data are discrete antibiotic concentrations with 24 possible values. 
The repetition of exact values in several cases means a triangulation does not occur. 
We extended our methods to this degenerate case by restricting the analysis to the faces that do have a triangulation, broadening the application of our approach. 
We focused on the piperacillin with clavanulate data from \cite{PhysRevLett.106.198102} as it is the better behaved.

\subsection{Computational analysis} The filtrations code is available as a \polymake \cite{DMV:polymake} package (c.f. \url{https://github.com/holgereble/EpistaticFiltration}) and the analysis pipeline is available as a \jupyter notebook. We also provide an online client, which processes raw csv data sheets, cf. {\small \url{https://www3.math.tu-berlin.de/combi/dmg/data/epistatic_filtrations/}}.

\section{Terminology}\label{Sec:Terminology}

{\bf Loci} (singular {\bf locus}) refer to individual sites in the genome where a mutation may occur, or in the microbiome sense, a locus is a particular bacterial species.
We write $[n]:=\{1,\dots,n\}$ for the set of all loci.

{\bf Genotypes}, $v=(v_{1},\dots,v_{n})$, are vectors of loci with 0/1-coordinates that form points in some fixed Euclidean space $\RR^n$, where $n$ is the number of genetic loci or bacterial species considered.
In this article we focus on {\bf biallelic} $n$-locus systems, i.e.\ genotype sets of the form $V=\{0,1\}^n$ where $n$ is the number of loci and each locus is either $0$, absent, or $1$, present.
For instance, $v=(1,0,1)$ denotes a genotype in a 3-locus system $\RR^3$, where the first and third loci are mutant and the second is wild type.
The set of all genotypes will be denoted by $V$.
The convex hull $P:=\conv(V)$ of all genotypes is called the {\bf genotope}.
In our setting $P$ is the $n$-dimensional unit cube $[0,1]^n$ (cf.\ (Fig.~\ref{sc5}) for a $2D$ projection of $[0,1]^5$).

A {\bf fitness function} (also called {\bf height function}) associates to each genotype $v\in V$ a quantified {\bf phenotype} describing the impact of the genotype on the organism.
For example, if the measured phenotype is fitness, $h$ encodes the reproductive output of the genotype.

The {\bf fitness landscape} is the pair $(V, h)$, which defines the fitness $h(v)$ for each genotype $v\in V$. 
Let $v=(v_{1},\dots,v_{n})\in V$ be a genotype.
Then its {\bf lift} is given by $(v,h(v))=(v_{1},\dots,v_{n},h(v))\in \RR^{n+1}$.

A set of points $W=\{w^{(1)},\dots,w^{(\ell)}\}$ is {\bf affinely independent} if for every point $x\in\RR^n$ which admits real scalars $\lambda_{i}$ with $\sum_{i=1}^{\ell}\lambda_{i}=1$ and $\sum_{i=1}^{\ell}\lambda_{i} w^{(i)}=x$ those scalars are uniquely determined.
Otherwise $W$ is {\bf affinely dependent}.

An {\bf interaction} with respect to a fitness function $h$ occurs between a collection of $k+2$ affinely dependent genotypes $v^{(1)},\dots,v^{(k+2)}\in V\subset\RR^{n}$, for $k\leq n$, whose lifts are affinely independent points in $\RR^{n+1}$.
This is in line with the standard concept of additive epistasis.
The number $k$ is the \textbf{dimension} of the interaction; throughout we assume that $k\geq 2$.

Let $U=\{v^{(1)},\dots,v^{(\ell)}\}$ be a set of genotypes.
Its {\bf support} is the set
\[ \supp(U) := \SetOf{ k\in[n] }{ \text{there are distinct } 1\leq i,j\leq \ell \text{ with } v^{(i)}_k\neq v^{(j)}_k } \enspace .\]
That is, the support is the set of loci where at least two of the given genotypes differ.
For example, if $n=3$ and $U=\{(0,0,0),(1,0,1),(1,0,0)\}$ then $\supp(U)=\{1,3\}$.

The number of loci that vary ($0$ vs $1$) in the support is called the {\bf order} of an interaction; this definition agrees with, cf., \cite{Weinreich:2018aa}: \enquote{We designate interactions among any subset of $k$ mutations as $k$th-order epistasis.}.
We give two examples: First, let $n=2$ and $U=\{(0,0),(0,1),(1,0),(1,1)\}=V$ such that $U$ is an interaction with respect to some fitness function.  Then $U$ is an interaction of dimension 2 and order 2.
Second, let $n=3$ and $U=\{(0,0,0),(0,1,1),(1,0,0),(1,1,1)\}$ such that, again, $U$ is an interaction with respect to some height function.  Then the dimension is 2 and the order is 3.
In general, the order is at least as large as the dimension, but the two quantities may differ.
We say that genes (corresponding to loci) {\bf interact} if they form the support set of an interaction of genotypes.

\begin{remark}
  The dimension $k$ of an interaction $v^{(1)},\dots,v^{(k+2)}$ with respect to some fitness function agrees with the dimension of the affine span of the given points in $\RR^n$.
  This can be seen as follows.
  By definition the lifted points $(v^{(1)},h(v^{(1)})),\dots,(v^{(k+2)},h(v^{(k+2)}))$ are affinely independent in $\RR^{n+1}$.
  So their affine span has dimension $k+1$.
  As $v^{(1)},\dots,v^{(k+2)}$ are affinely dependent, the dimension of their affine span is at most $k$.
  Now the affine dimension can only increase by at most one if one coordinate is appended.
\end{remark}

\subsection{A primer on epistatic filtrations}\label{primer+epistatic+filtrations}
We first explain the biallelic case with $n\geq 2$ loci.
In the geometric framework \cite{BPS:2007}, two interacting loci give rise to four possible genotypes, which form the vertices of a square and may be written as vectors of zeros and ones, indicating the absence (0, wildtype) or the presence (1, mutant) of each locus respectively (Box \ref{box:definition+epistatic+filtration}a) \cite{Eble2019,BPS:2007}. 
The measured phenotypes lift the genotype vertices into 3-space, and there is epistasis corresponding to the volume of the simplex enclosed by the lifted points (see green simplex in Box \ref{box:definition+epistatic+filtration}a).
Geometrically, the four genotypes involved are fully symmetric, meaning that the sign of the epistasis for $n=2$ is relative to the choice of a coordinate system. Thus, the sign of epistasis depends on which genotype is considered wildtype. By considering the simplex volume rather than the fold of the upper shell of the simplex, epistatic filtrations do not specify a sign and thus avoid this caveat. However, directionality is considered by parallel transport (see later section).
Returning to our explanation, by taking the upper convex hull of all $2^n$ lifted points and projecting back onto the genotope $[0,1]^{n}$ we induce a {\bf subdivision} $\subdivision(h)$; cf.\ \cite[\S2.1]{Eble2019,Triangulations}, into {\bf maximal cells} (Box \ref{box:definition+epistatic+filtration}b).
Generically, every maximal cell of $\subdivision(h)$ is an $n$-dimensional simplex, which is the convex hull of $(n+1)$ affinely independent genotypes. 
Importantly, these $n$-dimensional simplices
are the most elementary parts into which a fitness landscape can naturally be decomposed. 

Our framework generalizes to higher dimensions through a geometric shape called a {\bf bipyramid}, where two satellite vertices, each the apex of one pyramid, are joined to a common set of base vertices. 
The satellites correspond in the $2D$ example (Box \ref{box:definition+epistatic+filtration}) to $00$ and $11$ and the base to $10$ and $01$.
This is naturally associated with $\subdivision(h)$, set up by the {\bf ridge} (Box \ref{box:definition+epistatic+filtration}).
For an ordered sequence of $n+2$ genotypes $(v^{(1)},v^{(2)},\dots,v^{(n+2)})$  we let
\[
  s  =  \conv\{v^{(1)},\dots,v^{(n+1)}\} \quad \text{and} \quad t =  \conv\{v^{(2)},\dots,v^{(n+2)}\} \enspace.
\]
In other words, $s$ and $t$ form convex hulls. We call such a pair $(s,t)$ a bipyramid with vertices $v^{(1)},v^{(2)},\dots,v^{(n+2)}$.
Then we can find the volume of the lifted bipyramid by forming the $(n+2){\times}(n+2)$-matrix
\begin{equation}\label{eq:epistatic-matrix}
  E_h(s,t) \ := \
  \begin{pmatrix}
    1 & v_{1,1} & v_{1,2} & \dots & v_{1,n} & h(v^{(1)}) \\
    1 & v_{2,1} & v_{2,2} &  \dots & v_{2,n} & h(v^{(2)}) \\
    \vdots  & \vdots &  \vdots & \vdots & \vdots &\vdots \\
    1 & v_{n+2,1} & v_{n+2,2} & \dots & v_{n+2,n}  & h(v^{(n+2)})
  \end{pmatrix} \enspace ,
\end{equation}
where $v_{i,1},v_{i,2},\dots,v_{i,n}$ are the coordinates of $v^{(i)}\in\RR^n$.
The {\bf epistatic weight} of the bipyramid $(s,t)$ is 
\begin{equation}\label{eq:epistatic-weight}
  e_h(s,t) \ := \ \bigl|\det E_h(s,t)\bigr| \cdot \frac{\nvol(s\cap t)}{\nvol{(s)}\cdot\nvol{(t)}} \enspace .
\end{equation}
Here $\nvol$ denotes the dimensionally normalized volume.
The quantity $\nvol(s\cap t)$ is the relative $(n{-}1)$-dimensional normalized volume of the {\bf ridge} of the bipyramid, given by the intersection $s\cap t=\conv(v^{(2)},\dots,v^{(n+1)})$.
We use the notation
\begin{equation}\label{eq:bipyramid}
  \{v^{(1)}\} + \{v^{(2)},\dots,v^{(n+1)}\} + \{v^{(n+2)}\} 
\end{equation}
for the bipyramid $(s,t)$, where the first and last vertices are the satellites and the middle set forms the base.
Now the $n+2$ genotypes of the bipyramid form an interaction of dimension $n$ when $e_h(s,t)>0$.

In our regular triangulation $\subdivision(h)$, the two $n$-dimensional simplices, $s$ and $t$, are {\bf adjacent} because their intersection $s\cap t$ is a common face of dimension $n-1$.

\subsection{Constructing a filtration from the epistasis of adjacent simplices}
We visualize the topography of the {\bf epistatic landscape} by forming a {\bf dual graph} of $\subdivision(h)$, where the nodes are the maximal simplices and adjacent simplices form the dual edges. 
A rugged path is one with more blue edges (Box \ref{box:example+epistatic+filtration}d). 
To each such dual edge we associate an epistatic weight and a label, epistatic weights are in shades of blue and red, while labels are in black). 
In this way, we construct an epistatic landscape that corresponds to the underlying fitness landscape with the ruggedness specified along the dual graph.
	The {\bf epistatic filtration} of $h$ (Box \ref{box:example+epistatic+filtration}e) depicts the path from weakest to highest epistasis by merging adjacent simplices. 
These diagrams summarize the information contained in epistatic weights and dual graphs, and facilitate comparisons across data sets. 
But there is important new information contained in epistatic filtrations, which is not directly visible from the dual graph and its epistatic weights.
Indeed, a step in the epistatic filtration merges adjacent simplices. 
We build the complete fitness landscape by stepwise merging of maximal cells, starting from the lowest epistatic weight and stepwise merging adjacent simplices to form a connected {\bf cluster} cf. \cite{Eble2019}. 
In this sense, epistatic filtrations encode a global notion of epistasis in higher dimensions by connecting adjacent bipyramids.

To see this, notice that each row of the diagram has a number of bars and a black leftmost line. In the top row the black line marks the epistatic weight of zero ($x$-coordinate). Each bar is red and corresponds to one maximal simplex of $\subdivision(h)$.
In the second row (counting from the top), we see three things: (1) the value of the lowest epistatic weight moves the $x$-coordinate of the black line slightly to the right. (2) The two maximal simplices of $\subdivision(h)$ corresponding to this epistatic weight are merged into one. These correspond to the two bars in the previous row above the new, longer bar in the row. The lengths of the other bars remain unchanged but are shifted horizontally by the epistatic weight in (1). (3) The statistical significance of the epistatic weight giving rise to the merging step, encoded by the colors of the bars; cf.\ Section~\ref{subsec:stat}. 

The merging procedure is then repeated for each pair of maximal simplices arising in each epistatic weight until one reaches the highest epistatic weight and the last maximal simplex of $\subdivision(h)$ to be merged with the rest. In this way the indentation of the bar charts increases from top to bottom. The total width of the bars stays constant throughout.

Importantly, in the epistatic filtration diagram, not every merging step is displayed;
e.g., in Box \ref{box:example+epistatic+filtration}e there are fewer rows than dual edges in Box \ref{box:example+epistatic+filtration}d.
This is because some steps do not change the resulting fitness landscape (no actual new portion is merged to the previous one).
The reported steps are only the ones increasing the connected components of the fitness landscape obtained from the previous merging steps. The epistatic weights corresponding to these steps are the edges in the dual graph which we call {\bf critical} in \cite[\S.3.2]{Eble2019}. 

\subsection{Normalized epistatic weights}
To gain a perspective on the generality of higher-order interactions, it is desirable to compare epistatic landscapes. Different phenotypes have different metrics, making comparisons difficult for current approaches to epistasis. 
Filtrations are well-suited in this sense. Scaling the height function $h$ by a positive constant does not change the regular triangulation, and thus it does not change the dual graph.
In order to compare different data sets, we scale the height function to Euclidean norm one.
The epistatic weights are scaled accordingly.
The resulting {\bf normalized epistatic weights} are measured in {\bf epistatic units}, giving a generalized metric for epistasis.

Measuring the effect of context on epistatic interactions is also desirable, e.g.\ to detect the marginal or conditional effects of a locus \cite{Khan1193}, and these are a natural feature of filtrations. If we fix some $k$ loci and let the remaining $n-k$ loci vary, we obtain a height function, which is {\bf restricted} to a face of the genotope $[0,1]^n$.
That face has $2^{n-k}$ vertices, and it is an isomorphic copy of the cube $[0,1]^{n-k}$.
For instance, if $n=5$ and we fix the first and the fourth locus to~$0$, we obtain a $3$-dimensional face, which we denote ${0}{\ast}{\ast}{0}{\ast}$.
That is, such a face is written as a string of $n$ symbols in the alphabet $\{0,1,\ast\}$, where $0$ or $1$ mark the fixed choices, and $\ast$ stands for variation.
The number of $\ast$ symbols equals the dimension of the face.
Triangulations, their dual graphs, epistatic weights, etc. are well-defined for height functions restricted to faces.
This aspect of the theory allows the study of conditional epistatic effects.

\subsection{Statistics of epistatic weights}
\label{subsec:stat}
We developed a statistical test to quantify the significance of an interaction associated with a fixed bipyramid; cf.\ \cite[\S4.2]{Eble2019}.
Here we assume that $h(v)$ is the mean value of the individual phenotype measurements for some number of replicated experiments for the fixed genotype~$v$.
To each dual edge we associate a $p$-value, which is independent of the epistatic weight normalization.
If that $p$-value is below $0.05$ we call that dual edge {\bf significant}.
It is useful to also consider $p$-values, which are slightly higher because one can use the shape of the landscape to identify interesting locations for further statistical analysis.
To this end we call a dual edge {\bf semi-significant} if $0.05 \leq  p < 0.1$.

While it may be possible that this approach misses some biologically relevant interactions (e.g.\ if they do not correspond to a bipyramid selected by our method), those interactions that we identify  carry information that is robust and supported by a statistical model.
The fact that not all possible interactions can be approached is an inevitable consequence of the higher dimensional nature of fitness landscapes, also reflected by a very high number of possible regular triangulations of $[0,1]^n$.
That number equals 74 for $n=3$ and $87{,}959{,}448$ for $n=4$, whereas the precise numbers for $n\geq 5$ are unknown; cf.\ \cite[\S6.3]{Triangulations}.
Thus, filtrations use the data to greatly condense the number of possible interactions considered.

The bar colorings in the filtrations of epistatic weights, as in (Fig.~\ref{fig:Khan-5cube}), reflect the outcome of multiple simultaneous statistical tests (one for each epistatic weight) \cite{Eble2019}.

Significant dual edges at $p<0.05$ are shown in blue, $0.05\leq p<0.1$ in purple, and $p\geq 0.1$ in red.

It may happen that a triangulation has a significant dual edge, which is not critical, whence it does not show in the epistatic filtration.
In that case the next critical dual edge becomes blue; so a filtration encodes all significant interactions found by our method.

\begin{remark}
  % It would be possible to apply false discovery rate controlling procedures to our statistical analysis of significant epistatic weights.
  % We refrain from doing so for the following reasons: (\emph{i}) A filtration is a stepwise procedure built on a continuous filtration, rather than on a discrete classification of statistically significant outcomes where multiple simultaneous tests are performed, e.g. genome-wide association studies. (\emph{ii}) Established procedures to adjust $p$-values do not account for the multidimensional, geometric nature of our methods. For instance, we calculate multidimensional volumes, and it is unclear what statistical null model should be used. Future work will develop these methods, but they are beyond the scope of the present study. (\emph{iii}) Interesting epistatic weights detected within a filtration can be further validated statistically using traditional linear regression approaches, as demonstrated in Section~\ref{srm}.
  By funneling the analysis through the concept of regular triangulations our approach pre-selects interactions, which are most relevant with respect to fitness \cite[\S2.2]{Eble2019}.
  Via this major deviation from \cite{BPS:2007} we are able to detect interactions in many data sets, which are biologically plausible; this suggests strongly that our method is particularly good at avoiding false positives.
  Future work will investigate the relationship to other methods from statistics and signal processing.
  While most of this is beyond the scope of the present study, in Appendix B12 we offer a first step by comparing with traditional linear regression approaches.
\end{remark}

\subsection{A synthetic experiment examining how epistatic weights change as a function of the interaction order}
Our method calculates significance of detected interactions and normalizes the epistatic weight to the volume of the unit cube of the same dimensionality. We used synthetic data to analyze the method performance. We first examined 468 synthetic filtrations over the $4$-dimensional cube, producing 10011 critical dual edges. We found that the epistatic weight is indeed constant as a function of the interaction order, see (Fig.~\ref{fig:synthetic_MP}a). This indicates that the normalization method is effective. Furthermore, the number of significant interactions decreased as the standard deviation of the input data increased, indicating the statistical method is sensitive to noise, see (Fig.~\ref{fig:synthetic_MP}b).

%% R commands
%%library(ggplot2)
%%library(ggbeeswarm)
%% synth <- read.csv("drosophila/polymake/synthetic+experiments/synthexp_dim4_order.csv")
%% plot <- ggplot(synth,aes(order, eweight, colour=significance)) + geom_quasirandom(varwidth = TRUE) + scale_x_continuous(breaks=c(2,3,4,5))+ scale_color_manual(values=c("blue", "red"))
%% print (plot + labs(x ="interaction order", y = "epistatic weight") + theme(legend.position = "none"))

\subsection{A microbiome example in dimension 4}

Here $n=4$, and the fitness function $h$ is defined by assigning the following values to the 16 genotypes:
 
\begin{equation*}\label{eq:ttd_0ssss}
  \begin{array}{llll}
0 0 0 0 \mapsto 0.2484\ ; &
1 0 0 0 \mapsto 0.2320\ ; &
0 1 0 0 \mapsto 0.1618\ ; &
0 0 1 0 \mapsto 0.1698\ ; \\
0 0 0 1 \mapsto 0.1943\ ; &
1 1 0 0 \mapsto 0.1749\ ; &
1 0 1 0 \mapsto 0.1714\ ; &
1 0 0 1 \mapsto 0.1929\ ; \\
0 1 1 0 \mapsto 0.1668\ ; &
0 1 0 1 \mapsto 0.1608\ ; &
0 0 1 1 \mapsto 0.1617\ ; &
1 1 1 0 \mapsto 0.1643\ ; \\
1 1 0 1 \mapsto 0.1677\ ; &
1 0 1 1 \mapsto 0.1715\ ; &
0 1 1 1 \mapsto 0.1613\ ; &
1 1 1 1 \mapsto 0.1594\ \enspace .
  \end{array}
\end{equation*}

The vertices $U:=\{v^{(1)},\dots, v^{(6)}\}\in V$ given by
	\begin{equation*}
\begin{array}{lll}

v^{(1)}=(1, 1, 0, 0)\ ; & v^{(2)}=(0,0,0,0)\ ; & v^{(3)}=(1,0,0,0)\ ;\\ v^{(4)}=(1,1,0,1)\ ; & v^{(5)}=(1,1,1,1)\ ; & v^{(6)}=(1,0,0,1)
\end{array}
	\end{equation*}
form a bipyramid $(s,t)$ consisting of $4$-dimensional simplices $s$ and $t$ as above. The simplices $s$ and $t$ correspond to nodes in the dual graph of $\subdivision(h)$ that share a dual edge recording their adjacency relation as indicated in (Fig.~\ref{fig:Eble-0ssss}b). 

In this situation, equation (\ref{eq:epistatic-weight}) reads
\begin{equation*}
  	e_h(s,t) \ = \	\begin{vmatrix}
1 & 1 & 1 & 0 & 0 & 0.1749 \\
1 & 0 & 0 & 0 & 0 & 0.2484 \\
1 & 1 & 0 & 0 & 0 & 0.2320 \\
1 & 1 & 1 & 0 & 1 & 0.1677 \\
1 & 1 & 1 & 1 & 1 & 0.1594 \\
1 & 1 & 0 & 0 & 1 & 0.1929 \\
 	 \end{vmatrix} \ \cdot \ \frac{\nvol(s\cap t)}{\nvol{(s)}\cdot\nvol{(t)}} = 0.0318\  \cdot\  \frac{\sqrt{2}}{1\cdot 1}\approx 0.045 \enspace . 
	\end{equation*}
Since $e_{h}(s,t)>0$, the genotype set $U$ defines a $4$-dimensional interaction with full support $\{1,2,3,4\}$ and of order $4$, according to our terminology of Section Terminology. With a $p$-value of $0.0005<0.05$ the significance test established in \cite[\S.4]{Eble2019} rejects the zero hypothesis for $e_{h}(s,t)$ and therefore proves the effect of the interaction $U$ to be significant. We indicate this fact with the color \textcolor{blue}{blue} both in the dual graph of $\subdivision(h)$ in (Fig.~\ref{fig:Eble-0ssss}b) and in the epistatic filtration of $h$ in (Fig.~\ref{fig:Eble-0ssss}c).

This example illustrates the following fact of biological interest. For the bacterial combinations $v^{(1)},v^{(2)},\dots,v^{(6)}$ fitness, given by the fitness function $h$, varies significantly in a non-linear way. 

%\subsection{Interactions are sparse in higher-dimensions}
%We analyzed the few higher-order interactions in greater detail using a geometric approach. As we noted previously, the interactions in the Khan genetic data (Table~\ref{tab:complexity:gen-micro}) are based on a vertex split of the genotype ${0}{0}{0}{0}{1}$, meaning that the entire epistatic weight of the landscape is balanced by a single maximal cell (Fig.~\ref{fig:Eble-0ssss_khanraw}).

\subsection{Parallel transport of epistatic weights}\label{ss:parallel_transport}
The notion of parallel transport in a fitness landscape $(V,h)$ was introduced in \cite[\S6.6]{Eble2019} as a way to geometrically compare biological information between pairs of parallel facets of the convex polytope $\conv V$. In this work, we extended that notion to include the case of two fitness landscapes, $(V,h_1)$ and $(V,h_2)$, associated to different generic and normalized height functions 
$h_i: V\rightarrow \RR, i\in\{1,2\},$ defined on the same vertex set 
$V=\{0,1\}^n$ for some $n\in\NN$.
To enable meaningful comparisons, we assume that each $h_i$ is normalized and that there is a larger fitness landscape $(W,h)$ with a generic and normalized height function $h:W\rightarrow \RR$ restricting to $h_1$ and $h_2$ on the parallel facets $V$ in $W$, such that the partition of $\conv W$ induced by $h$ is compatible with the one of $\conv V$ induced by $h_1$, resp.~ by $h_2$. 
In this setting, we define {\bf normalized epistatic weights} as with Eq.~(\ref{eq:epistatic-weight}) with $h$ the normalized height function and $s,t$ any adjacent simplices forming a bipyramid.

Parallel transports enable us to transport epistatic filtrations along the reflection map 
\[
\phi\colon V\rightarrow V; v=(v_1,v_2,\dots,v_n)\mapsto(v'_1,v'_2,\dots,v'_n) \enspace,
\] with $v_i'=1-v_k$ if $i=k$ and $v_i'=v_i$ otherwise. More precisely, let $e_{h_1}(s,t)$ be the normalized epistatic weight associated to a bipyramid of $\cS(h_1)$ and let 
$\phi(e_{h_1}(s,t)) : =e_{h_2}(\phi(s),\phi(t))$ be the parallel normalized epistatic weight transported by $\phi$. Then the filtration of normalized epistatic weights induces a filtration of parallel normalized epistatic weights. Additionally, to $e_{h_1}(s,t)$ and to $\phi(e_{h_1}(s,t))$ a $p$-value can unambiguously be associated \cite[\S4.1-4.2]{Eble2019}.
Notice that by design epistatic filtrations for $\cS(h_1)$ only show normalized epistatic weights 
associated to critical dual edges, defined as in \cite{Eble2019}. But normalized epistatic weights and their significance can be defined for all bipyramids 
including the ones associated to noncritical dual edges. This explains the labelling of the parallel transport tables below. There a row is numbered only if the bipyramid corresponds to a critical dual edge in the dual graph of $\cS(h_1)$. 
Noncritical dual edges whose normalized epistatic weight remains 
non-significant after the parallel transport are omitted.
The normalized epistatic weight before 
(denoted by $e_o=e_{h_1}(s,t)$) and after (denoted by $e_p=\phi(e_{h_1}(s,t))$)
the parallel transport, as well as their $p$-values (denoted by $p_o$ and $p_p$) 
are also reported, as well as ratios of these quantities. 

These parallel transport tables are linked to the epistatic filtration diagrams.
Indeed, each numbered row in the table corresponds to the
row in the epistatic filtration diagram with the black line set at $e_o$.
It also corresponds to the row with
black line set at $e_p$ in the parallel transported filtration diagram.

Recall from Section \emph{Statistics of epistatic weights} that there may be dual edges of the triangulations 
which are significant but not critical.
Since only the critical dual edges are labeled (by the row number in the epistatic filtration), in our tables for parallel transport these show up as unlabelled rows.

\begin{figure}[ht]
  \begin{subfigure}{.85\textwidth}
    \begin{minipage}[t]{.45\linewidth}
    \caption{}
      \centering
      \newcommand\maxepi{0.0128}
\newcommand\maxclusters{6}
\newcommand\clusterlength{0.4pt}
\newcommand\clusterspace{0.1pt}
\newcommand\flagfacup{1.2}
\newcommand\flagfacdown{2}
\newcommand\lw{6pt}
\begin{tikzpicture}[yscale=0.5,xscale=50, scale = 0.8]
\tikzset{
        cluster/.style = {red, line width=\lw, join=round},
        significant/.style = {blue, line width=\lw, join=round},
        semisignificant/.style = {purple, line width=\lw, join=round},
        tick/.style = {black, thick},
        tick_thin/.style = {black, thin}}

\coordinate (offset) at (0, 6);
\foreach \name\nclusters\ngaps/\size in {B/0/0/1, C/1/1/1, A/2/2/1, F/3/3/1, E/4/4/1, D/5/5/1}{
        \draw[cluster] ($ (offset) + (\nclusters*\clusterlength,0) + (\ngaps*\clusterspace,0) $) 
        -- ($ (offset) + (\nclusters*\clusterlength,0) + (\ngaps*\clusterspace,0) + (\size*\clusterlength,0) $);
 \node at ($(offset) + (\nclusters*\clusterlength,0) + (\ngaps*\clusterspace,0)+(0.008,0)$) [above,black] {\footnotesize $\name$};        
        }
 \draw[tick] ($ (offset) + (0,\flagfacup*\lw) $) -- ($ (offset) - (0,\flagfacdown*\lw) $); 
 
\coordinate (offset) at (0.00091715658422571, 5);
\foreach \nclusters\ngaps/\size in {0/0/1, 1/1/1, 2/2/2, 4/3/1, 5/4/1}{
 
        \draw[cluster] ($ (offset) + (\nclusters*\clusterlength,0) + (\ngaps*\clusterspace,0) $) 
        -- ($ (offset) + (\nclusters*\clusterlength,0) + (\ngaps*\clusterspace,0) + (\size*\clusterlength,0) $);}
 \draw[tick] ($ (offset) + (0,\flagfacup*\lw) $) -- ($ (offset) - (0,\flagfacdown*\lw) $);
\coordinate (offset) at (0.00518240065726195, 4);
\foreach \nclusters\ngaps/\size in {0/0/1, 1/1/1, 2/2/3, 5/3/1}{
 
        \draw[cluster] ($ (offset) + (\nclusters*\clusterlength,0) + (\ngaps*\clusterspace,0) $) 
        -- ($ (offset) + (\nclusters*\clusterlength,0) + (\ngaps*\clusterspace,0) + (\size*\clusterlength,0) $);}
 \draw[tick] ($ (offset) + (0,\flagfacup*\lw) $) -- ($ (offset) - (0,\flagfacdown*\lw) $);
\coordinate (offset) at (0.00609955724148766, 3);
\foreach \nclusters\ngaps/\size in {0/0/2, 2/1/3, 5/2/1}{
 
        \draw[cluster] ($ (offset) + (\nclusters*\clusterlength,0) + (\ngaps*\clusterspace,0) $) 
        -- ($ (offset) + (\nclusters*\clusterlength,0) + (\ngaps*\clusterspace,0) + (\size*\clusterlength,0) $);}
 \draw[tick] ($ (offset) + (0,\flagfacup*\lw) $) -- ($ (offset) - (0,\flagfacdown*\lw) $);
\coordinate (offset) at (0.00960815174621709, 2);
\foreach \nclusters\ngaps/\size in {0/0/5, 5/1/1}{
 
        \draw[significant] ($ (offset) + (\nclusters*\clusterlength,0) + (\ngaps*\clusterspace,0) $) 
        -- ($ (offset) + (\nclusters*\clusterlength,0) + (\ngaps*\clusterspace,0) + (\size*\clusterlength,0) $);}
 \draw[tick] ($ (offset) + (0,\flagfacup*\lw) $) -- ($ (offset) - (0,\flagfacdown*\lw) $);
\coordinate (offset) at (0.0127792342183067, 1);
\foreach \nclusters\ngaps/\size in {0/0/6}{
 
        \draw[significant] ($ (offset) + (\nclusters*\clusterlength,0) + (\ngaps*\clusterspace,0) $) 
        -- ($ (offset) + (\nclusters*\clusterlength,0) + (\ngaps*\clusterspace,0) + (\size*\clusterlength,0) $);}
 \draw[tick] ($ (offset) + (0,\flagfacup*\lw) $) -- ($ (offset) - (0,\flagfacdown*\lw) $);
 \draw[thin, gray, dotted ,step=1] (0,0) grid ($ (0.05,\maxclusters) $);
        \foreach \x in {0.025, 0.05, 0.075} {
            \draw[tick,black,thin] (\x,-0.4) -- (\x,-0.1);
        }
        \foreach \x in {0.1} {
            \draw[tick,black,thin] (\x,-0.4) -- (\x,0.1);
        }
        \foreach \x in {0.1} {
            \node at (\x,-0.4) [below,black] { $\x$ };
        }
        \foreach \x in {0.05} {
            \draw[tick,black,thin] (\x,-0.4) -- (\x,0);
        }
        \foreach \x in {0.025, 0.05, 0.075} {
            \node at (\x,-0.4) [below,black] {\tiny $\x$ };
        }
        \foreach \y in {1,...,5} {
            \draw[tick,black,thin] (-0.0125,\y) -- (-0.01,\y);
            \node at (-0.0125,\y) [left,black] { $\y$ };
        }
        \draw[tick,black,thin]  ($ (0.005+0.1,\maxclusters+1) $) -- ($ (-0.01,\maxclusters+1) $) -- (-0.01,-0.4) -- ($ (0.005+0.1,-0.4) $) ;
        \node at ($(0.02+3*\maxepi,-2.2)$) [left,black] {\footnotesize epistatic units };
        \node at ($(-2.5*\maxepi+0.004,\maxclusters*0.2+0.5)$) [rotate=90] { \footnotesize dual edges };

\end{tikzpicture}
    \end{minipage}%
    \hfill%
    \begin{minipage}[t]{.45\linewidth}
    \caption{}
      \centering
      \newcommand\maxepi{0.0128}
\newcommand\maxclusters{6}
\newcommand\clusterlength{0.4pt}
\newcommand\clusterspace{0.1pt}
\newcommand\flagfacup{1.2}
\newcommand\flagfacdown{2}
\newcommand\lw{6pt}
\begin{tikzpicture}[yscale=0.5,xscale=50, scale = 0.8]
\tikzset{
        cluster/.style = {red, line width=\lw, join=round},
        significant/.style = {blue, line width=\lw, join=round},
        semisignificant/.style = {purple, line width=\lw, join=round},
        tick/.style = {black, thick},
        tick_thin/.style = {black, thin}}

  \coordinate (offset) at (0.00120070379309365, 5);

 \draw[cluster] ($ (offset) $)
        -- ($ (offset)  + (3*\clusterlength,0) $);

 % \node [draw] at ($  (offset)  + (3.5,0)$)  {1.309}; 
 \draw[tick] ($ (offset) + (0,\flagfacup*\lw) $) -- ($ (offset) - (0,\flagfacdown*\lw) $);

  \coordinate (offset) at (0.00139809385659613, 4);

 \draw[cluster] ($ (offset) $)
        -- ($ (offset)  + (3*\clusterlength,0) $);

 % \node [draw] at ($  (offset)  + (3.5,0)$)  {0.270}; 
 \draw[tick] ($ (offset) + (0,\flagfacup*\lw) $) -- ($ (offset) - (0,\flagfacdown*\lw) $);

  \coordinate (offset) at (0.000197390063502479, 3);

 \draw[cluster] ($ (offset) $)
        -- ($ (offset)  + (3*\clusterlength,0) $);

 % \node [draw] at ($  (offset)  + (3.5,0)$)  {0.032}; 
 \draw[tick] ($ (offset) + (0,\flagfacup*\lw) $) -- ($ (offset) - (0,\flagfacdown*\lw) $);

  \coordinate (offset) at (0.00248735630374838, 2);

 \draw[cluster] ($ (offset) $)
        -- ($ (offset)  + (3*\clusterlength,0) $);

 % \node [draw] at ($  (offset)  + (3.5,0)$)  {0.259}; 
 \draw[tick] ($ (offset) + (0,\flagfacup*\lw) $) -- ($ (offset) - (0,\flagfacdown*\lw) $);

  \coordinate (offset) at (0.0203295016984773, 1);

 \draw[significant] ($ (offset) $)
        -- ($ (offset)  + (3*\clusterlength,0) $);

 % \node [draw] at ($  (offset)  + (3.5,0)$)  {1.591}; 
 \draw[tick] ($ (offset) + (0,\flagfacup*\lw) $) -- ($ (offset) - (0,\flagfacdown*\lw) $);

\draw[thin, gray, dotted ,step=1] (0,0) grid ($ (0.05,\maxclusters) $);
        \foreach \x in {0.025, 0.05, 0.075} {
            \draw[tick,black,thin] (\x,-0.4) -- (\x,-0.1);
        }
        \foreach \x in {0.1} {
            \draw[tick,black,thin] (\x,-0.4) -- (\x,0.1);
        }
        \foreach \x in {0.1} {
            \node at (\x,-0.4) [below,black] { $\x$ };
        }
        \foreach \x in {0.05} {
            \draw[tick,black,thin] (\x,-0.4) -- (\x,0);
        }
        \foreach \x in {0.025, 0.05, 0.075} {
            \node at (\x,-0.4) [below,black] {\tiny $\x$ };
        }
        \foreach \y in {1,...,5} {
            \draw[tick,black,thin] (-0.0125,\y) -- (-0.01,\y);
            \node at (-0.0125,\y) [left,black] { $\y$ };
        }
        \draw[tick,black,thin]  ($ (0.005+0.1,\maxclusters+1) $) -- ($ (-0.01,\maxclusters+1) $) -- (-0.01,-0.4) -- ($ (0.005+0.1,-0.4) $) ;
        \node at ($(0.08,-2.2)$) [left,black] {\footnotesize parallel epistatic units };    
%\node at ($(-2.5*\maxepi+0.004,\maxclusters*0.2+0.5)$) [rotate=90] { \footnotesize dual edges };

\end{tikzpicture}
    \end{minipage}
    
  \end{subfigure}
  \begin{subfigure}{.85\textwidth}
  	\caption{}
    \vspace{5pt}
    \includegraphics[width=1.0\linewidth]{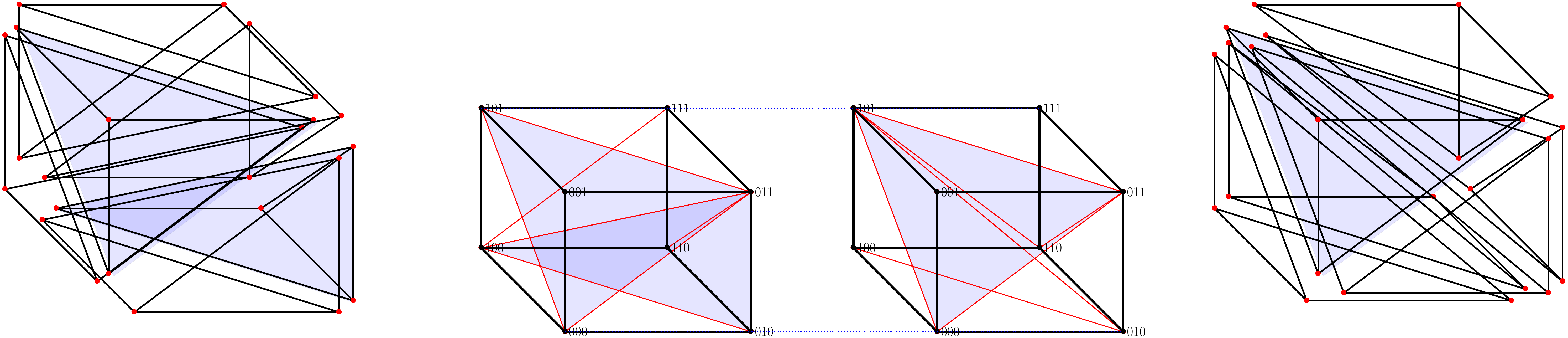}
  \end{subfigure}
  \caption[Parallel transport from $0{\ast}{\ast}0{\ast}$ to $1{\ast}{\ast}0{\ast}$ within the Khan data.]{{\bf Parallel transport from $0{\ast}{\ast}0{\ast}$ to $1{\ast}{\ast}0{\ast}$ within the Khan dataset.} (a) Filtration based on the triangulation of $0{\ast}{\ast}0{\ast}$. (b) Parallel epistatic weights computed from $1{\ast}{\ast}0{\ast}$ for the triangulation based on $0{\ast}{\ast}0{\ast}$. (c) The two parallel triangulations (and exploded copies) are depicted. The partitions in the node set are transferred from the cube on the middle left to the cube on the middle right. Exploded versions of these same triangulation on the far left and far right demonstrate the geometry of the simplices generated by the triangulations.}
  \label{fig:khan:0ss0s-1ss0s}
\end{figure}

Examples for the parallel transport of epistatic filtrations are shown in Figures \ref{fig:khan:0ss0s-1ss0s}, ~\ref{fig:khan_parallel_ssss0_ssss1}, ~\ref{fig:khan_parallel_ssss1_ssss0}, ~\ref{fig:weinreich:0ssss-1ssss}, and ~\ref{fig:weinreich:ss0ss-ss1ss}. 
The magnitude of the epistasis in the left panels are roughly comparable between data sets due to normalization of the input data. Compare each left panel with its corresponding right panel to observe the relative change in epistasis in the parallel path. Larger changes in epistasis indicate stronger context-dependence of the interaction. For instance, in the first Weinreich comparison (Fig.~\ref{fig:weinreich:0ssss-1ssss}), bar 10 in the right panel has a parallel epistasis greater than the original filtration on the left, indicating context-dependence.

\subsection{Meta-epistatic charts}\label{supplement:meta}
The {\bf meta-epistatic chart} is a diagram drawn on top of the induced epistatic filtrations for some selection of faces of a fixed cube; higher-order interactions induced by lower order interactions are marked as corresponding.

In (Fig.~\ref{fig4}b) and (Fig.~\ref{fig4}c) we exhibit an example for the Eble data set, with 5 loci, where we take the five 4-dimensional faces $0{\ast}{\ast}{\ast}{\ast}$, ${\ast}0{\ast}{\ast}{\ast}$, ${\ast}{\ast}0{\ast}{\ast}$, ${\ast}{\ast}{\ast}0{\ast}$ and ${\ast}{\ast}{\ast}{\ast}0$ into consideration.
Mathematically, these five 4-faces constitute the face figure of the wild type.
Fix one 4-face, say $0{\ast}{\ast}{\ast}{\ast}$.
The induced epistatic filtration on this face shows two blue bars corresponding to dual edges labeled $1$ and $2$.
Each of them refers to the ridge of a bipyramid, which is a $3$-dimensional simplex in this case.
These two ridges may intersect certain $3$-dimensional faces in the right dimension and thus may or may not descend to significant ridges within certain $3$-dimensional filtrations.
In case of an incidence with a lower dimensional significant ridge, the significant $4$-dimensional effect is induced by a lower dimensional effect and one may picture this fact as a directed assignment pointing from the lower towards the higher dimensional interaction.

\subsection{Comparison with a simple linear regression approach}\label{srm} 
In the theory of fitness landscapes many linear regression approaches have been proposed to study higher-order interactions, cf.~\cite{berringtondegonzalez2007,7,McCandlish,Sailer2017}. In this section, we compare our epistatic weight method to an elementary regression approach using an example from the data.   

The regression analysis we have in mind assumes that there is a linear relationship between the predictors $X_1,X_2,\dots,X_n$ (one associated to each locus/dimension of the genotope) and response, or dependent, variables $Y$ (associated to the biological measurements). That is, one assumes that $Y=f(X_1,X_2,\dots,X_n)+\epsilon$ where $f:\RR^n\rightarrow \RR;(X_1,X_2,\dots,X_n)\mapsto\beta_0+\beta_1 X_1+ \beta_2 X_2+\dots+\beta_n X_n$ and where $\epsilon$ is a random error term. The coefficients $\beta_1,\beta_2,\dots,\beta_n$ are unknown but can be estimated by minimizing the sum of squared residuals associated to the observations pairs $(x,y)$. These observations pairs 
consisting of a genotype and a measurement associated to it. 
Notice that more than one measurements are typically associated to a single genotype.
With the coefficient estimates one can make predictions for the dependent variable via
\begin{equation}
\label{reg:eq}
\hat{y}=\hat{\beta_0}+\hat{\beta_1} x_1+ \hat{\beta_2} x_2+\dots+\hat{\beta_n} x_n\enspace.
\end{equation}
The hat symbol $\hat{}$ indicates a prediction, for instance of $Y$ on the basis of $x_i=X_i$, or an estimate for an unknown coefficient.  

Below, we are interested in the differences between the observed measurements $y$ associated to the genotypes of $[0,1]^n$, expressed in terms of $x_1,x_2,\dots x_n$ and the predicated values $\hat{y}$ on the regression hyperplane (\ref{reg:eq}). Notice that the regression analysis remains unchanged after normalizing the height function to Euclidean norm one. Additionally, computing residues for all replicated measurements (when provided) and then take averages builds on the assumption that measurements associated to different genotypes are statistically independent from each other. This assumption is consistent with the one underlying the computation of statistical significances for epistatic weights, following \cite[\S.~4.2-4.3]{Eble2019}. 

\begin{remark}
In the regression setting of (\ref{reg:eq}) there are hypothesis tests (like the $F$-statistic, $t$-statistics and $p$-value) to answer if at least one regression coefficient $\beta_j,1\leq j\leq n$ is nonzero, see for example \cite{JWHT}. Such statistical approaches are different from the one  in \cite[\S.~4.2-4.3]{Eble2019}, where other hypothesis tests for each epistatic weight were proposed.
\end{remark}

\subsubsection{Regression for Eble data}
In the following, we perform a regression analysis focusing on the replicated measurements for the lifespan fitness landscape on $[0,1]^5$ obtained from Eble and subspaces thereof.
Numerical measures of model fit ($F$-statistic: 2357, with $p$-value essentially zero, and for 3840 observations and 5 predictors) show that the multiple linear regression model can be considered to be appropriated for this data.  
Since the epistatic weights of the dual edges are close to zero ($\leq 0.02$) and are mostly not significant, the above regression analysis conclusion is in line with what we see from the filtration of epistatic weights associated to the same fitness landscapes, see (Fig.~\ref{fig:Eble-5cube}).

From this example we see that the regression approach provides some general information on higher-order interactions. However, without further assumptions, only one interaction formula is given in terms of a regression hyperplane (\ref{reg:eq}) while the epistatic weight approach gives more fine grained information.
This example also illustrate that when the regression model fits the data well (essentially the higher the $F$-statistics and the more coefficients in the hyperplane equation are significantly non-zero) the epistatic filtration has little horizontal shifts and few significant epistatic weights.

We now proceed repeating the above analysis on some of the bipyramids considered in the
parallel analysis for the normalized lifespan Eble data.
Regressing over bipyramid 23 in Table \ref{tab:Eble:parallel:0to1****}
\[
\{0001\}+\{0000,1001,1011,0111\}+\{1111\}
\]
in ${0}{\ast}{\ast}{\ast}{\ast}$ and ${1}{\ast}{\ast}{\ast}{\ast}$ reveals that only two average residues over 
${0}{\ast}{\ast}{\ast}{\ast}$
are non-zero (associated to the microbiomes ${0}{0}{0}{0}{0}$ and ${0}{0}{0}{0}{1}$), and only one is non-zero over
${1}{\ast}{\ast}{\ast}{\ast}$ (associated to the microbiome ${1}{0}{0}{0}{0}$). This confirms the two non significant epistatic weights
over bipyramid 23 in Table \ref{tab:Eble:parallel:0to1****}. 

\begin{remark}
If minimally dependent sets of points in the genotope are fixed, as in the epistatic weight approach, and one regresses above these points, then the corresponding regression hyperplanes equations are learned from data and the equations generally differ from the epistatic weights given as in \eqref{eq:epistatic-weight}, but similar biological and geometric conclusions can be drawn. This idea could then be taken further by considering smoothing splines, instead of linear regression, and their relation to epistatic filtrations. From an application point of view, one would obtain an interesting new extension of the concept of epistasis because intermediate genotypes could be assessed, which would correspond to the case of genetically heterogeneous populations of organisms as occur in nature. 
\end{remark}

Other numerical results for the above regressions are summarized in Table \ref{reg:bipy1}.
Over ${0}{\ast}{\ast}{\ast}{\ast}$ two coefficients are significantly non-zero (for $x_1$ and $x_4$), see top part of Table \ref{reg:bipy1}.
Similarly, over ${1}{\ast}{\ast}{\ast}{\ast}$ four coefficients are significantly non-zero ($x_1,x_2,x_3,x_4$), see bottom part of Table \ref{reg:bipy1}.
The fit of the linear regression models is confirmed by the relatively high values of the $F$-statistic.
Over ${0}{\ast}{\ast}{\ast}{v}$  the $F$-statistics is $459.1$ for a $p$-value near zero and 720 observations.
Over ${1}{\ast}{\ast}{\ast}{\ast}$ the corresponding $F$-statistics (near zero) is $52.61$.
%Over ${1}{\ast}{\ast}{\ast}{\ast}$ the $F$-statistics is $52.61$ for a $p$-value essentially of zero and 720 observations.           

\subsection{Microbiome data sets}

In this work, \emph{Drosophila} microbiome fitness landscapes consist of experimental measurements on germ-free \emph{Drosophila} 
flies inoculated with different bacterial species. The lifespan of approximately 100 individual flies were measured for each combination of bacterial species, giving roughly 3,200 individual fly lifespans for each of the two data sets presented. The experimental methods are described in 
\cite{Gould232959,IMA:IMA2}. The first data set is the exact data presented in \cite{Gould232959,IMA:IMA2}. The second data set is the second set of species with exactly the same methods used in \cite{Gould232959,IMA:IMA2}. The bacterial compositions considered consist of
all possible combinations of five species. The species considered can all occur naturally in the gut of wild flies:
\emph{Lactobacillus plantarum} (LP), \emph{Lactobacillus brevis} (LB), 
\emph{Acetobacter pasteurianus} (APa), \emph{Acetobacter tropicalis} (AT),  \emph{Acetobacter orientalis} (AO),
\emph{Acetobacter cerevisiae}  (AC),
\emph{Acetobacter malorum} (AM).
The $5$-member communities both stably persist in the fly gut. For the purposes of this work, we define {\bf stable} as maintaining colonization of the gut when $\leq{}20$ flies are co-housed in a standard fly vial and transferred daily to fresh food containing 10\% glucose, 5\% live yeast that has subsequently been autoclaved, 1.2\% agar, and 0.42\% propionic acid, with a pH of 4.5. The total number of species found stably associated with an individual fly is typically between 3 and 8. 
Consistently, \emph{Lactobacillus plantarum} and \emph{Lactobacillus brevis}, are found with two to three \emph{Acetobacter} species. 
Less consistently, species of \emph{Enterobacteria} and \emph{Enterococci} occur, and these have been described as pathogens. 
While more strains may be present, for each of the two data sets in the present work, a set of five non pathogen species was chosen, including the two \emph{Lactobacilli} and three \emph{Acetobacter} species. 
The combinations of species are shown in Table \ref{tab:datasets}. Different strains of the same species were used in the two data sets.

% FigS2
\begin{figure}[ht]
\flushleft
  \begin{subfigure}[t]{.47\linewidth}\vspace{0pt}
    \subcaption{} 
    \input{Figures/filtration_khanraw_norm_ssss0.tikz}    
  \end{subfigure}
    \begin{subfigure}[t]{.47\linewidth}\vspace{0pt}
    \subcaption{} 
    \newcommand\maxepi{0.0138}
\newcommand\maxclusters{23}
\newcommand\clusterlength{0.09pt}
\newcommand\clusterspace{0.045pt}
\newcommand\flagfacup{2.5}
\newcommand\flagfacdown{4}
\newcommand\lw{3pt}
\begin{tikzpicture}[yscale=0.5,xscale=110, scale = 0.4]
\tikzset{
        cluster/.style = {red, line width=\lw, join=round},
        significant/.style = {blue, line width=\lw, join=round},
        semisignificant/.style = {purple, line width=\lw, join=round},
        tick/.style = {black, thick},
        tick_thin/.style = {black, thin}}

  \coordinate (offset) at (0.000655824560775758, 20);

 \draw[cluster] ($ (offset) $)-- ($ (offset)  + (10*\clusterlength,0) $);
 \draw[tick] ($ (offset) + (0,\flagfacup*\lw) $) -- ($ (offset) - (0,\flagfacdown*\lw) $);

  \coordinate (offset) at (0.00881410476736332, 19);

 \draw[cluster] ($ (offset) $)-- ($ (offset)  + (10*\clusterlength,0) $);
 \draw[tick] ($ (offset) + (0,\flagfacup*\lw) $) -- ($ (offset) - (0,\flagfacdown*\lw) $);

  \coordinate (offset) at (0.00881410476736332, 18);

 \draw[cluster] ($ (offset) $)-- ($ (offset)  + (10*\clusterlength,0) $);
 \draw[tick] ($ (offset) + (0,\flagfacup*\lw) $) -- ($ (offset) - (0,\flagfacdown*\lw) $);

  \coordinate (offset) at (0.00647061369431488, 17);

 \draw[cluster] ($ (offset) $)-- ($ (offset)  + (10*\clusterlength,0) $);
 \draw[tick] ($ (offset) + (0,\flagfacup*\lw) $) -- ($ (offset) - (0,\flagfacdown*\lw) $);

  \coordinate (offset) at (0.00346557833551319, 16);

 \draw[cluster] ($ (offset) $)-- ($ (offset)  + (10*\clusterlength,0) $);
 \draw[tick] ($ (offset) + (0,\flagfacup*\lw) $) -- ($ (offset) - (0,\flagfacdown*\lw) $);

  \coordinate (offset) at (0.0102136813319327, 15);

 \draw[semisignificant] ($ (offset) $)-- ($ (offset)  + (10*\clusterlength,0) $);
 \draw[tick] ($ (offset) + (0,\flagfacup*\lw) $) -- ($ (offset) - (0,\flagfacdown*\lw) $);

  \coordinate (offset) at (0.0109188846290803, 14);

 \draw[semisignificant] ($ (offset) $)-- ($ (offset)  + (10*\clusterlength,0) $);
 \draw[tick] ($ (offset) + (0,\flagfacup*\lw) $) -- ($ (offset) - (0,\flagfacdown*\lw) $);

  \coordinate (offset) at (0.00499855830913983, 13);

 \draw[cluster] ($ (offset) $)-- ($ (offset)  + (10*\clusterlength,0) $);
 \draw[tick] ($ (offset) + (0,\flagfacup*\lw) $) -- ($ (offset) - (0,\flagfacdown*\lw) $);

  \coordinate (offset) at (0.00499855830913983, 12);

 \draw[cluster] ($ (offset) $)-- ($ (offset)  + (10*\clusterlength,0) $);
 \draw[tick] ($ (offset) + (0,\flagfacup*\lw) $) -- ($ (offset) - (0,\flagfacdown*\lw) $);

  \coordinate (offset) at (0.0112335767610364, 11);

 \draw[significant] ($ (offset) $)-- ($ (offset)  + (10*\clusterlength,0) $);
 \draw[tick] ($ (offset) + (0,\flagfacup*\lw) $) -- ($ (offset) - (0,\flagfacdown*\lw) $);

  \coordinate (offset) at (0.0133728479507825, 10);

 \draw[semisignificant] ($ (offset) $)-- ($ (offset)  + (10*\clusterlength,0) $);
 \draw[tick] ($ (offset) + (0,\flagfacup*\lw) $) -- ($ (offset) - (0,\flagfacdown*\lw) $);

  \coordinate (offset) at (0.00858067549764409, 9);

 \draw[cluster] ($ (offset) $)-- ($ (offset)  + (10*\clusterlength,0) $);
 \draw[tick] ($ (offset) + (0,\flagfacup*\lw) $) -- ($ (offset) - (0,\flagfacdown*\lw) $);

  \coordinate (offset) at (0.0202870287256131, 8);

 \draw[significant] ($ (offset) $)-- ($ (offset)  + (10*\clusterlength,0) $);
 \draw[tick] ($ (offset) + (0,\flagfacup*\lw) $) -- ($ (offset) - (0,\flagfacdown*\lw) $);

  \coordinate (offset) at (0.00456687695383449, 7);

 \draw[cluster] ($ (offset) $)-- ($ (offset)  + (10*\clusterlength,0) $);
 \draw[tick] ($ (offset) + (0,\flagfacup*\lw) $) -- ($ (offset) - (0,\flagfacdown*\lw) $);

  \coordinate (offset) at (0.0192525063621299, 6);

 \draw[significant] ($ (offset) $)-- ($ (offset)  + (10*\clusterlength,0) $);
 \draw[tick] ($ (offset) + (0,\flagfacup*\lw) $) -- ($ (offset) - (0,\flagfacdown*\lw) $);

  \coordinate (offset) at (0.0206156259048855, 5);

 \draw[significant] ($ (offset) $)-- ($ (offset)  + (10*\clusterlength,0) $);
 \draw[tick] ($ (offset) + (0,\flagfacup*\lw) $) -- ($ (offset) - (0,\flagfacdown*\lw) $);

  \coordinate (offset) at (0.000882690725763545, 4);

 \draw[cluster] ($ (offset) $)-- ($ (offset)  + (10*\clusterlength,0) $);
 \draw[tick] ($ (offset) + (0,\flagfacup*\lw) $) -- ($ (offset) - (0,\flagfacdown*\lw) $);

  \coordinate (offset) at (0.00944803074926443, 3);

 \draw[cluster] ($ (offset) $)-- ($ (offset)  + (10*\clusterlength,0) $);
 \draw[tick] ($ (offset) + (0,\flagfacup*\lw) $) -- ($ (offset) - (0,\flagfacdown*\lw) $);

  \coordinate (offset) at (0.0274688034974158, 2);

 \draw[significant] ($ (offset) $)-- ($ (offset)  + (10*\clusterlength,0) $);
 \draw[tick] ($ (offset) + (0,\flagfacup*\lw) $) -- ($ (offset) - (0,\flagfacdown*\lw) $);

  \coordinate (offset) at (0.0303982029388763, 1);

 \draw[significant] ($ (offset) $)-- ($ (offset)  + (10*\clusterlength,0) $);
 \draw[tick] ($ (offset) + (0,\flagfacup*\lw) $) -- ($ (offset) - (0,\flagfacdown*\lw) $);

\draw[thin, gray, dotted ,step=1] (0,0) grid ($ (0.1,\maxclusters) $);
        \foreach \x in {0.025, 0.05,...,0.1} {
            \draw[tick,black,thin] (\x,-0.4) -- (\x,-0.1);
        }
        \foreach \x in {0.025,0.05,0.075} {
            \node at (\x,-0.4) [below,black] {\footnotesize $\x$ };
        }
        \foreach \x in {0.1} {
            \draw[tick,black,thin] (\x,-0.4) -- (\x,0.15);
            \node at (\x,-0.4) [below,black] { $\x$ };
        }
        \foreach \y in {1,5,10,...,\maxclusters} {
            \draw[tick,black,thin] (-0.005,\y) -- (-0.008,\y);
            \node at (-0.007,\y) [left,black] { $\y$ };
        }
        \draw[tick,black,thin]  ($ (0.107,\maxclusters+1) $) -- ($ (-0.005,\maxclusters+1) $) -- (-0.005,-0.4) -- ($ (0.107,-0.4) $) ;
        \node at ($(0.075,-2.5-1)$) [left,black] {\footnotesize parallel epistatic units };
        %\node at ($(-0.022,\maxclusters*0.2)$) [rotate=90] { \footnotesize dual edges };

\end{tikzpicture}    
  \end{subfigure}
  \caption[Epistatic filtration and parallel epistatic units for transport from ${\ast}{\ast}{\ast}{\ast}0$ to ${\ast}{\ast}{\ast}{\ast}1$ within the Khan data.]{Epistatic filtration and parallel epistatic units for transport from ${\ast}{\ast}{\ast}{\ast}0$ to ${\ast}{\ast}{\ast}{\ast}1$ within the Khan data.}
  \label{fig:khan_parallel_ssss0_ssss1}
\end{figure}

% FigS3
\begin{figure}[ht]
\flushleft
  \begin{subfigure}[t]{.47\linewidth}\vspace{0pt}
    \subcaption{} 
    \input{Figures/filtration_khanraw_norm_ssss1.tikz}    
  \end{subfigure}
    \begin{subfigure}[t]{.47\linewidth}\vspace{0pt}
    \subcaption{} 
    \newcommand\maxepi{0.0138}
\newcommand\maxclusters{23}
\newcommand\clusterlength{0.09pt}
\newcommand\clusterspace{0.045pt}
\newcommand\flagfacup{2.5}
\newcommand\flagfacdown{4}
\newcommand\lw{3pt}
\begin{tikzpicture}[yscale=0.5,xscale=110, scale = 0.4]
\tikzset{
        cluster/.style = {red, line width=\lw, join=round},
        significant/.style = {blue, line width=\lw, join=round},
        semisignificant/.style = {purple, line width=\lw, join=round},
        tick/.style = {black, thick},
        tick_thin/.style = {black, thin}}

 \coordinate (offset) at (0.000363805676314371, 22);

 \draw[cluster] ($ (offset) $)-- ($ (offset)  + (10*\clusterlength,0) $);
 \draw[tick] ($ (offset) + (0,\flagfacup*\lw) $) -- ($ (offset) - (0,\flagfacdown*\lw) $);

  \coordinate (offset) at (0.000445569136249175, 21);

 \draw[cluster] ($ (offset) $)-- ($ (offset)  + (10*\clusterlength,0) $);
 \draw[tick] ($ (offset) + (0,\flagfacup*\lw) $) -- ($ (offset) - (0,\flagfacdown*\lw) $);

  \coordinate (offset) at (0.0067174966532479, 20);

 \draw[cluster] ($ (offset) $)-- ($ (offset)  + (10*\clusterlength,0) $);
 \draw[tick] ($ (offset) + (0,\flagfacup*\lw) $) -- ($ (offset) - (0,\flagfacdown*\lw) $);

  \coordinate (offset) at (0.00723110526841708, 19);

 \draw[cluster] ($ (offset) $)-- ($ (offset)  + (10*\clusterlength,0) $);
 \draw[tick] ($ (offset) + (0,\flagfacup*\lw) $) -- ($ (offset) - (0,\flagfacdown*\lw) $);

  \coordinate (offset) at (0.000493938123736622, 18);

 \draw[cluster] ($ (offset) $)-- ($ (offset)  + (10*\clusterlength,0) $);
 \draw[tick] ($ (offset) + (0,\flagfacup*\lw) $) -- ($ (offset) - (0,\flagfacdown*\lw) $);

  \coordinate (offset) at (0.00304695777206327, 17);

 \draw[cluster] ($ (offset) $)-- ($ (offset)  + (10*\clusterlength,0) $);
 \draw[tick] ($ (offset) + (0,\flagfacup*\lw) $) -- ($ (offset) - (0,\flagfacdown*\lw) $);

  \coordinate (offset) at (0.000520734513225139, 16);

 \draw[cluster] ($ (offset) $)-- ($ (offset)  + (10*\clusterlength,0) $);
 \draw[tick] ($ (offset) + (0,\flagfacup*\lw) $) -- ($ (offset) - (0,\flagfacdown*\lw) $);

  \coordinate (offset) at (0.000637766924429084, 15);

 \draw[cluster] ($ (offset) $)-- ($ (offset)  + (10*\clusterlength,0) $);
 \draw[tick] ($ (offset) + (0,\flagfacup*\lw) $) -- ($ (offset) - (0,\flagfacdown*\lw) $);

  \coordinate (offset) at (0.00253180193533068, 14);

 \draw[cluster] ($ (offset) $)-- ($ (offset)  + (10*\clusterlength,0) $);
 \draw[tick] ($ (offset) + (0,\flagfacup*\lw) $) -- ($ (offset) - (0,\flagfacdown*\lw) $);

  \coordinate (offset) at (0.0029188898386073, 13);

 \draw[cluster] ($ (offset) $)-- ($ (offset)  + (10*\clusterlength,0) $);
 \draw[tick] ($ (offset) + (0,\flagfacup*\lw) $) -- ($ (offset) - (0,\flagfacdown*\lw) $);

  \coordinate (offset) at (0.0029188898386073, 12);

 \draw[cluster] ($ (offset) $)-- ($ (offset)  + (10*\clusterlength,0) $);
 \draw[tick] ($ (offset) + (0,\flagfacup*\lw) $) -- ($ (offset) - (0,\flagfacdown*\lw) $);

  \coordinate (offset) at (0.0029188898386073, 11);

 \draw[cluster] ($ (offset) $)-- ($ (offset)  + (10*\clusterlength,0) $);
 \draw[tick] ($ (offset) + (0,\flagfacup*\lw) $) -- ($ (offset) - (0,\flagfacdown*\lw) $);

  \coordinate (offset) at (0.00310081143567555, 10);

 \draw[cluster] ($ (offset) $)-- ($ (offset)  + (10*\clusterlength,0) $);
 \draw[tick] ($ (offset) + (0,\flagfacup*\lw) $) -- ($ (offset) - (0,\flagfacdown*\lw) $);

  \coordinate (offset) at (0.00514375181927853, 9);

 \draw[cluster] ($ (offset) $)-- ($ (offset)  + (10*\clusterlength,0) $);
 \draw[tick] ($ (offset) + (0,\flagfacup*\lw) $) -- ($ (offset) - (0,\flagfacdown*\lw) $);

  \coordinate (offset) at (0.00785721068930355, 8);

 \draw[cluster] ($ (offset) $)-- ($ (offset)  + (10*\clusterlength,0) $);
 \draw[tick] ($ (offset) + (0,\flagfacup*\lw) $) -- ($ (offset) - (0,\flagfacdown*\lw) $);

  \coordinate (offset) at (0.00250851970836844, 7);

 \draw[cluster] ($ (offset) $)-- ($ (offset)  + (10*\clusterlength,0) $);
 \draw[tick] ($ (offset) + (0,\flagfacup*\lw) $) -- ($ (offset) - (0,\flagfacdown*\lw) $);

  \coordinate (offset) at (0.0113186013857707, 6);

 \draw[significant] ($ (offset) $)-- ($ (offset)  + (10*\clusterlength,0) $);
 \draw[tick] ($ (offset) + (0,\flagfacup*\lw) $) -- ($ (offset) - (0,\flagfacdown*\lw) $);

  \coordinate (offset) at (0.00112919452904339, 5);

 \draw[cluster] ($ (offset) $)-- ($ (offset)  + (10*\clusterlength,0) $);
 \draw[tick] ($ (offset) + (0,\flagfacup*\lw) $) -- ($ (offset) - (0,\flagfacdown*\lw) $);

  \coordinate (offset) at (0.0010722131593783, 4);

 \draw[cluster] ($ (offset) $)-- ($ (offset)  + (10*\clusterlength,0) $);
 \draw[tick] ($ (offset) + (0,\flagfacup*\lw) $) -- ($ (offset) - (0,\flagfacdown*\lw) $);

  \coordinate (offset) at (0.0147165721434216, 3);

 \draw[significant] ($ (offset) $)-- ($ (offset)  + (10*\clusterlength,0) $);
 \draw[tick] ($ (offset) + (0,\flagfacup*\lw) $) -- ($ (offset) - (0,\flagfacdown*\lw) $);

  \coordinate (offset) at (0.0053535062636485, 2);

 \draw[cluster] ($ (offset) $)-- ($ (offset)  + (10*\clusterlength,0) $);
 \draw[tick] ($ (offset) + (0,\flagfacup*\lw) $) -- ($ (offset) - (0,\flagfacdown*\lw) $);

  \coordinate (offset) at (0.00574774659858735, 1);

 \draw[cluster] ($ (offset) $)-- ($ (offset)  + (10*\clusterlength,0) $);
 \draw[tick] ($ (offset) + (0,\flagfacup*\lw) $) -- ($ (offset) - (0,\flagfacdown*\lw) $);

\draw[thin, gray, dotted ,step=1] (0,0) grid ($ (0.1,\maxclusters) $);
        \foreach \x in {0.025, 0.05,...,0.1} {
            \draw[tick,black,thin] (\x,-0.4) -- (\x,-0.1);
        }
        \foreach \x in {0.025,0.05,0.075} {
            \node at (\x,-0.4) [below,black] {\footnotesize $\x$ };
        }
        \foreach \x in {0.1} {
            \draw[tick,black,thin] (\x,-0.4) -- (\x,0.15);
            \node at (\x,-0.4) [below,black] { $\x$ };
        }
        \foreach \y in {1,5,10,...,\maxclusters} {
            \draw[tick,black,thin] (-0.005,\y) -- (-0.008,\y);
            \node at (-0.007,\y) [left,black] { $\y$ };
        }
        \draw[tick,black,thin]  ($ (0.107,\maxclusters+1) $) -- ($ (-0.005,\maxclusters+1) $) -- (-0.005,-0.4) -- ($ (0.107,-0.4) $) ;
        \node at ($(0.075,-2.5-1)$) [left,black] {\footnotesize parallel epistatic units };
        %\node at ($(-0.022,\maxclusters*0.2)$) [rotate=90] { \footnotesize dual edges };

\end{tikzpicture}    
  \end{subfigure}
  \caption[Epistatic filtration and parallel epistatic units for transport from ${\ast}{\ast}{\ast}{\ast}1$ to ${\ast}{\ast}{\ast}{\ast}0$ within the Khan data.]{Epistatic filtration and parallel epistatic units for transport from ${\ast}{\ast}{\ast}{\ast}1$ to ${\ast}{\ast}{\ast}{\ast}0$ within the Khan data.}
  \label{fig:khan_parallel_ssss1_ssss0}
\end{figure}

% FigS4
\begin{figure}[ht]
  \centering
  \includegraphics[width=.95\linewidth]{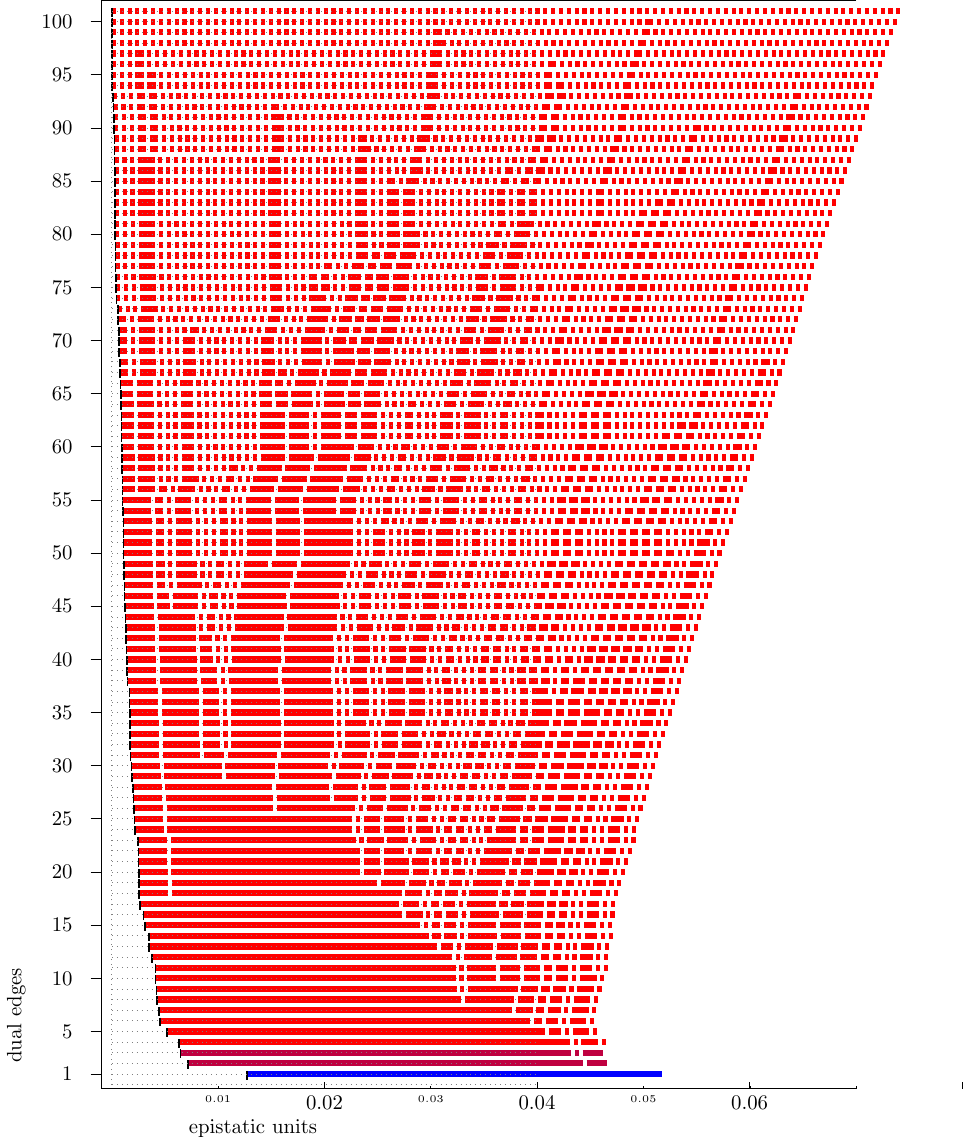}  
  \caption[Complete filtration of the Khan data over the whole $5$-cube.]{\bf Complete filtration of the Khan data over the whole $5$-cube.}
  \label{fig:Khan-5cube}
\end{figure}

% Fig S5
\begin{figure}[ht]
  \begin{minipage}[t]{.45\linewidth}
    \centering
    \input{Figures/filtration_weinreich_norm_0ssss.tikz}
  \end{minipage}%
  \hfill%
  \begin{minipage}[t]{.45\linewidth}
    \centering
    \newcommand\maxepi{0.1043}
            \newcommand\maxclusters{24}
            \newcommand\clusterlength{0.2pt}
            \newcommand\clusterspace{0.1pt}
            \newcommand\flagfacup{2.5}
            \newcommand\flagfacdown{4}
            \newcommand\lw{3pt}

\begin{tikzpicture}[yscale=0.5,xscale=45, scale = 0.4]
\tikzset{
        cluster/.style = {red, line width=\lw, join=round},
        significant/.style = {blue, line width=\lw, join=round},
        semisignificant/.style = {purple, line width=\lw, join=round},
        tick/.style = {black, thick},
        tick_thin/.style = {black, thin}}
        
 \coordinate (offset) at (5.04523836451822e-19, 23);

 \draw[cluster] ($ (offset) $)
        -- ($ (offset)  + (10*\clusterlength,0) $);

  %\node [draw] at ($  (offset)  + (3.5,0)$)  {0.000}; 
 \draw[tick] ($ (offset) + (0,\flagfacup*\lw) $) -- ($ (offset) - (0,\flagfacdown*\lw) $);

  \coordinate (offset) at (5.04523836451822e-19, 22);

 \draw[cluster] ($ (offset) $)
        -- ($ (offset)  + (10*\clusterlength,0) $);

  %\node [draw] at ($  (offset)  + (3.5,0)$)  {0.000}; 
 \draw[tick] ($ (offset) + (0,\flagfacup*\lw) $) -- ($ (offset) - (0,\flagfacdown*\lw) $);

  \coordinate (offset) at (0.0167231959431716, 21);

 \draw[cluster] ($ (offset) $)
        -- ($ (offset)  + (10*\clusterlength,0) $);

  %\node [draw] at ($  (offset)  + (3.5,0)$)  {5.750}; 
 \draw[tick] ($ (offset) + (0,\flagfacup*\lw) $) -- ($ (offset) - (0,\flagfacdown*\lw) $);

  \coordinate (offset) at (0.0181773868947517, 20);

 \draw[cluster] ($ (offset) $)
        -- ($ (offset)  + (10*\clusterlength,0) $);

  %\node [draw] at ($  (offset)  + (3.5,0)$)  {6.250}; 
 \draw[tick] ($ (offset) + (0,\flagfacup*\lw) $) -- ($ (offset) - (0,\flagfacdown*\lw) $);

  \coordinate (offset) at (0.0167231959431716, 19);

 \draw[cluster] ($ (offset) $)
        -- ($ (offset)  + (10*\clusterlength,0) $);

  %\node [draw] at ($  (offset)  + (3.5,0)$)  {5.412}; 
 \draw[tick] ($ (offset) + (0,\flagfacup*\lw) $) -- ($ (offset) - (0,\flagfacdown*\lw) $);

  \coordinate (offset) at (0.0181773868947517, 18);

 \draw[cluster] ($ (offset) $)
        -- ($ (offset)  + (10*\clusterlength,0) $);

  %\node [draw] at ($  (offset)  + (3.5,0)$)  {5.882}; 
 \draw[tick] ($ (offset) + (0,\flagfacup*\lw) $) -- ($ (offset) - (0,\flagfacdown*\lw) $);

  \coordinate (offset) at (0.00618031154421556, 17);

 \draw[cluster] ($ (offset) $)
        -- ($ (offset)  + (10*\clusterlength,0) $);

  %\node [draw] at ($  (offset)  + (3.5,0)$)  {1.030}; 
 \draw[tick] ($ (offset) + (0,\flagfacup*\lw) $) -- ($ (offset) - (0,\flagfacdown*\lw) $);

  \coordinate (offset) at (0.000445253227492946, 16);

 \draw[cluster] ($ (offset) $)
        -- ($ (offset)  + (10*\clusterlength,0) $);

  %\node [draw] at ($  (offset)  + (3.5,0)$)  {0.043}; 
 \draw[tick] ($ (offset) + (0,\flagfacup*\lw) $) -- ($ (offset) - (0,\flagfacdown*\lw) $);

  \coordinate (offset) at (0.000445253227492946, 15);

 \draw[cluster] ($ (offset) $)
        -- ($ (offset)  + (10*\clusterlength,0) $);

  %\node [draw] at ($  (offset)  + (3.5,0)$)  {0.043}; 
 \draw[tick] ($ (offset) + (0,\flagfacup*\lw) $) -- ($ (offset) - (0,\flagfacdown*\lw) $);

  \coordinate (offset) at (0.0243576984389672, 14);

 \draw[cluster] ($ (offset) $)
        -- ($ (offset)  + (10*\clusterlength,0) $);

  %\node [draw] at ($  (offset)  + (3.5,0)$)  {2.000}; 
 \draw[tick] ($ (offset) + (0,\flagfacup*\lw) $) -- ($ (offset) - (0,\flagfacdown*\lw) $);

  \coordinate (offset) at (0.00874028040552065, 13);

 \draw[cluster] ($ (offset) $)
        -- ($ (offset)  + (10*\clusterlength,0) $);

  %\node [draw] at ($  (offset)  + (3.5,0)$)  {0.680}; 
 \draw[tick] ($ (offset) + (0,\flagfacup*\lw) $) -- ($ (offset) - (0,\flagfacdown*\lw) $);

  \coordinate (offset) at (0.0947041857216562, 12);

 \draw[cluster] ($ (offset) $)
        -- ($ (offset)  + (10*\clusterlength,0) $);

  %\node [draw] at ($  (offset)  + (3.5,0)$)  {5.854}; 
 \draw[tick] ($ (offset) + (0,\flagfacup*\lw) $) -- ($ (offset) - (0,\flagfacdown*\lw) $);

  \coordinate (offset) at (0.0792534068611174, 11);

 \draw[cluster] ($ (offset) $)
        -- ($ (offset)  + (10*\clusterlength,0) $);

  %\node [draw] at ($  (offset)  + (3.5,0)$)  {4.317}; 
 \draw[tick] ($ (offset) + (0,\flagfacup*\lw) $) -- ($ (offset) - (0,\flagfacdown*\lw) $);

  \coordinate (offset) at (0.0712553566274265, 10);

 \draw[cluster] ($ (offset) $)
        -- ($ (offset)  + (10*\clusterlength,0) $);

  %\node [draw] at ($  (offset)  + (3.5,0)$)  {3.409}; 
 \draw[tick] ($ (offset) + (0,\flagfacup*\lw) $) -- ($ (offset) - (0,\flagfacdown*\lw) $);

  \coordinate (offset) at (0, 9);

 \draw[cluster] ($ (offset) $)
        -- ($ (offset)  + (10*\clusterlength,0) $);

  %\node [draw] at ($  (offset)  + (3.5,0)$)  {0.000}; 
 \draw[tick] ($ (offset) + (0,\flagfacup*\lw) $) -- ($ (offset) - (0,\flagfacdown*\lw) $);

  \coordinate (offset) at (0.100770291734238, 8);

 \draw[cluster] ($ (offset) $)
        -- ($ (offset)  + (10*\clusterlength,0) $);

  %\node [draw] at ($  (offset)  + (3.5,0)$)  {3.409}; 
 \draw[tick] ($ (offset) + (0,\flagfacup*\lw) $) -- ($ (offset) - (0,\flagfacdown*\lw) $);

  \coordinate (offset) at (0.0916140299495484, 7);

 \draw[cluster] ($ (offset) $)
        -- ($ (offset)  + (10*\clusterlength,0) $);

  %\node [draw] at ($  (offset)  + (3.5,0)$)  {3.055}; 
 \draw[tick] ($ (offset) + (0,\flagfacup*\lw) $) -- ($ (offset) - (0,\flagfacdown*\lw) $);

  \coordinate (offset) at (0.115971728388516, 6);

 \draw[cluster] ($ (offset) $)
        -- ($ (offset)  + (10*\clusterlength,0) $);

  %\node [draw] at ($  (offset)  + (3.5,0)$)  {2.750}; 
 \draw[tick] ($ (offset) + (0,\flagfacup*\lw) $) -- ($ (offset) - (0,\flagfacdown*\lw) $);

  \coordinate (offset) at (0.0422990566118308, 5);

 \draw[cluster] ($ (offset) $)
        -- ($ (offset)  + (10*\clusterlength,0) $);

  %\node [draw] at ($  (offset)  + (3.5,0)$)  {0.841}; 
 \draw[tick] ($ (offset) + (0,\flagfacup*\lw) $) -- ($ (offset) - (0,\flagfacdown*\lw) $);

  \coordinate (offset) at (0.0422990566118308, 4);

 \draw[cluster] ($ (offset) $)
        -- ($ (offset)  + (10*\clusterlength,0) $);

  %\node [draw] at ($  (offset)  + (3.5,0)$)  {0.841}; 
 \draw[tick] ($ (offset) + (0,\flagfacup*\lw) $) -- ($ (offset) - (0,\flagfacdown*\lw) $);

  \coordinate (offset) at (0.0230852813563346, 3);

 \draw[cluster] ($ (offset) $)
        -- ($ (offset)  + (10*\clusterlength,0) $);

  %\node [draw] at ($  (offset)  + (3.5,0)$)  {0.380}; 
 \draw[tick] ($ (offset) + (0,\flagfacup*\lw) $) -- ($ (offset) - (0,\flagfacdown*\lw) $);

  \coordinate (offset) at (0.0279931758179176, 2);

 \draw[cluster] ($ (offset) $)
        -- ($ (offset)  + (10*\clusterlength,0) $);

  %\node [draw] at ($  (offset)  + (3.5,0)$)  {0.448}; 
 \draw[tick] ($ (offset) + (0,\flagfacup*\lw) $) -- ($ (offset) - (0,\flagfacdown*\lw) $);

  \coordinate (offset) at (0.0181773868947517, 1);

 \draw[cluster] ($ (offset) $)
        -- ($ (offset)  + (10*\clusterlength,0) $);

  %\node [draw] at ($  (offset)  + (3.5,0)$)  {0.174}; 
 \draw[tick] ($ (offset) + (0,\flagfacup*\lw) $) -- ($ (offset) - (0,\flagfacdown*\lw) $);

\draw[thin, gray, dotted ,step=1] (0,0) grid ($ (0.2,\maxclusters) $);

        \foreach \x in {0.025,0.05,..., 0.225} {
            \draw[tick,black,thin] (\x,-0.4) -- (\x,-0.1);
        }
        \foreach \x in {0.05,0.15} {
            \node at (\x,-0.4) [below,black] {\footnotesize $\x$ };
        }
        \foreach \x in {0.1,0.2} {
            \draw[tick,black,thin] (\x,-0.4) -- (\x,0.15);
            \node at (\x,-0.4) [below,black] { $\x$ };
        }
        \foreach \y in {1,5,10,...,\maxclusters} {
            \draw[tick,black,thin] (-0.01,\y) -- (-0.022,\y);
            \node at (-0.02,\y) [left,black] { $\y$ };
        }
        \draw[tick,black,thin]  ($ (0.25,\maxclusters+1) $) -- ($ (-0.01,\maxclusters+1) $) -- (-0.01,-0.4) -- ($ (0.25,-0.4) $) ;
        \node at ($(0.2,-2.5-1)$) [left,black] { \footnotesize parallel epistatic units };
        %\node at ($(-0.06,\maxclusters*0.2)$) [rotate=90] { \footnotesize dual edges };

\end{tikzpicture}
  \end{minipage}
  \caption[Parallel transport from $0{\ast}{\ast}{\ast}{\ast}$ to $1{\ast}{\ast}{\ast}{\ast}$ within the Tan data.]{%
    Parallel transport from $0{\ast}{\ast}{\ast}{\ast}$ to $1{\ast}{\ast}{\ast}{\ast}$ within the Tan data.
    Analysis based on mean values only; hence there is no color coding for the significance.
  }
  \label{fig:weinreich:0ssss-1ssss}
\end{figure}

% Fig S6
\begin{figure}[ht]
  \begin{minipage}[t]{.45\linewidth}
    \centering
    \input{Figures/filtration_weinreich_norm_ss0ss.tikz}
  \end{minipage}%
  \hfill%
  \begin{minipage}[t]{.45\linewidth}
    \centering
     \newcommand\maxepi{0.0799}
            \newcommand\maxclusters{24}
            \newcommand\clusterlength{0.2pt}
            \newcommand\clusterspace{0.1pt}
            \newcommand\flagfacup{2.5}
            \newcommand\flagfacdown{4}
            \newcommand\lw{3pt}

\begin{tikzpicture}[yscale=0.5,xscale=45, scale = 0.4]
\tikzset{
        cluster/.style = {red, line width=\lw, join=round},
        significant/.style = {blue, line width=\lw, join=round},
        semisignificant/.style = {purple, line width=\lw, join=round},
        tick/.style = {black, thick},
        tick_thin/.style = {black, thin}}

 \coordinate (offset) at (0.00927046731632334, 23);

 \draw[cluster] ($ (offset) $)
        -- ($ (offset)  + (10*\clusterlength,0) $);

  %\node [draw] at ($  (offset)  + (3.5,0)$)  {3.000}; 
 \draw[tick] ($ (offset) + (0,\flagfacup*\lw) $) -- ($ (offset) - (0,\flagfacdown*\lw) $);

  \coordinate (offset) at (0.00927046731632334, 22);

 \draw[cluster] ($ (offset) $)
        -- ($ (offset)  + (10*\clusterlength,0) $);

  %\node [draw] at ($  (offset)  + (3.5,0)$)  {3.000}; 
 \draw[tick] ($ (offset) + (0,\flagfacup*\lw) $) -- ($ (offset) - (0,\flagfacdown*\lw) $);

  \coordinate (offset) at (0, 21);

 \draw[cluster] ($ (offset) $)
        -- ($ (offset)  + (10*\clusterlength,0) $);

  %\node [draw] at ($  (offset)  + (3.5,0)$)  {0.000}; 
 \draw[tick] ($ (offset) + (0,\flagfacup*\lw) $) -- ($ (offset) - (0,\flagfacdown*\lw) $);

  \coordinate (offset) at (0.141601843910115, 20);

 \draw[cluster] ($ (offset) $)
        -- ($ (offset)  + (10*\clusterlength,0) $);

  %\node [draw] at ($  (offset)  + (3.5,0)$)  {21.639}; 
 \draw[tick] ($ (offset) + (0,\flagfacup*\lw) $) -- ($ (offset) - (0,\flagfacdown*\lw) $);

  \coordinate (offset) at (0.173426132108506, 19);

 \draw[cluster] ($ (offset) $)
        -- ($ (offset)  + (10*\clusterlength,0) $);

  %\node [draw] at ($  (offset)  + (3.5,0)$)  {21.639}; 
 \draw[tick] ($ (offset) + (0,\flagfacup*\lw) $) -- ($ (offset) - (0,\flagfacdown*\lw) $);

  \coordinate (offset) at (0.013110420608281, 18);

 \draw[cluster] ($ (offset) $)
        -- ($ (offset)  + (10*\clusterlength,0) $);

  %\node [draw] at ($  (offset)  + (3.5,0)$)  {1.545}; 
 \draw[tick] ($ (offset) + (0,\flagfacup*\lw) $) -- ($ (offset) - (0,\flagfacdown*\lw) $);

  \coordinate (offset) at (0.0014541909515801, 17);

 \draw[cluster] ($ (offset) $)
        -- ($ (offset)  + (10*\clusterlength,0) $);

  %\node [draw] at ($  (offset)  + (3.5,0)$)  {0.170}; 
 \draw[tick] ($ (offset) + (0,\flagfacup*\lw) $) -- ($ (offset) - (0,\flagfacdown*\lw) $);

  \coordinate (offset) at (0.00290838190316027, 16);

 \draw[cluster] ($ (offset) $)
        -- ($ (offset)  + (10*\clusterlength,0) $);

  %\node [draw] at ($  (offset)  + (3.5,0)$)  {0.314}; 
 \draw[tick] ($ (offset) + (0,\flagfacup*\lw) $) -- ($ (offset) - (0,\flagfacdown*\lw) $);

  \coordinate (offset) at (0.00290838190316027, 15);

 \draw[cluster] ($ (offset) $)
        -- ($ (offset)  + (10*\clusterlength,0) $);

  %\node [draw] at ($  (offset)  + (3.5,0)$)  {0.239}; 
 \draw[tick] ($ (offset) + (0,\flagfacup*\lw) $) -- ($ (offset) - (0,\flagfacdown*\lw) $);

  \coordinate (offset) at (0.0187006355547041, 14);

 \draw[cluster] ($ (offset) $)
        -- ($ (offset)  + (10*\clusterlength,0) $);

  %\node [draw] at ($  (offset)  + (3.5,0)$)  {1.012}; 
 \draw[tick] ($ (offset) + (0,\flagfacup*\lw) $) -- ($ (offset) - (0,\flagfacdown*\lw) $);

  \coordinate (offset) at (0.0187006355547041, 13);

 \draw[cluster] ($ (offset) $)
        -- ($ (offset)  + (10*\clusterlength,0) $);

  %\node [draw] at ($  (offset)  + (3.5,0)$)  {1.012}; 
 \draw[tick] ($ (offset) + (0,\flagfacup*\lw) $) -- ($ (offset) - (0,\flagfacdown*\lw) $);

  \coordinate (offset) at (0.00145419095158017, 12);

 \draw[cluster] ($ (offset) $)
        -- ($ (offset)  + (10*\clusterlength,0) $);

  %\node [draw] at ($  (offset)  + (3.5,0)$)  {0.071}; 
 \draw[tick] ($ (offset) + (0,\flagfacup*\lw) $) -- ($ (offset) - (0,\flagfacdown*\lw) $);

  \coordinate (offset) at (0.00145419095158017, 11);

 \draw[cluster] ($ (offset) $)
        -- ($ (offset)  + (10*\clusterlength,0) $);

  %\node [draw] at ($  (offset)  + (3.5,0)$)  {0.071}; 
 \draw[tick] ($ (offset) + (0,\flagfacup*\lw) $) -- ($ (offset) - (0,\flagfacdown*\lw) $);

  \coordinate (offset) at (0.142703659411492, 10);

 \draw[cluster] ($ (offset) $)
        -- ($ (offset)  + (10*\clusterlength,0) $);

  %\node [draw] at ($  (offset)  + (3.5,0)$)  {6.163}; 
 \draw[tick] ($ (offset) + (0,\flagfacup*\lw) $) -- ($ (offset) - (0,\flagfacdown*\lw) $);

  \coordinate (offset) at (0.0107246582679035, 9);

 \draw[cluster] ($ (offset) $)
        -- ($ (offset)  + (10*\clusterlength,0) $);

  %\node [draw] at ($  (offset)  + (3.5,0)$)  {0.457}; 
 \draw[tick] ($ (offset) + (0,\flagfacup*\lw) $) -- ($ (offset) - (0,\flagfacdown*\lw) $);

  \coordinate (offset) at (0.0107246582679035, 8);

 \draw[cluster] ($ (offset) $)
        -- ($ (offset)  + (10*\clusterlength,0) $);

  %\node [draw] at ($  (offset)  + (3.5,0)$)  {0.457}; 
 \draw[tick] ($ (offset) + (0,\flagfacup*\lw) $) -- ($ (offset) - (0,\flagfacdown*\lw) $);

  \coordinate (offset) at (0.0289414597870421, 7);

 \draw[cluster] ($ (offset) $)
        -- ($ (offset)  + (10*\clusterlength,0) $);

  %\node [draw] at ($  (offset)  + (3.5,0)$)  {1.130}; 
 \draw[tick] ($ (offset) + (0,\flagfacup*\lw) $) -- ($ (offset) - (0,\flagfacdown*\lw) $);

  \coordinate (offset) at (0.144484672321464, 6);

 \draw[cluster] ($ (offset) $)
        -- ($ (offset)  + (10*\clusterlength,0) $);

  %\node [draw] at ($  (offset)  + (3.5,0)$)  {4.298}; 
 \draw[tick] ($ (offset) + (0,\flagfacup*\lw) $) -- ($ (offset) - (0,\flagfacdown*\lw) $);

  \coordinate (offset) at (0.020481648464676, 5);

 \draw[cluster] ($ (offset) $)
        -- ($ (offset)  + (10*\clusterlength,0) $);

  %\node [draw] at ($  (offset)  + (3.5,0)$)  {0.484}; 
 \draw[tick] ($ (offset) + (0,\flagfacup*\lw) $) -- ($ (offset) - (0,\flagfacdown*\lw) $);

  \coordinate (offset) at (0.00356202581994364, 4);

 \draw[cluster] ($ (offset) $)
        -- ($ (offset)  + (10*\clusterlength,0) $);

  %\node [draw] at ($  (offset)  + (3.5,0)$)  {0.071}; 
 \draw[tick] ($ (offset) + (0,\flagfacup*\lw) $) -- ($ (offset) - (0,\flagfacdown*\lw) $);

  \coordinate (offset) at (0.136330401710637, 3);

 \draw[cluster] ($ (offset) $)
        -- ($ (offset)  + (10*\clusterlength,0) $);

  %\node [draw] at ($  (offset)  + (3.5,0)$)  {2.358}; 
 \draw[tick] ($ (offset) + (0,\flagfacup*\lw) $) -- ($ (offset) - (0,\flagfacdown*\lw) $);

  \coordinate (offset) at (0.162517428034929, 2);

 \draw[cluster] ($ (offset) $)
        -- ($ (offset)  + (10*\clusterlength,0) $);

  %\node [draw] at ($  (offset)  + (3.5,0)$)  {2.260}; 
 \draw[tick] ($ (offset) + (0,\flagfacup*\lw) $) -- ($ (offset) - (0,\flagfacdown*\lw) $);

  \coordinate (offset) at (0.0109087040735774, 1);

 \draw[cluster] ($ (offset) $)
        -- ($ (offset)  + (10*\clusterlength,0) $);

  %\node [draw] at ($  (offset)  + (3.5,0)$)  {0.136}; 
 \draw[tick] ($ (offset) + (0,\flagfacup*\lw) $) -- ($ (offset) - (0,\flagfacdown*\lw) $);

\draw[thin, gray, dotted ,step=1] (0,0) grid ($ (0.2,\maxclusters) $);

        \foreach \x in {0.025,0.05,..., 0.225} {
            \draw[tick,black,thin] (\x,-0.4) -- (\x,-0.1);
        }
        \foreach \x in {0.05,0.15} {
            \node at (\x,-0.4) [below,black] {\footnotesize $\x$ };
        }
        \foreach \x in {0.1,0.2} {
            \draw[tick,black,thin] (\x,-0.4) -- (\x,0.15);
            \node at (\x,-0.4) [below,black] { $\x$ };
        }
        \foreach \y in {1,5,10,...,\maxclusters} {
            \draw[tick,black,thin] (-0.01,\y) -- (-0.022,\y);
            \node at (-0.02,\y) [left,black] { $\y$ };
        }
        \draw[tick,black,thin]  ($ (0.25,\maxclusters+1) $) -- ($ (-0.01,\maxclusters+1) $) -- (-0.01,-0.4) -- ($ (0.25,-0.4) $) ;
        \node at ($(0.2,-2.5-1)$) [left,black] { \footnotesize parallel epistatic units };
        %\node at ($(-0.06,\maxclusters*0.2)$) [rotate=90] { \footnotesize dual edges };

\end{tikzpicture}
  \end{minipage}
  \caption[Parallel transport from ${\ast}{\ast}0{\ast}{\ast}$ to ${\ast}{\ast}1{\ast}{\ast}$ within the Tan data.]{%
    Parallel transport from the face ${\ast}{\ast}0{\ast}{\ast}$ to the face ${\ast}{\ast}1{\ast}{\ast}$ within the Tan data.
    Analysis based on mean values only; hence there is no color coding for the significance.
  }
  \label{fig:weinreich:ss0ss-ss1ss}
\end{figure}

%% Fig S 7
%\begin{figure}[b]
%  \begin{minipage}[t]{.45\linewidth}
%    \centering
%    \input{Figures/filtration_khanraw_norm_0ssss.tikz}
%  \end{minipage}%
%  \hfill%
%  \begin{minipage}[t]{.45\linewidth}
%    \centering
%    \input{Figures/filtration_khanraw_norm_1ssss_parallel.tikz}
%  \end{minipage}
%  \caption[Parallel transport from $0{\ast}{\ast}{\ast}{\ast}$ to $1{\ast}{\ast}{\ast}{\ast}$ within the Khan data.]{Epistatic filtration and parallel epistatic units for transport from $0{\ast}{\ast}{\ast}{\ast}$ to $1{\ast}{\ast}{\ast}{\ast}$ within the Khan data.}
%  \label{fig:khan:0ssss-1ssss}
%\end{figure}

%\input{Figures/paralleltransports/analyze_khanrawNormalizedstarstarstarstar0TOkhanrawNormalizedstarstarstarstar1_180tables.tex}
%
%\input{Figures/paralleltransports/analyze_khanrawNormalizedstarstarstarstar1TOkhanrawNormalizedstarstarstarstar0_180tables.tex}

%Eble 1**** to 0****
%%Fig S8
%\begin{figure}[b]
%  \begin{minipage}[t]{.45\linewidth}
%    \centering
%    \input{Figures/paralleltransports/singles/filtrationbars_Lud2018SurvDataNormalized1starstarstarstar.tex}
%  \end{minipage}%
%  \hfill%
%  \begin{minipage}[t]{.45\linewidth}
%    \centering
%    \input{Figures/paralleltransports/singles/targetbars_Lud2018SurvDataNormalized1starstarstarstarTOLud2018SurvDataNormalized0starstarstarstar.tex}
%  \end{minipage}
%  \caption[$1{\ast}{\ast}{\ast}{\ast}$(Eble) to $0{\ast}{\ast}{\ast}{\ast}$(Eble).]{$1{\ast}{\ast}{\ast}{\ast}$(Eble) to $0{\ast}{\ast}{\ast}{\ast}$(Eble).}
%  \label{fig:Eble:parallel:1to0****}
%\end{figure}

\begin{samepage}

%%Eble 0**** to 1****.
% Fig S7
\begin{figure}[b]
  \begin{minipage}[t]{.3\linewidth}
    \centering
    \input{Figures/paralleltransports/singles/filtrationbars_Lud2018SurvDataNormalized0starstarstarstar.tex}
  \end{minipage}%
  \hfill%
  \begin{minipage}[t]{.3\linewidth}
    \centering
    \newcommand\maxepi{1}
            \newcommand\maxclusters{24}
            \newcommand\clusterlength{0.33pt}
            \newcommand\clusterspace{0.13pt}
            \newcommand\flagfacup{2.5}
            \newcommand\flagfacdown{4}
            \newcommand\lw{3pt}

\begin{tikzpicture}[yscale=0.5,xscale=90, scale = 0.4]
\tikzset{
        cluster/.style = {red, line width=\lw, join=round},
        significant/.style = {blue, line width=\lw, join=round},
        semisignificant/.style = {purple, line width=\lw, join=round},
        tick/.style = {black, thick},
        tick_thin/.style = {black, thin}}

  \coordinate (offset) at (0.0118428031592889, 23);

 \draw[cluster] ($ (offset) $)-- ($ (offset)  + (3*\clusterlength,0) $);
 \draw[tick] ($ (offset) + (0,\flagfacup*\lw) $) -- ($ (offset) - (0,\flagfacdown*\lw) $);

  \coordinate (offset) at (0.0118428031592889, 22);

 \draw[cluster] ($ (offset) $)-- ($ (offset)  + (3*\clusterlength,0) $);
 \draw[tick] ($ (offset) + (0,\flagfacup*\lw) $) -- ($ (offset) - (0,\flagfacdown*\lw) $);

  \coordinate (offset) at (0.0246231555315973, 21);

 \draw[significant] ($ (offset) $)-- ($ (offset)  + (3*\clusterlength,0) $);
 \draw[tick] ($ (offset) + (0,\flagfacup*\lw) $) -- ($ (offset) - (0,\flagfacdown*\lw) $);

  \coordinate (offset) at (0.034822400501207, 20);

 \draw[significant] ($ (offset) $)-- ($ (offset)  + (3*\clusterlength,0) $);
 \draw[tick] ($ (offset) + (0,\flagfacup*\lw) $) -- ($ (offset) - (0,\flagfacdown*\lw) $);

  \coordinate (offset) at (0.0118360232598236, 19);

 \draw[cluster] ($ (offset) $)-- ($ (offset)  + (3*\clusterlength,0) $);
 \draw[tick] ($ (offset) + (0,\flagfacup*\lw) $) -- ($ (offset) - (0,\flagfacdown*\lw) $);

  \coordinate (offset) at (0.0144961087851405, 18);

 \draw[cluster] ($ (offset) $)-- ($ (offset)  + (3*\clusterlength,0) $);
 \draw[tick] ($ (offset) + (0,\flagfacup*\lw) $) -- ($ (offset) - (0,\flagfacdown*\lw) $);

  \coordinate (offset) at (0.0127871322717737, 17);

 \draw[cluster] ($ (offset) $)-- ($ (offset)  + (3*\clusterlength,0) $);
 \draw[tick] ($ (offset) + (0,\flagfacup*\lw) $) -- ($ (offset) - (0,\flagfacdown*\lw) $);

  \coordinate (offset) at (0.00284744993993832, 16);

 \draw[cluster] ($ (offset) $)-- ($ (offset)  + (3*\clusterlength,0) $);
 \draw[tick] ($ (offset) + (0,\flagfacup*\lw) $) -- ($ (offset) - (0,\flagfacdown*\lw) $);

  \coordinate (offset) at (0.00951307139024847, 15);

 \draw[cluster] ($ (offset) $)-- ($ (offset)  + (3*\clusterlength,0) $);
 \draw[tick] ($ (offset) + (0,\flagfacup*\lw) $) -- ($ (offset) - (0,\flagfacdown*\lw) $);

  \coordinate (offset) at (0.00507258139706726, 14);

 \draw[cluster] ($ (offset) $)-- ($ (offset)  + (3*\clusterlength,0) $);
 \draw[tick] ($ (offset) + (0,\flagfacup*\lw) $) -- ($ (offset) - (0,\flagfacdown*\lw) $);

  \coordinate (offset) at (0.0239444654040268, 13);

 \draw[semisignificant] ($ (offset) $)-- ($ (offset)  + (3*\clusterlength,0) $);
 \draw[tick] ($ (offset) + (0,\flagfacup*\lw) $) -- ($ (offset) - (0,\flagfacdown*\lw) $);

  \coordinate (offset) at (0.0239444654040268, 12);

 \draw[semisignificant] ($ (offset) $)-- ($ (offset)  + (3*\clusterlength,0) $);
 \draw[tick] ($ (offset) + (0,\flagfacup*\lw) $) -- ($ (offset) - (0,\flagfacdown*\lw) $);

  \coordinate (offset) at (0.0183079144266003, 11);

 \draw[cluster] ($ (offset) $)-- ($ (offset)  + (3*\clusterlength,0) $);
 \draw[tick] ($ (offset) + (0,\flagfacup*\lw) $) -- ($ (offset) - (0,\flagfacdown*\lw) $);

  \coordinate (offset) at (0.0141078201569671, 10);

 \draw[cluster] ($ (offset) $)-- ($ (offset)  + (3*\clusterlength,0) $);
 \draw[tick] ($ (offset) + (0,\flagfacup*\lw) $) -- ($ (offset) - (0,\flagfacdown*\lw) $);

  \coordinate (offset) at (0.025950623316256, 9);

 \draw[semisignificant] ($ (offset) $)-- ($ (offset)  + (3*\clusterlength,0) $);
 \draw[tick] ($ (offset) + (0,\flagfacup*\lw) $) -- ($ (offset) - (0,\flagfacdown*\lw) $);

  \coordinate (offset) at (0.00280048764629755, 8);

 \draw[cluster] ($ (offset) $)-- ($ (offset)  + (3*\clusterlength,0) $);
 \draw[tick] ($ (offset) + (0,\flagfacup*\lw) $) -- ($ (offset) - (0,\flagfacdown*\lw) $);

  \coordinate (offset) at (0.00280048764629755, 7);

 \draw[cluster] ($ (offset) $)-- ($ (offset)  + (3*\clusterlength,0) $);
 \draw[tick] ($ (offset) + (0,\flagfacup*\lw) $) -- ($ (offset) - (0,\flagfacdown*\lw) $);

  \coordinate (offset) at (0.0054360407944277, 6);

 \draw[cluster] ($ (offset) $)-- ($ (offset)  + (3*\clusterlength,0) $);
 \draw[tick] ($ (offset) + (0,\flagfacup*\lw) $) -- ($ (offset) - (0,\flagfacdown*\lw) $);

  \coordinate (offset) at (0.0112019505851902, 5);

 \draw[cluster] ($ (offset) $)-- ($ (offset)  + (3*\clusterlength,0) $);
 \draw[tick] ($ (offset) + (0,\flagfacup*\lw) $) -- ($ (offset) - (0,\flagfacdown*\lw) $);

  \coordinate (offset) at (0.0112019505851902, 4);

 \draw[cluster] ($ (offset) $)-- ($ (offset)  + (3*\clusterlength,0) $);
 \draw[tick] ($ (offset) + (0,\flagfacup*\lw) $) -- ($ (offset) - (0,\flagfacdown*\lw) $);

  \coordinate (offset) at (0.0134212049464071, 3);

 \draw[cluster] ($ (offset) $)-- ($ (offset)  + (3*\clusterlength,0) $);
 \draw[tick] ($ (offset) + (0,\flagfacup*\lw) $) -- ($ (offset) - (0,\flagfacdown*\lw) $);

  \coordinate (offset) at (0.0370932514660543, 2);

 \draw[significant] ($ (offset) $)-- ($ (offset)  + (3*\clusterlength,0) $);
 \draw[tick] ($ (offset) + (0,\flagfacup*\lw) $) -- ($ (offset) - (0,\flagfacdown*\lw) $);

  \coordinate (offset) at (0.0243061191942806, 1);

 \draw[semisignificant] ($ (offset) $)-- ($ (offset)  + (3*\clusterlength,0) $);
 \draw[tick] ($ (offset) + (0,\flagfacup*\lw) $) -- ($ (offset) - (0,\flagfacdown*\lw) $);

  \draw[thin, gray, dotted ,step=1] (0,0) grid ($ (0.05,\maxclusters) $);
        \foreach \x in {0.025, 0.05,...,0.1} {
            \draw[tick,black,thin] (\x,-0.4) -- (\x,-0.1);
        }
        \foreach \x in {0.025,0.05,0.075} {
            \node at (\x,-0.4) [below,black] {\footnotesize $\x$ };
        }
        \foreach \x in {0.1} {
            \draw[tick,black,thin] (\x,-0.4) -- (\x,0.15);
            \node at (\x,-0.4) [below,black] { $\x$ };
        }
        \foreach \y in {1,5,10,...,\maxclusters} {
            \draw[tick,black,thin] (-0.005,\y) -- (-0.008,\y);
            \node at (-0.007,\y) [left,black] { $\y$ };
        }
        \draw[tick,black,thin]  ($ (0.125,\maxclusters+1) $) -- ($ (-0.005,\maxclusters+1) $) -- (-0.005,-0.4) -- ($ (0.125,-0.4) $) ;
        \node at ($(0.06,-2.5-1)$) [left,black] {\footnotesize epistatic units };
        \node at ($(-0.024,\maxclusters*0.2)$) [rotate=90] { \footnotesize dual edges };
\end{tikzpicture}
  \end{minipage}
%  \hfill
  \begin{minipage}[t]{.3\linewidth}
    \centering
    \includegraphics{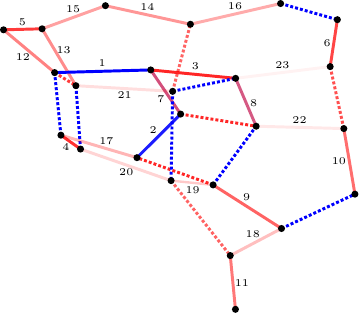}
  \end{minipage}
  \caption[$0{\ast}{\ast}{\ast}{\ast}$(Eble) to $1{\ast}{\ast}{\ast}{\ast}$(Eble).]{{Effect of \emph{L. plantarum}. Comparing $0{\ast}{\ast}{\ast}{\ast}$ to $1{\ast}{\ast}{\ast}{\ast}$ for Eble data.} \emph{Left}. Filtration of $0{\ast}{\ast}{\ast}{\ast}$. \emph{Middle}. Parallel filtration of $1{\ast}{\ast}{\ast}{\ast}$. \emph{Right}. Dual graph of $0{\ast}{\ast}{\ast}{\ast}$.}
  \label{fig:Eble:parallel:0to1****}
\end{figure}

%%Eble *1*** to *0***.
%% Fig S
%\begin{figure}[b]
%  \begin{minipage}[t]{.45\linewidth}
%    \centering
%    \input{Figures/paralleltransports/singles/filtrationbars_Lud2018SurvDataNormalizedstar1starstarstar.tex}
%  \end{minipage}%
%  \hfill%
%  \begin{minipage}[t]{.45\linewidth}
%    \centering
%    \input{Figures/paralleltransports/singles/targetbars_Lud2018SurvDataNormalizedstar1starstarstarTOLud2018SurvDataNormalizedstar0starstarstar.tex}
%  \end{minipage}
%  \caption[${\ast}1{\ast}{\ast}{\ast}$(Eble) to ${\ast}0{\ast}{\ast}{\ast}$(Eble).]{${\ast}1{\ast}{\ast}{\ast}$(Eble) to ${\ast}0{\ast}{\ast}{\ast}$(Eble).}
%  \label{fig:Eble:parallel:*1to0***}
%\end{figure}

%%Eble *0*** to *1***.
%Fig S8
\begin{figure}[b]
  \begin{minipage}[t]{.45\linewidth}
    \centering
    \input{Figures/paralleltransports/singles/filtrationbars_Lud2018SurvDataNormalizedstar0starstarstar.tex}
  \end{minipage}%
  \hfill%
  \begin{minipage}[t]{.45\linewidth}
    \centering
    \newcommand\maxepi{1}
            \newcommand\maxclusters{24}
            \newcommand\clusterlength{0.33pt}
            \newcommand\clusterspace{0.13pt}
            \newcommand\flagfacup{2.5}
            \newcommand\flagfacdown{4}
            \newcommand\lw{3pt}

\begin{tikzpicture}[yscale=0.48,xscale=90, scale = 0.4]
\tikzset{
        cluster/.style = {red, line width=\lw, join=round},
        significant/.style = {blue, line width=\lw, join=round},
        semisignificant/.style = {purple, line width=\lw, join=round},
        tick/.style = {black, thick},
        tick_thin/.style = {black, thin}}

  \coordinate (offset) at (0.0195505741345301, 23);

 \draw[semisignificant] ($ (offset) $)-- ($ (offset)  + (3*\clusterlength,0) $);
 \draw[tick] ($ (offset) + (0,\flagfacup*\lw) $) -- ($ (offset) - (0,\flagfacdown*\lw) $);

  \coordinate (offset) at (0.0276486870932331, 22);

 \draw[semisignificant] ($ (offset) $)-- ($ (offset)  + (3*\clusterlength,0) $);
 \draw[tick] ($ (offset) + (0,\flagfacup*\lw) $) -- ($ (offset) - (0,\flagfacdown*\lw) $);

  \coordinate (offset) at (0.0132626867777488, 21);

 \draw[cluster] ($ (offset) $)-- ($ (offset)  + (3*\clusterlength,0) $);
 \draw[tick] ($ (offset) + (0,\flagfacup*\lw) $) -- ($ (offset) - (0,\flagfacdown*\lw) $);

  \coordinate (offset) at (0.00860706459117718, 20);

 \draw[cluster] ($ (offset) $)-- ($ (offset)  + (3*\clusterlength,0) $);
 \draw[tick] ($ (offset) + (0,\flagfacup*\lw) $) -- ($ (offset) - (0,\flagfacdown*\lw) $);

  \coordinate (offset) at (0.0158551189837474, 19);

 \draw[cluster] ($ (offset) $)-- ($ (offset)  + (3*\clusterlength,0) $);
 \draw[tick] ($ (offset) + (0,\flagfacup*\lw) $) -- ($ (offset) - (0,\flagfacdown*\lw) $);

  \coordinate (offset) at (0.0158551189837474, 18);

 \draw[cluster] ($ (offset) $)-- ($ (offset)  + (3*\clusterlength,0) $);
 \draw[tick] ($ (offset) + (0,\flagfacup*\lw) $) -- ($ (offset) - (0,\flagfacdown*\lw) $);

  \coordinate (offset) at (0.00414174536718301, 17);

 \draw[cluster] ($ (offset) $)-- ($ (offset)  + (3*\clusterlength,0) $);
 \draw[tick] ($ (offset) + (0,\flagfacup*\lw) $) -- ($ (offset) - (0,\flagfacdown*\lw) $);

  \coordinate (offset) at (0.0189614280091347, 16);

 \draw[cluster] ($ (offset) $)-- ($ (offset)  + (3*\clusterlength,0) $);
 \draw[tick] ($ (offset) + (0,\flagfacup*\lw) $) -- ($ (offset) - (0,\flagfacdown*\lw) $);

  \coordinate (offset) at (0.0189614280091347, 15);

 \draw[cluster] ($ (offset) $)-- ($ (offset)  + (3*\clusterlength,0) $);
 \draw[tick] ($ (offset) + (0,\flagfacup*\lw) $) -- ($ (offset) - (0,\flagfacdown*\lw) $);

  \coordinate (offset) at (5.2839389552784e-05, 14);

 \draw[cluster] ($ (offset) $)-- ($ (offset)  + (3*\clusterlength,0) $);
 \draw[tick] ($ (offset) + (0,\flagfacup*\lw) $) -- ($ (offset) - (0,\flagfacdown*\lw) $);

  \coordinate (offset) at (5.2839389552784e-05, 13);

 \draw[cluster] ($ (offset) $)-- ($ (offset)  + (3*\clusterlength,0) $);
 \draw[tick] ($ (offset) + (0,\flagfacup*\lw) $) -- ($ (offset) - (0,\flagfacdown*\lw) $);

  \coordinate (offset) at (0.0098809658463706, 12);

 \draw[cluster] ($ (offset) $)-- ($ (offset)  + (3*\clusterlength,0) $);
 \draw[tick] ($ (offset) + (0,\flagfacup*\lw) $) -- ($ (offset) - (0,\flagfacdown*\lw) $);

  \coordinate (offset) at (0.0127871322717737, 11);

 \draw[cluster] ($ (offset) $)-- ($ (offset)  + (3*\clusterlength,0) $);
 \draw[tick] ($ (offset) + (0,\flagfacup*\lw) $) -- ($ (offset) - (0,\flagfacdown*\lw) $);

  \coordinate (offset) at (0.0103543634179575, 10);

 \draw[cluster] ($ (offset) $)-- ($ (offset)  + (3*\clusterlength,0) $);
 \draw[tick] ($ (offset) + (0,\flagfacup*\lw) $) -- ($ (offset) - (0,\flagfacdown*\lw) $);

  \coordinate (offset) at (0.0180837358826011, 9);

 \draw[cluster] ($ (offset) $)-- ($ (offset)  + (3*\clusterlength,0) $);
 \draw[tick] ($ (offset) + (0,\flagfacup*\lw) $) -- ($ (offset) - (0,\flagfacdown*\lw) $);

  \coordinate (offset) at (0.0166444077091269, 8);

 \draw[semisignificant] ($ (offset) $)-- ($ (offset)  + (3*\clusterlength,0) $);
 \draw[tick] ($ (offset) + (0,\flagfacup*\lw) $) -- ($ (offset) - (0,\flagfacdown*\lw) $);

  \coordinate (offset) at (0.00280048764629755, 7);

 \draw[cluster] ($ (offset) $)-- ($ (offset)  + (3*\clusterlength,0) $);
 \draw[tick] ($ (offset) + (0,\flagfacup*\lw) $) -- ($ (offset) - (0,\flagfacdown*\lw) $);

  \coordinate (offset) at (0.00280048764629755, 6);

 \draw[cluster] ($ (offset) $)-- ($ (offset)  + (3*\clusterlength,0) $);
 \draw[tick] ($ (offset) + (0,\flagfacup*\lw) $) -- ($ (offset) - (0,\flagfacdown*\lw) $);

  \coordinate (offset) at (0.0370932514660543, 5);

 \draw[significant] ($ (offset) $)-- ($ (offset)  + (3*\clusterlength,0) $);
 \draw[tick] ($ (offset) + (0,\flagfacup*\lw) $) -- ($ (offset) - (0,\flagfacdown*\lw) $);

  \coordinate (offset) at (0.0240947616360695, 4);

 \draw[semisignificant] ($ (offset) $)-- ($ (offset)  + (3*\clusterlength,0) $);
 \draw[tick] ($ (offset) + (0,\flagfacup*\lw) $) -- ($ (offset) - (0,\flagfacdown*\lw) $);

  \coordinate (offset) at (0.0240947616360695, 3);

 \draw[semisignificant] ($ (offset) $)-- ($ (offset)  + (3*\clusterlength,0) $);
 \draw[tick] ($ (offset) + (0,\flagfacup*\lw) $) -- ($ (offset) - (0,\flagfacdown*\lw) $);

  \coordinate (offset) at (0.0438566933288107, 2);

 \draw[significant] ($ (offset) $)-- ($ (offset)  + (3*\clusterlength,0) $);
 \draw[tick] ($ (offset) + (0,\flagfacup*\lw) $) -- ($ (offset) - (0,\flagfacdown*\lw) $);

  \coordinate (offset) at (0.0212414346002192, 1);

 \draw[cluster] ($ (offset) $)-- ($ (offset)  + (3*\clusterlength,0) $);
 \draw[tick] ($ (offset) + (0,\flagfacup*\lw) $) -- ($ (offset) - (0,\flagfacdown*\lw) $);

  \draw[thin, gray, dotted ,step=1] (0,0) grid ($ (0.05,\maxclusters) $);
        \foreach \x in {0.025, 0.05,...,0.1} {
            \draw[tick,black,thin] (\x,-0.4) -- (\x,-0.1);
        }
        \foreach \x in {0.025,0.05,0.075} {
            \node at (\x,-0.4) [below,black] {\footnotesize $\x$ };
        }
        \foreach \x in {0.1} {
            \draw[tick,black,thin] (\x,-0.4) -- (\x,0.15);
            \node at (\x,-0.4) [below,black] { $\x$ };
        }
        \foreach \y in {1,5,10,...,\maxclusters} {
            \draw[tick,black,thin] (-0.005,\y) -- (-0.008,\y);
            \node at (-0.007,\y) [left,black] { $\y$ };
        }
        \draw[tick,black,thin]  ($ (0.125,\maxclusters+1) $) -- ($ (-0.005,\maxclusters+1) $) -- (-0.005,-0.4) -- ($ (0.125,-0.4) $) ;
        \node at ($(0.06,-2.5-1)$) [left,black] {\footnotesize epistatic units };
        \node at ($(-0.024,\maxclusters*0.2)$) [rotate=90] { \footnotesize dual edges };

\end{tikzpicture}
  \end{minipage}
  \caption[${\ast}0{\ast}{\ast}{\ast}$(Eble) to ${\ast}1{\ast}{\ast}{\ast}$(Eble).]{Effect of \emph{L. brevis}. Comparing ${\ast}0{\ast}{\ast}{\ast}$ to ${\ast}1{\ast}{\ast}{\ast}$ for Eble data. \emph{Left}. Filtration of ${\ast}0{\ast}{\ast}{\ast}$. \emph{Right}. Parallel filtration of ${\ast}1{\ast}{\ast}{\ast}$.}
  \label{fig:Eble:parallel:*0to1***}
\end{figure}
\end{samepage}

\begin{samepage}
%Fig S12
\begin{figure}[t]
  \begin{minipage}[t]{.45\linewidth}
    \centering
    \input{Figures/paralleltransports/GouldTTDvsCFU/filtrationbars_Lud2017FlygutCFUsDataNormalized0starstarstarstar.tex}
  \end{minipage}%
  \hfill%
  \begin{minipage}[t]{.45\linewidth}
    \centering
    \newcommand\maxepi{1}
            \newcommand\maxclusters{23}
            \newcommand\clusterlength{0.33pt}
            \newcommand\clusterspace{0.13pt}
            \newcommand\flagfacup{2.5}
            \newcommand\flagfacdown{4}
            \newcommand\lw{3pt}

\begin{tikzpicture}[yscale=0.5,xscale=95, scale = 0.4]
\tikzset{
        cluster/.style = {red, line width=\lw, join=round},
        significant/.style = {blue, line width=\lw, join=round},
        semisignificant/.style = {purple, line width=\lw, join=round},
        tick/.style = {black, thick},
        tick_thin/.style = {black, thin}}

  \coordinate (offset) at (0.0272363548629681, 22);

 \draw[significant] ($ (offset) $)-- ($ (offset)  + (3*\clusterlength,0) $);
 \draw[tick] ($ (offset) + (0,\flagfacup*\lw) $) -- ($ (offset) - (0,\flagfacdown*\lw) $);

  \coordinate (offset) at (0.0272363548629681, 21);

 \draw[significant] ($ (offset) $)-- ($ (offset)  + (3*\clusterlength,0) $);
 \draw[tick] ($ (offset) + (0,\flagfacup*\lw) $) -- ($ (offset) - (0,\flagfacdown*\lw) $);

  \coordinate (offset) at (0.0272363548629681, 20);

 \draw[significant] ($ (offset) $)-- ($ (offset)  + (3*\clusterlength,0) $);
 \draw[tick] ($ (offset) + (0,\flagfacup*\lw) $) -- ($ (offset) - (0,\flagfacdown*\lw) $);

  \coordinate (offset) at (0.0385180224368159, 19);

 \draw[significant] ($ (offset) $)-- ($ (offset)  + (3*\clusterlength,0) $);
 \draw[tick] ($ (offset) + (0,\flagfacup*\lw) $) -- ($ (offset) - (0,\flagfacdown*\lw) $);

  \coordinate (offset) at (0.0060650156243312, 18);

 \draw[cluster] ($ (offset) $)-- ($ (offset)  + (3*\clusterlength,0) $);
 \draw[tick] ($ (offset) + (0,\flagfacup*\lw) $) -- ($ (offset) - (0,\flagfacdown*\lw) $);

  \coordinate (offset) at (0.0131408671860509, 17);

 \draw[cluster] ($ (offset) $)-- ($ (offset)  + (3*\clusterlength,0) $);
 \draw[tick] ($ (offset) + (0,\flagfacup*\lw) $) -- ($ (offset) - (0,\flagfacdown*\lw) $);

  \coordinate (offset) at (0.0174124642117896, 16);

 \draw[cluster] ($ (offset) $)-- ($ (offset)  + (3*\clusterlength,0) $);
 \draw[tick] ($ (offset) + (0,\flagfacup*\lw) $) -- ($ (offset) - (0,\flagfacdown*\lw) $);

  \coordinate (offset) at (0.0165068817351322, 15);

 \draw[cluster] ($ (offset) $)-- ($ (offset)  + (3*\clusterlength,0) $);
 \draw[tick] ($ (offset) + (0,\flagfacup*\lw) $) -- ($ (offset) - (0,\flagfacdown*\lw) $);

  \coordinate (offset) at (0.0116613132257869, 14);

 \draw[cluster] ($ (offset) $)-- ($ (offset)  + (3*\clusterlength,0) $);
 \draw[tick] ($ (offset) + (0,\flagfacup*\lw) $) -- ($ (offset) - (0,\flagfacdown*\lw) $);

  \coordinate (offset) at (0.0108583344242059, 13);

 \draw[cluster] ($ (offset) $)-- ($ (offset)  + (3*\clusterlength,0) $);
 \draw[tick] ($ (offset) + (0,\flagfacup*\lw) $) -- ($ (offset) - (0,\flagfacdown*\lw) $);

  \coordinate (offset) at (0.0193556500345824, 12);

 \draw[cluster] ($ (offset) $)-- ($ (offset)  + (3*\clusterlength,0) $);
 \draw[tick] ($ (offset) + (0,\flagfacup*\lw) $) -- ($ (offset) - (0,\flagfacdown*\lw) $);

  \coordinate (offset) at (0.0193556500345824, 11);

 \draw[cluster] ($ (offset) $)-- ($ (offset)  + (3*\clusterlength,0) $);
 \draw[tick] ($ (offset) + (0,\flagfacup*\lw) $) -- ($ (offset) - (0,\flagfacdown*\lw) $);

  \coordinate (offset) at (0.0179994012076926, 10);

 \draw[cluster] ($ (offset) $)-- ($ (offset)  + (3*\clusterlength,0) $);
 \draw[tick] ($ (offset) + (0,\flagfacup*\lw) $) -- ($ (offset) - (0,\flagfacdown*\lw) $);

  \coordinate (offset) at (0.0354305563177359, 9);

 \draw[semisignificant] ($ (offset) $)-- ($ (offset)  + (3*\clusterlength,0) $);
 \draw[tick] ($ (offset) + (0,\flagfacup*\lw) $) -- ($ (offset) - (0,\flagfacdown*\lw) $);

  \coordinate (offset) at (0.0249466906222885, 8);

 \draw[semisignificant] ($ (offset) $)-- ($ (offset)  + (3*\clusterlength,0) $);
 \draw[tick] ($ (offset) + (0,\flagfacup*\lw) $) -- ($ (offset) - (0,\flagfacdown*\lw) $);

  \coordinate (offset) at (0.0165863798066894, 7);

 \draw[cluster] ($ (offset) $)-- ($ (offset)  + (3*\clusterlength,0) $);
 \draw[tick] ($ (offset) + (0,\flagfacup*\lw) $) -- ($ (offset) - (0,\flagfacdown*\lw) $);

  \coordinate (offset) at (0.0103833610914541, 6);

 \draw[cluster] ($ (offset) $)-- ($ (offset)  + (3*\clusterlength,0) $);
 \draw[tick] ($ (offset) + (0,\flagfacup*\lw) $) -- ($ (offset) - (0,\flagfacdown*\lw) $);

  \coordinate (offset) at (0.0288604190336828, 5);

 \draw[significant] ($ (offset) $)-- ($ (offset)  + (3*\clusterlength,0) $);
 \draw[tick] ($ (offset) + (0,\flagfacup*\lw) $) -- ($ (offset) - (0,\flagfacdown*\lw) $);

  \coordinate (offset) at (0.0279121149162769, 4);

 \draw[significant] ($ (offset) $)-- ($ (offset)  + (3*\clusterlength,0) $);
 \draw[tick] ($ (offset) + (0,\flagfacup*\lw) $) -- ($ (offset) - (0,\flagfacdown*\lw) $);

  \coordinate (offset) at (0.0144125640105075, 3);

 \draw[cluster] ($ (offset) $)-- ($ (offset)  + (3*\clusterlength,0) $);
 \draw[tick] ($ (offset) + (0,\flagfacup*\lw) $) -- ($ (offset) - (0,\flagfacdown*\lw) $);

  \coordinate (offset) at (0.0187819838688966, 2);

 \draw[cluster] ($ (offset) $)-- ($ (offset)  + (3*\clusterlength,0) $);
 \draw[tick] ($ (offset) + (0,\flagfacup*\lw) $) -- ($ (offset) - (0,\flagfacdown*\lw) $);

  \coordinate (offset) at (0.0319228510549475, 1);

 \draw[significant] ($ (offset) $)-- ($ (offset)  + (3*\clusterlength,0) $);
 \draw[tick] ($ (offset) + (0,\flagfacup*\lw) $) -- ($ (offset) - (0,\flagfacdown*\lw) $);
\draw[thin, gray, dotted ,step=1] (0,0) grid ($ (0.05,\maxclusters) $);
        \foreach \x in {0.025, 0.05,...,0.1} {
            \draw[tick,black,thin] (\x,-0.4) -- (\x,-0.1);
        }
        \foreach \x in {0.025,0.05,0.075} {
            \node at (\x,-0.4) [below,black] {\footnotesize $\x$ };
        }
        \foreach \x in {0.1} {
            \draw[tick,black,thin] (\x,-0.4) -- (\x,0.15);
            \node at (\x,-0.4) [below,black] { $\x$ };
        }
        \foreach \y in {1,5,10,...,\maxclusters} {
            \draw[tick,black,thin] (-0.005,\y) -- (-0.008,\y);
            \node at (-0.007,\y) [left,black] { $\y$ };
        }
        \draw[tick,black,thin]  ($ (0.125,\maxclusters+1) $) -- ($ (-0.005,\maxclusters+1) $) -- (-0.005,-0.4) -- ($ (0.125,-0.4) $) ;
        \node at ($(0.06,-2.5-1)$) [left,black] {\footnotesize epistatic units };
        \node at ($(-0.024,\maxclusters*0.2)$) [rotate=90] { \footnotesize dual edges };

\end{tikzpicture}
  \end{minipage}
  \caption[$0{\ast}{\ast}{\ast}{\ast}$(GouldCFU) to $0{\ast}{\ast}{\ast}{\ast}$(GouldTTD).]{Comparing $0{\ast}{\ast}{\ast}{\ast}$(Gould bacterial CFU counts) to $0{\ast}{\ast}{\ast}{\ast}$(Gould lifespans). \emph{Left}. Filtration of $0{\ast}{\ast}{\ast}{\ast}$ CFU counts. \emph{Right}. Parallel filtration of $0{\ast}{\ast}{\ast}{\ast}$ lifespans.}
  \label{fig:Gould:parallel:CFUtoTTD:0****}
\end{figure}

% FigS13
\begin{figure}[t]
  \begin{minipage}[t]{.45\linewidth}
    \centering
    \input{Figures/paralleltransports/GouldTTDvsCFU/filtrationbars_Lud2017FlygutCFUsDataNormalized1starstarstarstar.tex}
  \end{minipage}%
  \hfill%
  \begin{minipage}[t]{.45\linewidth}
    \centering
    \newcommand\maxepi{1}
            \newcommand\maxclusters{24}
            \newcommand\clusterlength{0.33pt}
            \newcommand\clusterspace{0.13pt}
            \newcommand\flagfacup{2.5}
            \newcommand\flagfacdown{4}
            \newcommand\lw{3pt}

\begin{tikzpicture}[yscale=0.5,xscale=90, scale = 0.4]
\tikzset{
        cluster/.style = {red, line width=\lw, join=round},
        significant/.style = {blue, line width=\lw, join=round},
        semisignificant/.style = {purple, line width=\lw, join=round},
        tick/.style = {black, thick},
        tick_thin/.style = {black, thin}}

  \coordinate (offset) at (0.0513243795306307, 23);

 \draw[significant] ($ (offset) $)-- ($ (offset)  + (3*\clusterlength,0) $);
 \draw[tick] ($ (offset) + (0,\flagfacup*\lw) $) -- ($ (offset) - (0,\flagfacdown*\lw) $);

  \coordinate (offset) at (0.0061302308460982, 22);

 \draw[cluster] ($ (offset) $)-- ($ (offset)  + (3*\clusterlength,0) $);
 \draw[tick] ($ (offset) + (0,\flagfacup*\lw) $) -- ($ (offset) - (0,\flagfacdown*\lw) $);

  \coordinate (offset) at (0.00212992022388803, 21);

 \draw[cluster] ($ (offset) $)-- ($ (offset)  + (3*\clusterlength,0) $);
 \draw[tick] ($ (offset) + (0,\flagfacup*\lw) $) -- ($ (offset) - (0,\flagfacdown*\lw) $);

  \coordinate (offset) at (0.00212992022388803, 20);

 \draw[cluster] ($ (offset) $)-- ($ (offset)  + (3*\clusterlength,0) $);
 \draw[tick] ($ (offset) + (0,\flagfacup*\lw) $) -- ($ (offset) - (0,\flagfacdown*\lw) $);

  \coordinate (offset) at (0.00212992022388803, 19);

 \draw[cluster] ($ (offset) $)-- ($ (offset)  + (3*\clusterlength,0) $);
 \draw[tick] ($ (offset) + (0,\flagfacup*\lw) $) -- ($ (offset) - (0,\flagfacdown*\lw) $);

  \coordinate (offset) at (0.00212992022388803, 18);

 \draw[cluster] ($ (offset) $)-- ($ (offset)  + (3*\clusterlength,0) $);
 \draw[tick] ($ (offset) + (0,\flagfacup*\lw) $) -- ($ (offset) - (0,\flagfacdown*\lw) $);

  \coordinate (offset) at (0.00212992022388803, 17);

 \draw[cluster] ($ (offset) $)-- ($ (offset)  + (3*\clusterlength,0) $);
 \draw[tick] ($ (offset) + (0,\flagfacup*\lw) $) -- ($ (offset) - (0,\flagfacdown*\lw) $);

  \coordinate (offset) at (0.00212992022388803, 16);

 \draw[cluster] ($ (offset) $)-- ($ (offset)  + (3*\clusterlength,0) $);
 \draw[tick] ($ (offset) + (0,\flagfacup*\lw) $) -- ($ (offset) - (0,\flagfacdown*\lw) $);

  \coordinate (offset) at (0.0386309785143433, 15);

 \draw[significant] ($ (offset) $)-- ($ (offset)  + (3*\clusterlength,0) $);
 \draw[tick] ($ (offset) + (0,\flagfacup*\lw) $) -- ($ (offset) - (0,\flagfacdown*\lw) $);

  \coordinate (offset) at (0.0124067853041478, 14);

 \draw[cluster] ($ (offset) $)-- ($ (offset)  + (3*\clusterlength,0) $);
 \draw[tick] ($ (offset) + (0,\flagfacup*\lw) $) -- ($ (offset) - (0,\flagfacdown*\lw) $);

  \coordinate (offset) at (0.0309903392575708, 13);

 \draw[significant] ($ (offset) $)-- ($ (offset)  + (3*\clusterlength,0) $);
 \draw[tick] ($ (offset) + (0,\flagfacup*\lw) $) -- ($ (offset) - (0,\flagfacdown*\lw) $);

  \coordinate (offset) at (0.0170863881029546, 12);

 \draw[cluster] ($ (offset) $)-- ($ (offset)  + (3*\clusterlength,0) $);
 \draw[tick] ($ (offset) + (0,\flagfacup*\lw) $) -- ($ (offset) - (0,\flagfacdown*\lw) $);

  \coordinate (offset) at (0.0170863881029546, 11);

 \draw[cluster] ($ (offset) $)-- ($ (offset)  + (3*\clusterlength,0) $);
 \draw[tick] ($ (offset) + (0,\flagfacup*\lw) $) -- ($ (offset) - (0,\flagfacdown*\lw) $);

  \coordinate (offset) at (0.0684107676335852, 10);

 \draw[significant] ($ (offset) $)-- ($ (offset)  + (3*\clusterlength,0) $);
 \draw[tick] ($ (offset) + (0,\flagfacup*\lw) $) -- ($ (offset) - (0,\flagfacdown*\lw) $);

  \coordinate (offset) at (0.0118210572425785, 9);

 \draw[cluster] ($ (offset) $)-- ($ (offset)  + (3*\clusterlength,0) $);
 \draw[tick] ($ (offset) + (0,\flagfacup*\lw) $) -- ($ (offset) - (0,\flagfacdown*\lw) $);

  \coordinate (offset) at (0.0210098304200814, 8);

 \draw[cluster] ($ (offset) $)-- ($ (offset)  + (3*\clusterlength,0) $);
 \draw[tick] ($ (offset) + (0,\flagfacup*\lw) $) -- ($ (offset) - (0,\flagfacdown*\lw) $);

  \coordinate (offset) at (0.0205580061099721, 7);

 \draw[cluster] ($ (offset) $)-- ($ (offset)  + (3*\clusterlength,0) $);
 \draw[tick] ($ (offset) + (0,\flagfacup*\lw) $) -- ($ (offset) - (0,\flagfacdown*\lw) $);

  \coordinate (offset) at (0.035037187682958, 6);

 \draw[significant] ($ (offset) $)-- ($ (offset)  + (3*\clusterlength,0) $);
 \draw[tick] ($ (offset) + (0,\flagfacup*\lw) $) -- ($ (offset) - (0,\flagfacdown*\lw) $);

  \coordinate (offset) at (0.0282381910251119, 5);

 \draw[semisignificant] ($ (offset) $)-- ($ (offset)  + (3*\clusterlength,0) $);
 \draw[tick] ($ (offset) + (0,\flagfacup*\lw) $) -- ($ (offset) - (0,\flagfacdown*\lw) $);

  \coordinate (offset) at (0.0282381910251119, 4);

 \draw[semisignificant] ($ (offset) $)-- ($ (offset)  + (3*\clusterlength,0) $);
 \draw[tick] ($ (offset) + (0,\flagfacup*\lw) $) -- ($ (offset) - (0,\flagfacdown*\lw) $);

  \coordinate (offset) at (0.0429116159226874, 3);

 \draw[significant] ($ (offset) $)-- ($ (offset)  + (3*\clusterlength,0) $);
 \draw[tick] ($ (offset) + (0,\flagfacup*\lw) $) -- ($ (offset) - (0,\flagfacdown*\lw) $);

  \coordinate (offset) at (0.00681574471644168, 2);

 \draw[cluster] ($ (offset) $)-- ($ (offset)  + (3*\clusterlength,0) $);
 \draw[tick] ($ (offset) + (0,\flagfacup*\lw) $) -- ($ (offset) - (0,\flagfacdown*\lw) $);

  \coordinate (offset) at (0.029607710682219, 1);

 \draw[cluster] ($ (offset) $)-- ($ (offset)  + (3*\clusterlength,0) $);
 \draw[tick] ($ (offset) + (0,\flagfacup*\lw) $) -- ($ (offset) - (0,\flagfacdown*\lw) $);

  \draw[thin, gray, dotted ,step=1] (0,0) grid ($ (0.075,\maxclusters) $);
        \foreach \x in {0.025, 0.05,...,0.1} {
            \draw[tick,black,thin] (\x,-0.4) -- (\x,-0.1);
        }
        \foreach \x in {0.025,0.05,0.075} {
            \node at (\x,-0.4) [below,black] {\footnotesize $\x$ };
        }
        \foreach \x in {0.1} {
            \draw[tick,black,thin] (\x,-0.4) -- (\x,0.15);
            \node at (\x,-0.4) [below,black] { $\x$ };
        }
        \foreach \y in {1,5,10,...,\maxclusters} {
            \draw[tick,black,thin] (-0.005,\y) -- (-0.008,\y);
            \node at (-0.007,\y) [left,black] { $\y$ };
        }
        \draw[tick,black,thin]  ($ (0.125,\maxclusters+1) $) -- ($ (-0.005,\maxclusters+1) $) -- (-0.005,-0.4) -- ($ (0.125,-0.4) $) ;
        \node at ($(0.06,-2.5-1)$) [left,black] {\footnotesize epistatic units };
        \node at ($(-0.024,\maxclusters*0.2)$) [rotate=90] { \footnotesize dual edges };

\end{tikzpicture}
  \end{minipage}
  \caption[$1{\ast}{\ast}{\ast}{\ast}$(GouldCFU) to $1{\ast}{\ast}{\ast}{\ast}$(GouldTTD).]{Comparing $1{\ast}{\ast}{\ast}{\ast}$(Gould bacterial CFU counts) to $1{\ast}{\ast}{\ast}{\ast}$(Gould lifespans). \emph{Left}. Filtration of $1{\ast}{\ast}{\ast}{\ast}$ CFU counts. \emph{Right}. Parallel filtration of $1{\ast}{\ast}{\ast}{\ast}$ lifespans.}
  \label{fig:Gould:parallel:CFUtoTTD:1****}
\end{figure}

% Fig S14
\begin{figure}[t]
  \begin{minipage}[t]{.45\linewidth}
    \centering
    \input{Figures/paralleltransports/GouldTTDvsCFU/filtrationbars_Lud2017FlygutCFUsDataNormalizedstar0starstarstar.tex}
  \end{minipage}%
  \hfill%
  \begin{minipage}[t]{.45\linewidth}
    \centering
     \newcommand\maxepi{1}
            \newcommand\maxclusters{22}
            \newcommand\clusterlength{0.33pt}
            \newcommand\clusterspace{0.13pt}
            \newcommand\flagfacup{2.5}
            \newcommand\flagfacdown{4}
            \newcommand\lw{3pt}

\begin{tikzpicture}[yscale=0.5,xscale=90, scale = 0.4]
\tikzset{
        cluster/.style = {red, line width=\lw, join=round},
        significant/.style = {blue, line width=\lw, join=round},
        semisignificant/.style = {purple, line width=\lw, join=round},
        tick/.style = {black, thick},
        tick_thin/.style = {black, thin}}

  \coordinate (offset) at (0.024398512745901, 21);

 \draw[significant] ($ (offset) $)-- ($ (offset)  + (3*\clusterlength,0) $);
 \draw[tick] ($ (offset) + (0,\flagfacup*\lw) $) -- ($ (offset) - (0,\flagfacdown*\lw) $);

  \coordinate (offset) at (0.0123002892929533, 20);

 \draw[cluster] ($ (offset) $)-- ($ (offset)  + (3*\clusterlength,0) $);
 \draw[tick] ($ (offset) + (0,\flagfacup*\lw) $) -- ($ (offset) - (0,\flagfacdown*\lw) $);

  \coordinate (offset) at (0.00707858085837869, 19);

 \draw[cluster] ($ (offset) $)-- ($ (offset)  + (3*\clusterlength,0) $);
 \draw[tick] ($ (offset) + (0,\flagfacup*\lw) $) -- ($ (offset) - (0,\flagfacdown*\lw) $);

  \coordinate (offset) at (0.0124067853041478, 18);

 \draw[cluster] ($ (offset) $)-- ($ (offset)  + (3*\clusterlength,0) $);
 \draw[tick] ($ (offset) + (0,\flagfacup*\lw) $) -- ($ (offset) - (0,\flagfacdown*\lw) $);

  \coordinate (offset) at (0.0176783378582706, 17);

 \draw[cluster] ($ (offset) $)-- ($ (offset)  + (3*\clusterlength,0) $);
 \draw[tick] ($ (offset) + (0,\flagfacup*\lw) $) -- ($ (offset) - (0,\flagfacdown*\lw) $);

  \coordinate (offset) at (0.0216514536266447, 16);

 \draw[cluster] ($ (offset) $)-- ($ (offset)  + (3*\clusterlength,0) $);
 \draw[tick] ($ (offset) + (0,\flagfacup*\lw) $) -- ($ (offset) - (0,\flagfacdown*\lw) $);

  \coordinate (offset) at (0.0259798978312835, 15);

 \draw[significant] ($ (offset) $)-- ($ (offset)  + (3*\clusterlength,0) $);
 \draw[tick] ($ (offset) + (0,\flagfacup*\lw) $) -- ($ (offset) - (0,\flagfacdown*\lw) $);

  \coordinate (offset) at (0.0084664328899549, 14);

 \draw[cluster] ($ (offset) $)-- ($ (offset)  + (3*\clusterlength,0) $);
 \draw[tick] ($ (offset) + (0,\flagfacup*\lw) $) -- ($ (offset) - (0,\flagfacdown*\lw) $);

  \coordinate (offset) at (0.0479686809232683, 13);

 \draw[significant] ($ (offset) $)-- ($ (offset)  + (3*\clusterlength,0) $);
 \draw[tick] ($ (offset) + (0,\flagfacup*\lw) $) -- ($ (offset) - (0,\flagfacdown*\lw) $);

  \coordinate (offset) at (0.0230563864235879, 12);

 \draw[semisignificant] ($ (offset) $)-- ($ (offset)  + (3*\clusterlength,0) $);
 \draw[tick] ($ (offset) + (0,\flagfacup*\lw) $) -- ($ (offset) - (0,\flagfacdown*\lw) $);

  \coordinate (offset) at (0.0230563864235879, 11);

 \draw[semisignificant] ($ (offset) $)-- ($ (offset)  + (3*\clusterlength,0) $);
 \draw[tick] ($ (offset) + (0,\flagfacup*\lw) $) -- ($ (offset) - (0,\flagfacdown*\lw) $);

  \coordinate (offset) at (0.000753040516848796, 10);

 \draw[cluster] ($ (offset) $)-- ($ (offset)  + (3*\clusterlength,0) $);
 \draw[tick] ($ (offset) + (0,\flagfacup*\lw) $) -- ($ (offset) - (0,\flagfacdown*\lw) $);

  \coordinate (offset) at (0.0166421954223584, 9);

 \draw[cluster] ($ (offset) $)-- ($ (offset)  + (3*\clusterlength,0) $);
 \draw[tick] ($ (offset) + (0,\flagfacup*\lw) $) -- ($ (offset) - (0,\flagfacdown*\lw) $);

  \coordinate (offset) at (0.0328007714478756, 8);

 \draw[cluster] ($ (offset) $)-- ($ (offset)  + (3*\clusterlength,0) $);
 \draw[tick] ($ (offset) + (0,\flagfacup*\lw) $) -- ($ (offset) - (0,\flagfacdown*\lw) $);

  \coordinate (offset) at (0.0328007714478756, 7);

 \draw[cluster] ($ (offset) $)-- ($ (offset)  + (3*\clusterlength,0) $);
 \draw[tick] ($ (offset) + (0,\flagfacup*\lw) $) -- ($ (offset) - (0,\flagfacdown*\lw) $);

  \coordinate (offset) at (0.00410009643098445, 6);

 \draw[cluster] ($ (offset) $)-- ($ (offset)  + (3*\clusterlength,0) $);
 \draw[tick] ($ (offset) + (0,\flagfacup*\lw) $) -- ($ (offset) - (0,\flagfacdown*\lw) $);

  \coordinate (offset) at (0.00410009643098445, 5);

 \draw[cluster] ($ (offset) $)-- ($ (offset)  + (3*\clusterlength,0) $);
 \draw[tick] ($ (offset) + (0,\flagfacup*\lw) $) -- ($ (offset) - (0,\flagfacdown*\lw) $);

  \coordinate (offset) at (0.00933770240892508, 4);

 \draw[cluster] ($ (offset) $)-- ($ (offset)  + (3*\clusterlength,0) $);
 \draw[tick] ($ (offset) + (0,\flagfacup*\lw) $) -- ($ (offset) - (0,\flagfacdown*\lw) $);

  \coordinate (offset) at (0.0238551065075459, 3);

 \draw[significant] ($ (offset) $)-- ($ (offset)  + (3*\clusterlength,0) $);
 \draw[tick] ($ (offset) + (0,\flagfacup*\lw) $) -- ($ (offset) - (0,\flagfacdown*\lw) $);

  \coordinate (offset) at (0.0239616025187403, 2);

 \draw[significant] ($ (offset) $)-- ($ (offset)  + (3*\clusterlength,0) $);
 \draw[tick] ($ (offset) + (0,\flagfacup*\lw) $) -- ($ (offset) - (0,\flagfacdown*\lw) $);

  \coordinate (offset) at (0.00790692542691236, 1);

 \draw[cluster] ($ (offset) $)-- ($ (offset)  + (3*\clusterlength,0) $);
 \draw[tick] ($ (offset) + (0,\flagfacup*\lw) $) -- ($ (offset) - (0,\flagfacdown*\lw) $);

  \draw[thin, gray, dotted ,step=1] (0,0) grid ($ (0.05,\maxclusters) $);
        \foreach \x in {0.025, 0.05,...,0.1} {
            \draw[tick,black,thin] (\x,-0.4) -- (\x,-0.1);
        }
        \foreach \x in {0.025,0.05,0.075} {
            \node at (\x,-0.4) [below,black] {\footnotesize $\x$ };
        }
        \foreach \x in {0.1} {
            \draw[tick,black,thin] (\x,-0.4) -- (\x,0.15);
            \node at (\x,-0.4) [below,black] { $\x$ };
        }
        \foreach \y in {1,5,10,...,\maxclusters} {
            \draw[tick,black,thin] (-0.005,\y) -- (-0.008,\y);
            \node at (-0.007,\y) [left,black] { $\y$ };
        }
        \draw[tick,black,thin]  ($ (0.125,\maxclusters+1) $) -- ($ (-0.005,\maxclusters+1) $) -- (-0.005,-0.4) -- ($ (0.125,-0.4) $) ;
        \node at ($(0.06,-2.5-1)$) [left,black] {\footnotesize epistatic units };
        \node at ($(-0.024,\maxclusters*0.2)$) [rotate=90] { \footnotesize dual edges };

\end{tikzpicture}
  \end{minipage}
  \caption[${\ast}0{\ast}{\ast}{\ast}$(GouldCFU) to ${\ast}0{\ast}{\ast}{\ast}$(GouldTTD).]{Comparing ${\ast}0{\ast}{\ast}{\ast}$(Gould bacterial CFU counts) to ${\ast}0{\ast}{\ast}{\ast}$(Gould lifespans). \emph{Left}. Filtration of ${\ast}0{\ast}{\ast}{\ast}$ CFU counts. \emph{Right}. Parallel filtration of ${\ast}0{\ast}{\ast}{\ast}$ lifespans.}
 \label{fig:Gould:parallel:CFUtoTTD:*0***}
\end{figure}

\end{samepage}

% Fig S15
\begin{figure}[t]
  \begin{minipage}[t]{.45\linewidth}
    \centering
    \input{Figures/paralleltransports/GouldTTDvsCFU/filtrationbars_Lud2017FlygutCFUsDataNormalizedstar1starstarstar.tex}
  \end{minipage}%
  \hfill%
  \begin{minipage}[t]{.45\linewidth}
    \centering
    \newcommand\maxepi{1}
            \newcommand\maxclusters{24}
            \newcommand\clusterlength{0.33pt}
            \newcommand\clusterspace{0.13pt}
            \newcommand\flagfacup{2.5}
            \newcommand\flagfacdown{4}
            \newcommand\lw{3pt}

\begin{tikzpicture}[yscale=0.48,xscale=90, scale = 0.4]
\tikzset{
        cluster/.style = {red, line width=\lw, join=round},
        significant/.style = {blue, line width=\lw, join=round},
        semisignificant/.style = {purple, line width=\lw, join=round},
        tick/.style = {black, thick},
        tick_thin/.style = {black, thin}}

  \coordinate (offset) at (0.0178689707641586, 23);

 \draw[cluster] ($ (offset) $)-- ($ (offset)  + (3*\clusterlength,0) $);
 \draw[tick] ($ (offset) + (0,\flagfacup*\lw) $) -- ($ (offset) - (0,\flagfacdown*\lw) $);

  \coordinate (offset) at (0.0178689707641586, 22);

 \draw[cluster] ($ (offset) $)-- ($ (offset)  + (3*\clusterlength,0) $);
 \draw[tick] ($ (offset) + (0,\flagfacup*\lw) $) -- ($ (offset) - (0,\flagfacdown*\lw) $);

  \coordinate (offset) at (0.0178689707641586, 21);

 \draw[cluster] ($ (offset) $)-- ($ (offset)  + (3*\clusterlength,0) $);
 \draw[tick] ($ (offset) + (0,\flagfacup*\lw) $) -- ($ (offset) - (0,\flagfacdown*\lw) $);

  \coordinate (offset) at (0.0592642886760003, 20);

 \draw[significant] ($ (offset) $)-- ($ (offset)  + (3*\clusterlength,0) $);
 \draw[tick] ($ (offset) + (0,\flagfacup*\lw) $) -- ($ (offset) - (0,\flagfacdown*\lw) $);

  \coordinate (offset) at (0.0092623983572402, 19);

 \draw[cluster] ($ (offset) $)-- ($ (offset)  + (3*\clusterlength,0) $);
 \draw[tick] ($ (offset) + (0,\flagfacup*\lw) $) -- ($ (offset) - (0,\flagfacdown*\lw) $);

  \coordinate (offset) at (0.0233442560223127, 18);

 \draw[cluster] ($ (offset) $)-- ($ (offset)  + (3*\clusterlength,0) $);
 \draw[tick] ($ (offset) + (0,\flagfacup*\lw) $) -- ($ (offset) - (0,\flagfacdown*\lw) $);

  \coordinate (offset) at (0.0260915227426283, 17);

 \draw[cluster] ($ (offset) $)-- ($ (offset)  + (3*\clusterlength,0) $);
 \draw[tick] ($ (offset) + (0,\flagfacup*\lw) $) -- ($ (offset) - (0,\flagfacdown*\lw) $);

  \coordinate (offset) at (0.0260915227426283, 16);

 \draw[cluster] ($ (offset) $)-- ($ (offset)  + (3*\clusterlength,0) $);
 \draw[tick] ($ (offset) + (0,\flagfacup*\lw) $) -- ($ (offset) - (0,\flagfacdown*\lw) $);

  \coordinate (offset) at (0.0140864879016725, 15);

 \draw[cluster] ($ (offset) $)-- ($ (offset)  + (3*\clusterlength,0) $);
 \draw[tick] ($ (offset) + (0,\flagfacup*\lw) $) -- ($ (offset) - (0,\flagfacdown*\lw) $);

  \coordinate (offset) at (0.0195420180541726, 14);

 \draw[cluster] ($ (offset) $)-- ($ (offset)  + (3*\clusterlength,0) $);
 \draw[tick] ($ (offset) + (0,\flagfacup*\lw) $) -- ($ (offset) - (0,\flagfacdown*\lw) $);

  \coordinate (offset) at (0.0195420180541726, 13);

 \draw[cluster] ($ (offset) $)-- ($ (offset)  + (3*\clusterlength,0) $);
 \draw[tick] ($ (offset) + (0,\flagfacup*\lw) $) -- ($ (offset) - (0,\flagfacdown*\lw) $);

  \coordinate (offset) at (0.0151737664145032, 12);

 \draw[cluster] ($ (offset) $)-- ($ (offset)  + (3*\clusterlength,0) $);
 \draw[tick] ($ (offset) + (0,\flagfacup*\lw) $) -- ($ (offset) - (0,\flagfacdown*\lw) $);

  \coordinate (offset) at (0.0170863881029546, 11);

 \draw[cluster] ($ (offset) $)-- ($ (offset)  + (3*\clusterlength,0) $);
 \draw[tick] ($ (offset) + (0,\flagfacup*\lw) $) -- ($ (offset) - (0,\flagfacdown*\lw) $);

  \coordinate (offset) at (0.0170863881029546, 10);

 \draw[cluster] ($ (offset) $)-- ($ (offset)  + (3*\clusterlength,0) $);
 \draw[tick] ($ (offset) + (0,\flagfacup*\lw) $) -- ($ (offset) - (0,\flagfacdown*\lw) $);

  \coordinate (offset) at (0.0170863881029546, 9);

 \draw[cluster] ($ (offset) $)-- ($ (offset)  + (3*\clusterlength,0) $);
 \draw[tick] ($ (offset) + (0,\flagfacup*\lw) $) -- ($ (offset) - (0,\flagfacdown*\lw) $);

  \coordinate (offset) at (0.0179994012076926, 8);

 \draw[cluster] ($ (offset) $)-- ($ (offset)  + (3*\clusterlength,0) $);
 \draw[tick] ($ (offset) + (0,\flagfacup*\lw) $) -- ($ (offset) - (0,\flagfacdown*\lw) $);

  \coordinate (offset) at (0.0216719382780607, 7);

 \draw[cluster] ($ (offset) $)-- ($ (offset)  + (3*\clusterlength,0) $);
 \draw[tick] ($ (offset) + (0,\flagfacup*\lw) $) -- ($ (offset) - (0,\flagfacdown*\lw) $);

  \coordinate (offset) at (0.0216719382780607, 6);

 \draw[cluster] ($ (offset) $)-- ($ (offset)  + (3*\clusterlength,0) $);
 \draw[tick] ($ (offset) + (0,\flagfacup*\lw) $) -- ($ (offset) - (0,\flagfacdown*\lw) $);

  \coordinate (offset) at (6.52152217670021e-05, 5);

 \draw[cluster] ($ (offset) $)-- ($ (offset)  + (3*\clusterlength,0) $);
 \draw[tick] ($ (offset) + (0,\flagfacup*\lw) $) -- ($ (offset) - (0,\flagfacdown*\lw) $);

  \coordinate (offset) at (0.0299253791456268, 4);

 \draw[significant] ($ (offset) $)-- ($ (offset)  + (3*\clusterlength,0) $);
 \draw[tick] ($ (offset) + (0,\flagfacup*\lw) $) -- ($ (offset) - (0,\flagfacdown*\lw) $);

  \coordinate (offset) at (0.0299253791456268, 3);

 \draw[significant] ($ (offset) $)-- ($ (offset)  + (3*\clusterlength,0) $);
 \draw[tick] ($ (offset) + (0,\flagfacup*\lw) $) -- ($ (offset) - (0,\flagfacdown*\lw) $);

  \coordinate (offset) at (0.0127169682445654, 2);

 \draw[cluster] ($ (offset) $)-- ($ (offset)  + (3*\clusterlength,0) $);
 \draw[tick] ($ (offset) + (0,\flagfacup*\lw) $) -- ($ (offset) - (0,\flagfacdown*\lw) $);

  \coordinate (offset) at (0.0127169682445654, 1);

 \draw[cluster] ($ (offset) $)-- ($ (offset)  + (3*\clusterlength,0) $);
 \draw[tick] ($ (offset) + (0,\flagfacup*\lw) $) -- ($ (offset) - (0,\flagfacdown*\lw) $);

  \draw[thin, gray, dotted ,step=1] (0,0) grid ($ (0.075,\maxclusters) $);
        \foreach \x in {0.025, 0.05,...,0.1} {
            \draw[tick,black,thin] (\x,-0.4) -- (\x,-0.1);
        }
        \foreach \x in {0.025,0.05,0.075} {
            \node at (\x,-0.4) [below,black] {\footnotesize $\x$ };
        }
        \foreach \x in {0.1} {
            \draw[tick,black,thin] (\x,-0.4) -- (\x,0.15);
            \node at (\x,-0.4) [below,black] { $\x$ };
        }
        \foreach \y in {1,5,10,...,\maxclusters} {
            \draw[tick,black,thin] (-0.005,\y) -- (-0.008,\y);
            \node at (-0.007,\y) [left,black] { $\y$ };
        }
        \draw[tick,black,thin]  ($ (0.125,\maxclusters+1) $) -- ($ (-0.005,\maxclusters+1) $) -- (-0.005,-0.4) -- ($ (0.125,-0.4) $) ;
        \node at ($(0.06,-2.5-1)$) [left,black] {\footnotesize epistatic units };
        \node at ($(-0.024,\maxclusters*0.2)$) [rotate=90] { \footnotesize dual edges };

\end{tikzpicture}
  \end{minipage}
  \caption[${\ast}1{\ast}{\ast}{\ast}$(GouldCFU) to ${\ast}1{\ast}{\ast}{\ast}$(GouldTTD).]{Comparing ${\ast}1{\ast}{\ast}{\ast}$(Gould bacterial CFU counts) to ${\ast}1{\ast}{\ast}{\ast}$(Gould lifespans). \emph{Left}. Filtration of ${\ast}1{\ast}{\ast}{\ast}$ CFU counts. \emph{Right}. Parallel filtration of ${\ast}1{\ast}{\ast}{\ast}$ lifespans.}
  \label{fig:Gould:parallel:CFUtoTTD:*1***}
\end{figure}

\begin{figure}[t]\centering
    \begin{minipage}[t]{.9\linewidth}
    \subcaption{}
    \includegraphics[scale=0.5]{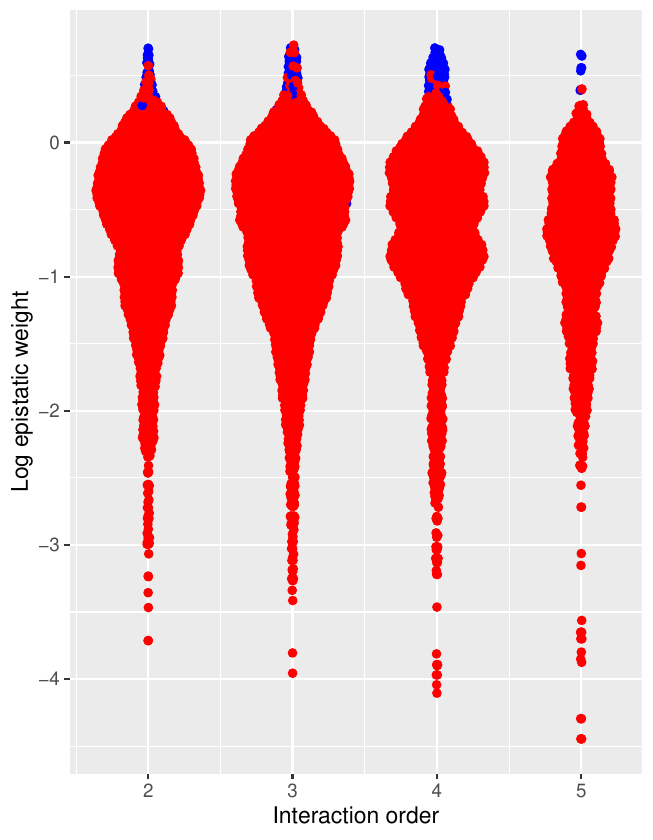}
  \end{minipage}
  \begin{minipage}[t]{.9\linewidth} 
    \subcaption{}
    \scalebox{0.7}{    \begin{tikzpicture}[yscale=1, scale=1, xscale=1]
        \tikzset{
            tick/.style = {black, thick},
                tick_thin/.style = {black, thin}}
    \draw[line width=0.2mm, gray ,step=1, dotted] (0,0) grid (12,10);
    \foreach \x in {0,...,12} {
        \draw[tick,black,thin] (\x,-0.05) -- (\x,0); 
        \node at (\x,-0.1) [below,black] { \small $\x$ };}

    \foreach \y in {0,...,10} {
        \draw[tick,black,thin] (0,\y) -- (0.1,\y);
        \node at (-0.1,\y) [left,black] {\small $\y$ };}
    \node at (4,-1) [left,black] {standard deviation};
    \node at (-1,3) [rotate=90] {\# of significant critical  interactions};\draw [blue, thick] (0.1 , 4) node {\tiny $\bullet$};
\draw [blue, thick] (0.2 , 5) node {\tiny $\bullet$};
\draw [blue, thick] (0.3 , 5) node {\tiny $\bullet$};
\draw [blue, thick] (0.4 , 4) node {\tiny $\bullet$};
\draw [blue, thick] (0.5 , 5) node {\tiny $\bullet$};
\draw [blue, thick] (0.6 , 5) node {\tiny $\bullet$};
\draw [blue, thick] (0.7 , 5) node {\tiny $\bullet$};
\draw [blue, thick] (0.8 , 4) node {\tiny $\bullet$};
\draw [blue, thick] (0.9 , 4) node {\tiny $\bullet$};
\draw [blue, thick] (1 , 4) node {\tiny $\bullet$};
\draw [blue, thick] (1.1 , 4) node {\tiny $\bullet$};
\draw [blue, thick] (1.2 , 4) node {\tiny $\bullet$};
\draw [blue, thick] (1.3 , 4) node {\tiny $\bullet$};
\draw [blue, thick] (1.4 , 4) node {\tiny $\bullet$};
\draw [blue, thick] (1.5 , 4) node {\tiny $\bullet$};
\draw [blue, thick] (1.6 , 4) node {\tiny $\bullet$};
\draw [blue, thick] (1.7 , 4) node {\tiny $\bullet$};
\draw [blue, thick] (1.8 , 4) node {\tiny $\bullet$};
\draw [blue, thick] (1.9 , 4) node {\tiny $\bullet$};
\draw [blue, thick] (2 , 4) node {\tiny $\bullet$};
\draw [blue, thick] (2.1 , 4) node {\tiny $\bullet$};
\draw [blue, thick] (2.2 , 4) node {\tiny $\bullet$};
\draw [blue, thick] (2.3 , 4) node {\tiny $\bullet$};
\draw [blue, thick] (2.4 , 6) node {\tiny $\bullet$};
\draw [blue, thick] (2.5 , 4) node {\tiny $\bullet$};
\draw [blue, thick] (2.6 , 4) node {\tiny $\bullet$};
\draw [blue, thick] (2.7 , 4) node {\tiny $\bullet$};
\draw [blue, thick] (2.8 , 4) node {\tiny $\bullet$};
\draw [blue, thick] (2.9 , 5) node {\tiny $\bullet$};
\draw [blue, thick] (3 , 4) node {\tiny $\bullet$};
\draw [blue, thick] (3.1 , 4) node {\tiny $\bullet$};
\draw [blue, thick] (3.2 , 4) node {\tiny $\bullet$};
\draw [blue, thick] (3.3 , 3) node {\tiny $\bullet$};
\draw [blue, thick] (3.4 , 4) node {\tiny $\bullet$};
\draw [blue, thick] (3.5 , 4) node {\tiny $\bullet$};
\draw [blue, thick] (3.6 , 4) node {\tiny $\bullet$};
\draw [blue, thick] (3.7 , 2) node {\tiny $\bullet$};
\draw [blue, thick] (3.8 , 4) node {\tiny $\bullet$};
\draw [blue, thick] (3.9 , 8) node {\tiny $\bullet$};
\draw [blue, thick] (4 , 4) node {\tiny $\bullet$};
\draw [blue, thick] (4.1 , 5) node {\tiny $\bullet$};
\draw [red, thick] (4.2 , 2) node {\tiny $\bullet$};
\draw [blue, thick] (4.3 , 4) node {\tiny $\bullet$};
\draw [blue, thick] (4.4 , 4) node {\tiny $\bullet$};
\draw [blue, thick] (4.5 , 3) node {\tiny $\bullet$};
\draw [blue, thick] (4.6 , 3) node {\tiny $\bullet$};
\draw [blue, thick] (4.7 , 2) node {\tiny $\bullet$};
\draw [blue, thick] (4.8 , 2) node {\tiny $\bullet$};
\draw [blue, thick] (4.9 , 2) node {\tiny $\bullet$};
\draw [blue, thick] (5 , 2) node {\tiny $\bullet$};
\draw [blue, thick] (5.1 , 2) node {\tiny $\bullet$};
\draw [red, thick] (5.2 , 3) node {\tiny $\bullet$};
\draw [blue, thick] (5.3 , 2) node {\tiny $\bullet$};
\draw [blue, thick] (5.4 , 3) node {\tiny $\bullet$};
\draw [blue, thick] (5.5 , 2) node {\tiny $\bullet$};
\draw [blue, thick] (5.6 , 2) node {\tiny $\bullet$};
\draw [red, thick] (5.7 , 0) node {\tiny $\bullet$};
\draw [red, thick] (5.8 , 2) node {\tiny $\bullet$};
\draw [red, thick] (5.9 , 0) node {\tiny $\bullet$};
\draw [red, thick] (5.99999999999999 , 0) node {\tiny $\bullet$};
\draw [blue, thick] (6.09999999999999 , 2) node {\tiny $\bullet$};
\draw [red, thick] (6.19999999999999 , 0) node {\tiny $\bullet$};
\draw [red, thick] (6.29999999999999 , 2) node {\tiny $\bullet$};
\draw [blue, thick] (6.39999999999999 , 2) node {\tiny $\bullet$};
\draw [blue, thick] (6.49999999999999 , 3) node {\tiny $\bullet$};
\draw [red, thick] (6.59999999999999 , 0) node {\tiny $\bullet$};
\draw [blue, thick] (6.69999999999999 , 2) node {\tiny $\bullet$};
\draw [red, thick] (6.79999999999999 , 0) node {\tiny $\bullet$};
\draw [blue, thick] (6.89999999999999 , 2) node {\tiny $\bullet$};
\draw [red, thick] (6.99999999999999 , 0) node {\tiny $\bullet$};
\draw [blue, thick] (7.09999999999999 , 2) node {\tiny $\bullet$};
\draw [blue, thick] (7.19999999999999 , 2) node {\tiny $\bullet$};
\draw [blue, thick] (7.29999999999999 , 2) node {\tiny $\bullet$};
\draw [red, thick] (7.39999999999999 , 0) node {\tiny $\bullet$};
\draw [red, thick] (7.49999999999999 , 0) node {\tiny $\bullet$};
\draw [blue, thick] (7.59999999999999 , 2) node {\tiny $\bullet$};
\draw [red, thick] (7.69999999999999 , 2) node {\tiny $\bullet$};
\draw [red, thick] (7.79999999999999 , 0) node {\tiny $\bullet$};
\draw [blue, thick] (7.89999999999999 , 2) node {\tiny $\bullet$};
\draw [blue, thick] (7.99999999999999 , 3) node {\tiny $\bullet$};
\draw [red, thick] (8.09999999999999 , 2) node {\tiny $\bullet$};
\draw [red, thick] (8.19999999999999 , 0) node {\tiny $\bullet$};
\draw [red, thick] (8.29999999999999 , 0) node {\tiny $\bullet$};
\draw [blue, thick] (8.39999999999999 , 2) node {\tiny $\bullet$};
\draw [blue, thick] (8.49999999999999 , 2) node {\tiny $\bullet$};
\draw [red, thick] (8.59999999999999 , 2) node {\tiny $\bullet$};
\draw [blue, thick] (8.69999999999999 , 2) node {\tiny $\bullet$};
\draw [blue, thick] (8.79999999999998 , 2) node {\tiny $\bullet$};
\draw [red, thick] (8.89999999999998 , 0) node {\tiny $\bullet$};
\draw [red, thick] (8.99999999999998 , 0) node {\tiny $\bullet$};
\draw [red, thick] (9.09999999999998 , 0) node {\tiny $\bullet$};
\draw [red, thick] (9.19999999999998 , 0) node {\tiny $\bullet$};
\draw [red, thick] (9.29999999999998 , 0) node {\tiny $\bullet$};
\draw [blue, thick] (9.39999999999998 , 2) node {\tiny $\bullet$};
\draw [blue, thick] (9.49999999999998 , 2) node {\tiny $\bullet$};
\draw [red, thick] (9.59999999999998 , 2) node {\tiny $\bullet$};
\draw [red, thick] (9.69999999999998 , 0) node {\tiny $\bullet$};
\draw [red, thick] (9.79999999999998 , 0) node {\tiny $\bullet$};
\draw [red, thick] (9.89999999999998 , 0) node {\tiny $\bullet$};
\draw [red, thick] (9.99999999999998 , 0) node {\tiny $\bullet$};
\draw [blue, thick] (10.1 , 2) node {\tiny $\bullet$};
\draw [red, thick] (10.2 , 0) node {\tiny $\bullet$};
\draw [blue, thick] (10.3 , 3) node {\tiny $\bullet$};
\draw [red, thick] (10.4 , 0) node {\tiny $\bullet$};
\draw [red, thick] (10.5 , 0) node {\tiny $\bullet$};
\draw [red, thick] (10.6 , 0) node {\tiny $\bullet$};
\draw [red, thick] (10.7 , 0) node {\tiny $\bullet$};
\draw [red, thick] (10.8 , 2) node {\tiny $\bullet$};
\draw [red, thick] (10.9 , 0) node {\tiny $\bullet$};
\draw [red, thick] (11 , 0) node {\tiny $\bullet$};
\draw [red, thick] (11.1 , 0) node {\tiny $\bullet$};
\draw [red, thick] (11.2 , 0) node {\tiny $\bullet$};
\draw [red, thick] (11.3 , 0) node {\tiny $\bullet$};
\draw [blue, thick] (11.4 , 2) node {\tiny $\bullet$};
\draw [red, thick] (11.5 , 0) node {\tiny $\bullet$};
\draw [red, thick] (11.6 , 0) node {\tiny $\bullet$};
\draw [red, thick] (11.7 , 0) node {\tiny $\bullet$};
\draw [red, thick] (11.8 , 0) node {\tiny $\bullet$};
\draw [red, thick] (11.9 , 0) node {\tiny $\bullet$};

  \end{tikzpicture}}
  \end{minipage}
  
  \caption[Synthetic data demonstrate method performance.]{{\bf Synthetic data demonstrate method performance.}
    Synthetic height functions over the $4$-dimensional cube are generated with $100$ replicates each and standard deviation as indicated.
    The heights of the wild type $0000$ and $0001$ are sampled with mean $53$, all the other vertices with mean $50$.
    (a) The distribution of $\log_{10}$-transformed epistatic weights is roughly constant as a function of interaction order, indicating the dimensional normalization is effective.
    (b) The number of significant interactions decreases as the standard deviation of the input data for each genotype increases.
    A \textcolor{blue}{blue} dot is drawn if the interaction is significant and a \textcolor{red}{red} dot is drawn otherwise.}
  \label{fig:synthetic_MP}
\end{figure}

%Fig S23
\begin{figure}[ht]
\begin{subfigure}{.75\textwidth}
  \centering
  \includegraphics[width=.38\linewidth]{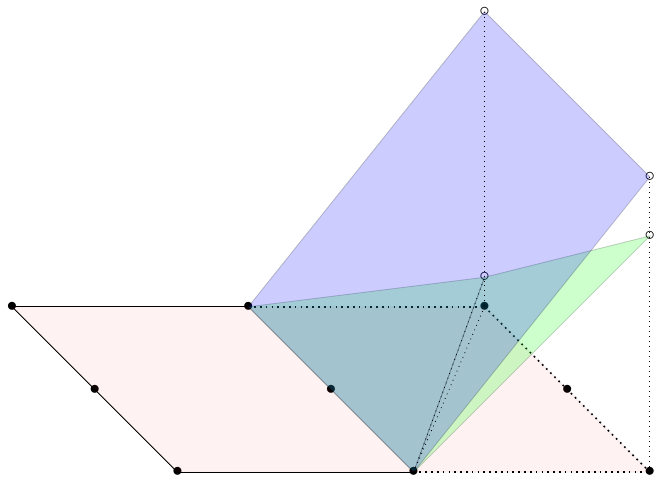}  
  \caption{}
  \label{fig:4d}
\end{subfigure}
\begin{subfigure}{.95\textwidth}
  \centering
  \caption{}
  \includegraphics[width=.9\linewidth]{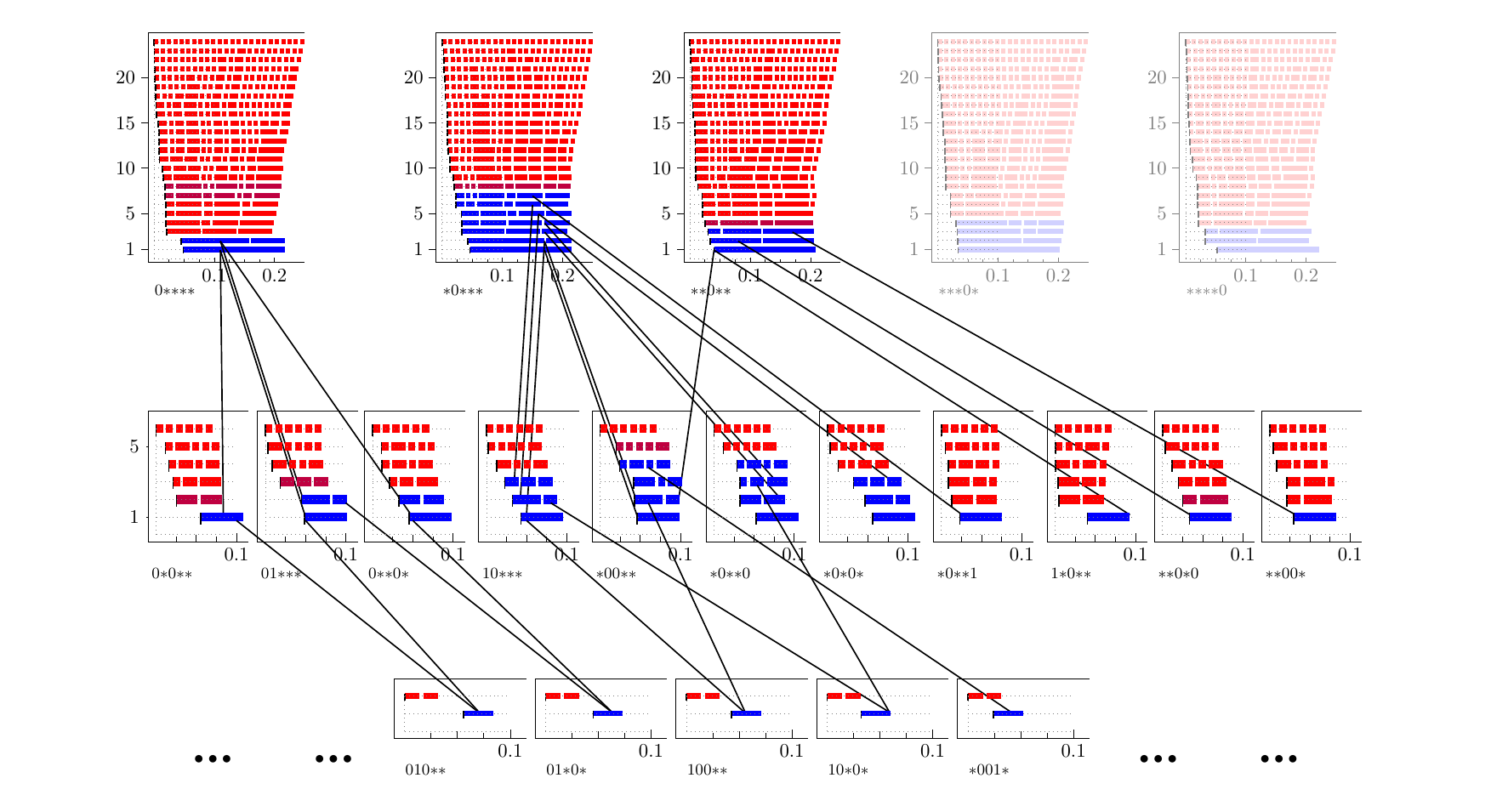}  
  \label{fig:4a}
\end{subfigure}
\begin{subfigure}{.95\textwidth}
  \centering
  \caption{}
  \includegraphics[width=.9\linewidth]{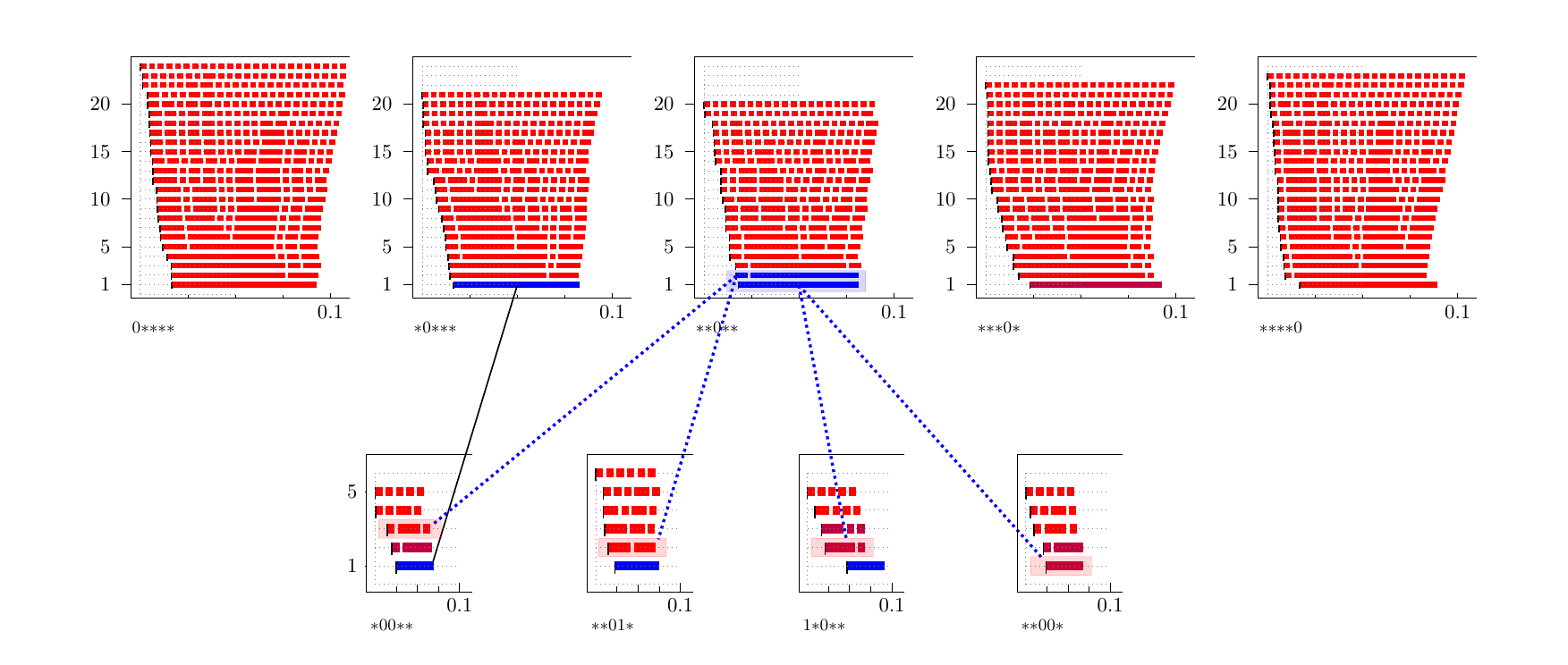}  
\end{subfigure}
\caption[Meta-epistatic charts illustrate whether higher-order interactions arise from lower-order interactions.]{%Figure 4. 
{\bf Meta-epistatic charts illustrate whether or not higher-order interactions arise from lower-order interactions.}  (a) Cartoon of the principle underlying meta-epistatic charts. The important loci in the interaction are depicted as black dots in a hyperplane through the genotypes, where the true dimensions of the genotypes are flattened onto the cartoon plane (pink). Higher-order interactions that derive from lower-order interactions occur in a new hyperplane (blue), which magnifies the weights of a subset of the landscape. In contrast, novel higher-order interactions that only arise in higher dimensions do not lie in a single additional hyperplane but instead require at least two additional hyperplanes (green). In (b) and (c) two meta-epistatic charts are represented. In each chart we identify the source of a higher-order interaction for the Eble and Gould data respectively. The results are compiled in Table~ \ref{Tab:special+bipyramids}. }
\label{fig4}
\end{figure}

% Fig S24
\begin{figure}[hb]
\centering
  \includegraphics[scale=0.5]{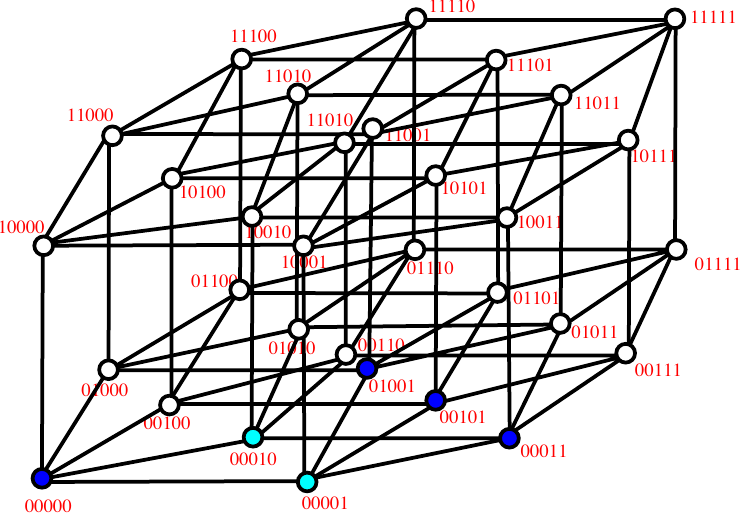}
  \caption[The bipyramid arising for the Khan data.]{{\bf Vertices of the bipyramid $\{00001\} + \{00000, 01001, 00101, 00011\} + \{00010\}$ arising for the Khan data set \cite{Khan1193} restricted to $n=4$ loci.} Dark blue dots correspond to common face $s\cap t$ of the bipyramid and light blue dots correspond to the satellite vertices of $s$ and $t$.}
  \label{sc5}
\end{figure}

% Fig S25
\begin{figure}[ht]
  \includegraphics[width=.95\linewidth]{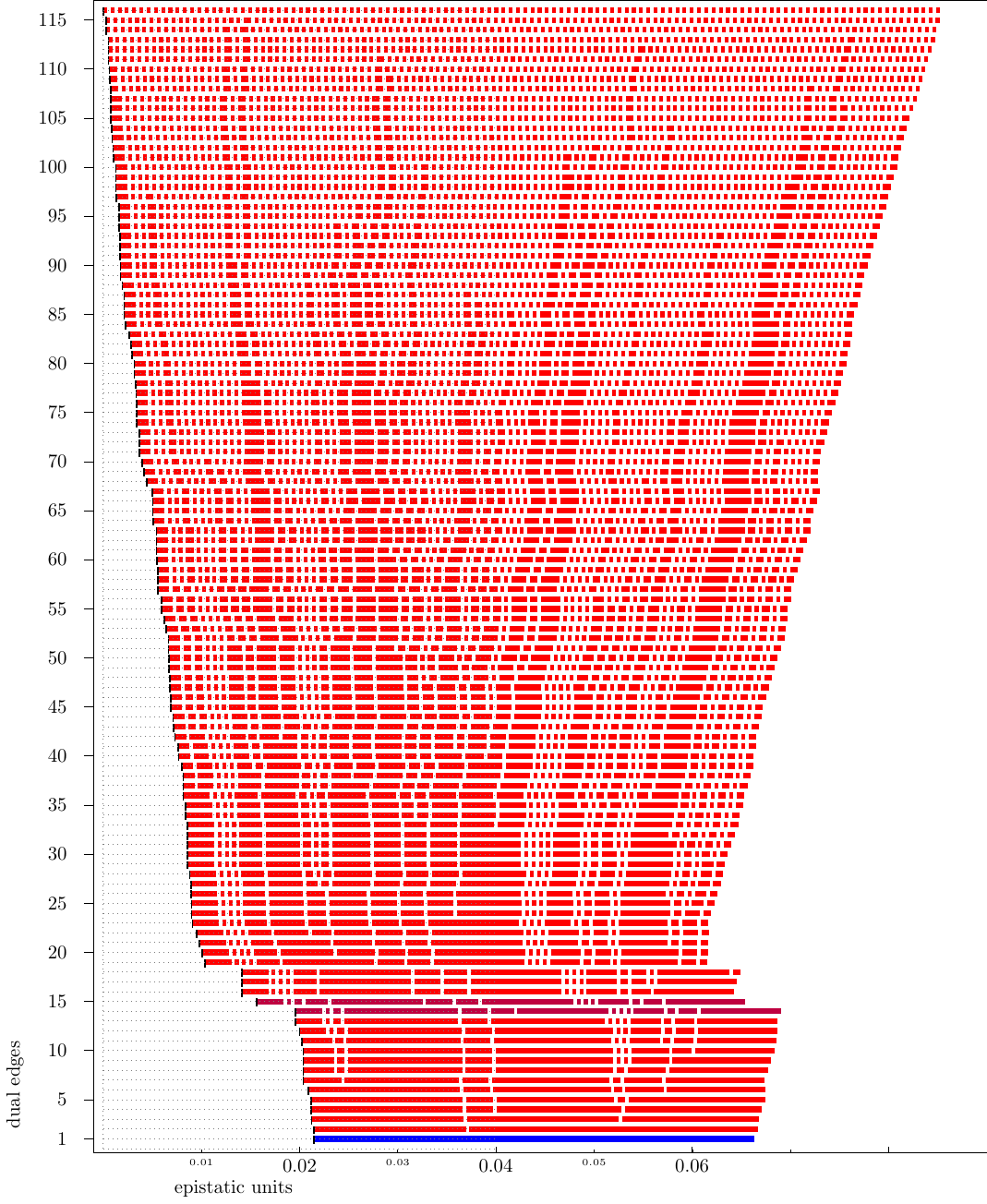}  
  \caption[Complete filtration of the Eble  data over the whole $5$-cube.]{\bf Complete filtration of the Eble fitness landscape over the whole $5$-cube.}
  \label{fig:Eble-5cube}
\end{figure}

\goodbreak
\FloatBarrier

%%%% tables
%Table S1
\begin{table}[!ht]\centering
  \caption{Number of circuits of $[0,1]^n$ and bipyramids among these. This indicates that bipyramids can analyze the majority of all possible interactions, which circuits exhaustively cover. Compare with Table~\ref{tab:complexity:gen-micro}, which shows the actual number of bipyramids for three datasets, indicating significantly fewer sectors are needed to cover the landscape.}
  \label{tab:bipyramids}
  \begin{tabular*}{.667\linewidth}{@{\extracolsep{\fill}}rrrr@{}}
    \toprule
    dimensions & circuits & bipyramids & percentage \\
    \midrule
    2 & 1 &  1 & 100.00\% \\
    3 & 20 & 8 &  40.00\% \\
    4 & 1348 & 1088 & 80.71\%\\
    5 & 353616 & 309056 & 87.40\%\\
    \bottomrule
  \end{tabular*}
\end{table}

% Tab S2-S5
\begin{table}[b]
                          \caption{\small Parallel analysis GouldCFU 0${\ast}$${\ast}$${\ast}$${\ast}$ $\rightarrow$ Gould 0${\ast}$${\ast}$${\ast}$${\ast}$, non-critical red/red-case omitted.}
\label{tab:Gould:parallel:CFUtoTTD:0****}
\begin{tabular*}{\linewidth}{@{\extracolsep{\fill}}lllllllll@{}}
        \toprule  \multicolumn{1}{l}{\scriptsize No.} &  \multicolumn{1}{l}{\scriptsize bipyramid} & \multicolumn{1}{l}{\scriptsize type} &  \multicolumn{1}{l}{\scriptsize $e_o$} & \multicolumn{1}{l}{\scriptsize $e_p$} & \multicolumn{1}{l}{\scriptsize $e_o/e_p$} & \multicolumn{1}{l}{\scriptsize $p_o$} & \multicolumn{1}{l}{\scriptsize $p_p$} & \multicolumn{1}{l}{\scriptsize $p_o/p_p$} \\ 
 \midrule  
\multicolumn{1}{r}{\scriptsize \textbf{22}}& \multicolumn{1}{l}{\scriptsize \{01001\}+\{01000,01100,01010,00111\}+\{00110\}} &\multicolumn{1}{l}{\scriptsize red/blue}& \multicolumn{1}{l}{\scriptsize 0.010} & \multicolumn{1}{l}{\scriptsize 0.027} & \multicolumn{1}{l}{\scriptsize 0.357} & \multicolumn{1}{l}{\scriptsize 0.978} & \multicolumn{1}{l}{\scriptsize 0.038} & \multicolumn{1}{l}{\scriptsize 25.873} \\ 
\multicolumn{1}{r}{\scriptsize \textbf{21}}& \multicolumn{1}{l}{\scriptsize \{01001\}+\{01000,00100,01100,00111\}+\{00110\}} &\multicolumn{1}{l}{\scriptsize red/blue}& \multicolumn{1}{l}{\scriptsize 0.010} & \multicolumn{1}{l}{\scriptsize 0.027} & \multicolumn{1}{l}{\scriptsize 0.357} & \multicolumn{1}{l}{\scriptsize 0.978} & \multicolumn{1}{l}{\scriptsize 0.038} & \multicolumn{1}{l}{\scriptsize 25.873} \\ 
\multicolumn{1}{r}{\scriptsize \textbf{20}}& \multicolumn{1}{l}{\scriptsize \{01001\}+\{01000,00010,01010,00111\}+\{00110\}} &\multicolumn{1}{l}{\scriptsize red/blue}& \multicolumn{1}{l}{\scriptsize 0.010} & \multicolumn{1}{l}{\scriptsize 0.027} & \multicolumn{1}{l}{\scriptsize 0.357} & \multicolumn{1}{l}{\scriptsize 0.978} & \multicolumn{1}{l}{\scriptsize 0.038} & \multicolumn{1}{l}{\scriptsize 25.873} \\
\bottomrule \end{tabular*}   \end{table}
                  
\newpage

\begin{table}[b]
                          \caption{\small Parallel analysis GouldCFU 1${\ast}$${\ast}$${\ast}$${\ast}$ $\rightarrow$ Gould 1${\ast}$${\ast}$${\ast}$${\ast}$, non-critical red/red-case omitted.}
\label{tab:Gould:parallel:CFUtoTTD:1****}
\begin{tabular*}{\linewidth}{@{\extracolsep{\fill}}lllllllll@{}}
        \toprule  \multicolumn{1}{l}{\scriptsize No.} &  \multicolumn{1}{l}{\scriptsize bipyramid} & \multicolumn{1}{l}{\scriptsize type} &  \multicolumn{1}{l}{\scriptsize $e_o$} & \multicolumn{1}{l}{\scriptsize $e_p$} & \multicolumn{1}{l}{\scriptsize $e_o/e_p$} & \multicolumn{1}{l}{\scriptsize $p_o$} & \multicolumn{1}{l}{\scriptsize $p_p$} & \multicolumn{1}{l}{\scriptsize $p_o/p_p$} \\ 
 \midrule  
\multicolumn{1}{r}{\scriptsize \textbf{23}}& \multicolumn{1}{l}{\scriptsize \{11001\}+\{11000,10101,11101,11011\}+\{11111\}} &\multicolumn{1}{l}{\scriptsize red/blue}& \multicolumn{1}{l}{\scriptsize 0.002} & \multicolumn{1}{l}{\scriptsize 0.051} & \multicolumn{1}{l}{\scriptsize 0.033} & \multicolumn{1}{l}{\scriptsize 0.962} & \multicolumn{1}{l}{\scriptsize 0.001} & \multicolumn{1}{l}{\scriptsize 1286.096} \\ 
\multicolumn{1}{r}{\scriptsize \textbf{22}}& \multicolumn{1}{l}{\scriptsize \{11100\}+\{11000,10101,11110,11101\}+\{11111\}} &\multicolumn{1}{l}{\scriptsize red/red}& \multicolumn{1}{l}{\scriptsize 0.017} & \multicolumn{1}{l}{\scriptsize 0.006} & \multicolumn{1}{l}{\scriptsize 2.799} & \multicolumn{1}{l}{\scriptsize 0.773} & \multicolumn{1}{l}{\scriptsize 0.689} & \multicolumn{1}{l}{\scriptsize 1.122} \\ 
\multicolumn{1}{r}{\scriptsize \textbf{21}}& \multicolumn{1}{l}{\scriptsize \{10000\}+\{11010,10101,10011,10111\}+\{11111\}} &\multicolumn{1}{l}{\scriptsize red/red}& \multicolumn{1}{l}{\scriptsize 0.023} & \multicolumn{1}{l}{\scriptsize 0.002} & \multicolumn{1}{l}{\scriptsize 10.615} & \multicolumn{1}{l}{\scriptsize 0.967} & \multicolumn{1}{l}{\scriptsize 0.875} & \multicolumn{1}{l}{\scriptsize 1.105} \\
\bottomrule \end{tabular*}   \end{table}
                  
\newpage

\begin{table}[b]
                          \caption{\small Parallel analysis GouldCFU ${\ast}$0${\ast}$${\ast}$${\ast}$ $\rightarrow$ Gould ${\ast}$0${\ast}$${\ast}$${\ast}$, non-critical red/red-case omitted.}
\label{tab:Gould:parallel:CFUtoTTD:*0***}
\begin{tabular*}{\linewidth}{@{\extracolsep{\fill}}lllllllll@{}}
        \toprule  \multicolumn{1}{l}{\scriptsize No.} &  \multicolumn{1}{l}{\scriptsize bipyramid} & \multicolumn{1}{l}{\scriptsize type} &  \multicolumn{1}{l}{\scriptsize $e_o$} & \multicolumn{1}{l}{\scriptsize $e_p$} & \multicolumn{1}{l}{\scriptsize $e_o/e_p$} & \multicolumn{1}{l}{\scriptsize $p_o$} & \multicolumn{1}{l}{\scriptsize $p_p$} & \multicolumn{1}{l}{\scriptsize $p_o/p_p$} \\ 
 \midrule  
\multicolumn{1}{r}{\scriptsize \textbf{21}}& \multicolumn{1}{l}{\scriptsize \{10001\}+\{10000,00001,10101,10011\}+\{00111\}} &\multicolumn{1}{l}{\scriptsize red/blue}& \multicolumn{1}{l}{\scriptsize 0.012} & \multicolumn{1}{l}{\scriptsize 0.024} & \multicolumn{1}{l}{\scriptsize 0.481} & \multicolumn{1}{l}{\scriptsize 0.963} & \multicolumn{1}{l}{\scriptsize 0.026} & \multicolumn{1}{l}{\scriptsize 36.756} \\ 
\multicolumn{1}{r}{\scriptsize \textbf{20}}& \multicolumn{1}{l}{\scriptsize \{00010\}+\{10000,10010,00011,00111\}+\{10011\}} &\multicolumn{1}{l}{\scriptsize red/red}& \multicolumn{1}{l}{\scriptsize 0.021} & \multicolumn{1}{l}{\scriptsize 0.012} & \multicolumn{1}{l}{\scriptsize 1.714} & \multicolumn{1}{l}{\scriptsize 0.797} & \multicolumn{1}{l}{\scriptsize 0.270} & \multicolumn{1}{l}{\scriptsize 2.952} \\ 
\multicolumn{1}{r}{\scriptsize \textbf{19}}& \multicolumn{1}{l}{\scriptsize \{10100\}+\{10000,00100,10110,10101\}+\{00110\}} &\multicolumn{1}{l}{\scriptsize red/red}& \multicolumn{1}{l}{\scriptsize 0.022} & \multicolumn{1}{l}{\scriptsize 0.007} & \multicolumn{1}{l}{\scriptsize 3.155} & \multicolumn{1}{l}{\scriptsize 0.869} & \multicolumn{1}{l}{\scriptsize 0.717} & \multicolumn{1}{l}{\scriptsize 1.212} \\
\bottomrule \end{tabular*}   \end{table}
                  
\newpage

\begin{table}[b]
                          \caption{\small Parallel analysis GouldCFU ${\ast}$1${\ast}$${\ast}$${\ast}$ $\rightarrow$ Gould ${\ast}$1${\ast}$${\ast}$${\ast}$, non-critical red/red-case omitted.}
\label{tab:Gould:parallel:CFUtoTTD:*1***}
\begin{tabular*}{\linewidth}{@{\extracolsep{\fill}}lllllllll@{}}
        \toprule  \multicolumn{1}{l}{\scriptsize No.} &  \multicolumn{1}{l}{\scriptsize bipyramid} & \multicolumn{1}{l}{\scriptsize type} &  \multicolumn{1}{l}{\scriptsize $e_o$} & \multicolumn{1}{l}{\scriptsize $e_p$} & \multicolumn{1}{l}{\scriptsize $e_o/e_p$} & \multicolumn{1}{l}{\scriptsize $p_o$} & \multicolumn{1}{l}{\scriptsize $p_p$} & \multicolumn{1}{l}{\scriptsize $p_o/p_p$} \\ 
 \midrule  
\multicolumn{1}{r}{\scriptsize \textbf{23}}& \multicolumn{1}{l}{\scriptsize \{01100\}+\{11000,01110,01101,11110\}+\{11010\}} &\multicolumn{1}{l}{\scriptsize red/red}& \multicolumn{1}{l}{\scriptsize 0.001} & \multicolumn{1}{l}{\scriptsize 0.018} & \multicolumn{1}{l}{\scriptsize 0.054} & \multicolumn{1}{l}{\scriptsize 0.998} & \multicolumn{1}{l}{\scriptsize 0.292} & \multicolumn{1}{l}{\scriptsize 3.418} \\ 
\multicolumn{1}{r}{\scriptsize \textbf{22}}& \multicolumn{1}{l}{\scriptsize \{01100\}+\{11000,01010,01001,01110\}+\{11010\}} &\multicolumn{1}{l}{\scriptsize red/red}& \multicolumn{1}{l}{\scriptsize 0.001} & \multicolumn{1}{l}{\scriptsize 0.018} & \multicolumn{1}{l}{\scriptsize 0.054} & \multicolumn{1}{l}{\scriptsize 0.998} & \multicolumn{1}{l}{\scriptsize 0.292} & \multicolumn{1}{l}{\scriptsize 3.418} \\ 
\multicolumn{1}{r}{\scriptsize \textbf{21}}& \multicolumn{1}{l}{\scriptsize \{01100\}+\{11000,01001,01110,01101\}+\{11010\}} &\multicolumn{1}{l}{\scriptsize red/red}& \multicolumn{1}{l}{\scriptsize 0.001} & \multicolumn{1}{l}{\scriptsize 0.018} & \multicolumn{1}{l}{\scriptsize 0.054} & \multicolumn{1}{l}{\scriptsize 0.998} & \multicolumn{1}{l}{\scriptsize 0.292} & \multicolumn{1}{l}{\scriptsize 3.418} \\
\bottomrule \end{tabular*}   \end{table}
                  
\newpage

\FloatBarrier
% TabS6
\begin{table}[ht]
\caption[Significant $4$-dimensional interactions, which cannot be seen in lower dimensions.]{Significant $4$-dimensional interactions, which cannot be seen in lower dimensions, cf.\ (Fig.~\ref{fig4}). The value $p\uparrow$ refers to the $p$-value of the $4$-dimensional bipyramid in question whereas $p\downarrow$ is the $p$-value of its ridge intersected with the $\cap$\,-\,face, cf. (Fig.~\ref{fig4}c) for the Gould data.}
\label{Tab:special+bipyramids}                          
\begin{tabular*}{0.95\linewidth}{@{\extracolsep{\fill}}lllll@{}}
        \toprule   \multicolumn{1}{l}{\small Data} &  \multicolumn{1}{c}{\small significant  bipyramid} & \multicolumn{1}{c}{\small  $\cap$\,-\,face} &\multicolumn{1}{c}{\small  $p\uparrow$} &   \multicolumn{1}{c}{\small  $p\downarrow$} \\ 
 \midrule  
\multicolumn{1}{l}{\small Eble} & \multicolumn{1}{c}{-} & \multicolumn{1}{c}{-}& \multicolumn{1}{c}{-}& \multicolumn{1}{c}{-} \\ \midrule
\multicolumn{1}{l}{\small Gould} \\
\multicolumn{1}{l}{\small ${\ast}{\ast}0{\ast}{\ast}$} & \multicolumn{1}{l}{\scriptsize $\{00010\}+\{00000,10010,00011,11011\}+\{10001\}$} & \multicolumn{1}{c}{\small ${\ast}{\ast}01{\ast}$} & \multicolumn{1}{l}{\small $0.041$} & \multicolumn{1}{r}{\small $0.270$}\\
 &  & \multicolumn{1}{c}{\small ${\ast}00{\ast}{\ast}$} & \multicolumn{1}{l}{\small $0.041$} & \multicolumn{1}{r}{\small $0.149$}\\
  & \multicolumn{1}{l}{\scriptsize $\{10010\}+\{00000,11000,10001,11011\}+\{01001\}$} & \multicolumn{1}{c}{\small $1{\ast}0{\ast}{\ast}$} & \multicolumn{1}{l}{\small $0.041$} & \multicolumn{1}{r}{\small $0.076$}\\
   &  & \multicolumn{1}{c}{\small ${\ast}{\ast}00{\ast}$} & \multicolumn{1}{l}{\small $0.041$} & \multicolumn{1}{r}{\small $0.063$}\\
  \midrule
\multicolumn{1}{l}{\small Khan} \\
\multicolumn{1}{l}{\small $0{\ast}{\ast}{\ast}{\ast}$} & \multicolumn{1}{l}{\scriptsize $\{00010\}+\{00000,01001,00101,00011\}+\{00001\}$} & \multicolumn{1}{c}{\small $0{\ast}{\ast}{\ast}1$} & \multicolumn{1}{l}{\small $0.009$} &  \multicolumn{1}{r}{\small $0.052$} \\
\bottomrule 
\end{tabular*}   
\end{table}

\newpage
\FloatBarrier

%Tab S7
\begin{table}[ht]
\caption[Bacterial species considered in the two microbiome data sets.]{Bacterial species considered in the two microbiome data sets.}\label{tab:datasets}
\begin{tabular*}{\linewidth}{@{\extracolsep{\fill}} r  r  r @{}}
  \toprule 
             &Gould data set & Eble data set \\
\midrule
Species 1 & \emph{L.~plantarum}  & \emph{L.~plantarum} \\ 
Species 2 & \emph{L.~brevis}  & \emph{L.~brevis} \\ 
Species 3 & \emph{A.~pasteurianus} & \emph{A.~cerevisiae} \\ 
Species 4 & \emph{A.~tropicalis}  & \emph{A.~malorum} \\
Species 5 & \emph{A.~orientalis} &  \emph{A.~orientalis} \\
\bottomrule
\end{tabular*}
\end{table}
%%% single tables

%
%\input{Figures/paralleltransports/GouldTTDvsCFU/analyze_Lud2017FlygutCFUsData_log10Normalized0starstarstarstarTOLud2017SurvDataNormalized0starstarstarstar_180tables.tex}
%\newpage
%
%\input{Figures/paralleltransports/GouldTTDvsCFU/analyze_Lud2017FlygutCFUsData_log10Normalized1starstarstarstarTOLud2017SurvDataNormalized1starstarstarstar_180tables.tex}
%\newpage
%
%\input{Figures/paralleltransports/GouldTTDvsCFU/analyze_Lud2017FlygutCFUsData_log10Normalizedstar0starstarstarTOLud2017SurvDataNormalizedstar0starstarstar_180tables.tex}
%\newpage
%
%\input{Figures/paralleltransports/GouldTTDvsCFU/analyze_Lud2017FlygutCFUsData_log10Normalizedstar1starstarstarTOLud2017SurvDataNormalizedstar1starstarstar_180tables.tex}
%

%\input{Figures/paralleltransports/singles/analyze_khanrawNormalizedstarstarstar0starTOkhanrawNormalizedstarstarstar1star_180tables.tex}
%\newpage
%\input{Figures/paralleltransports/singles/analyze_khanrawNormalizedstarstarstar1starTOkhanrawNormalizedstarstarstar0star_180tables.tex}
%\newpage

% Tab S8
%%Eble 0**** to 1****.
\begin{table}[b]
                          \caption{\small Parallel analysis Eble 0${\ast}$${\ast}$${\ast}$${\ast}$ $\rightarrow$ 1${\ast}$${\ast}$${\ast}$${\ast}$, non-critical red/red-case omitted.}
\label{tab:Eble:parallel:0to1****}
\begin{tabular*}{\linewidth}{@{\extracolsep{\fill}}lllllllll@{}}
        \toprule  \multicolumn{1}{l}{\scriptsize No.} &  \multicolumn{1}{l}{\scriptsize $\text{bipyramid}_s$} & \multicolumn{1}{l}{\scriptsize type} &  \multicolumn{1}{l}{\scriptsize $e_o$} & \multicolumn{1}{l}{\scriptsize $e_p$} & \multicolumn{1}{l}{\scriptsize $e_o/e_p$} & \multicolumn{1}{l}{\scriptsize $p_o$} & \multicolumn{1}{l}{\scriptsize $p_p$} & \multicolumn{1}{l}{\scriptsize $p_o/p_p$} \\ 
 \midrule  
\multicolumn{1}{r}{\scriptsize \textbf{23}}& \multicolumn{1}{l}{\scriptsize \{00001\}+\{00000,01001,01011,00111\}+\{01111\}} &\multicolumn{1}{l}{\scriptsize red/red}& \multicolumn{1}{l}{\scriptsize 0.001} & \multicolumn{1}{l}{\scriptsize 0.012} & \multicolumn{1}{l}{\scriptsize 0.066} & \multicolumn{1}{l}{\scriptsize 0.953} & \multicolumn{1}{l}{\scriptsize 0.390} & \multicolumn{1}{l}{\scriptsize 2.444} \\ 
\multicolumn{1}{r}{\scriptsize \textbf{22}}& \multicolumn{1}{l}{\scriptsize \{00001\}+\{00000,01001,01101,00111\}+\{01111\}} &\multicolumn{1}{l}{\scriptsize red/red}& \multicolumn{1}{l}{\scriptsize 0.001} & \multicolumn{1}{l}{\scriptsize 0.012} & \multicolumn{1}{l}{\scriptsize 0.066} & \multicolumn{1}{l}{\scriptsize 0.953} & \multicolumn{1}{l}{\scriptsize 0.390} & \multicolumn{1}{l}{\scriptsize 2.444} \\ 
\multicolumn{1}{r}{\scriptsize \textbf{21}}& \multicolumn{1}{l}{\scriptsize \{01110\}+\{00000,00110,01011,01111\}+\{00111\}} &\multicolumn{1}{l}{\scriptsize red/blue}& \multicolumn{1}{l}{\scriptsize 0.001} & \multicolumn{1}{l}{\scriptsize 0.025} & \multicolumn{1}{l}{\scriptsize 0.041} & \multicolumn{1}{l}{\scriptsize 0.923} & \multicolumn{1}{l}{\scriptsize 0.038} & \multicolumn{1}{l}{\scriptsize 24.226} \\ 
\multicolumn{1}{r}{\scriptsize \textbf{20}}& \multicolumn{1}{l}{\scriptsize \{01110\}+\{00000,01100,00110,01111\}+\{00111\}} &\multicolumn{1}{l}{\scriptsize red/blue}& \multicolumn{1}{l}{\scriptsize 0.001} & \multicolumn{1}{l}{\scriptsize 0.035} & \multicolumn{1}{l}{\scriptsize 0.041} & \multicolumn{1}{l}{\scriptsize 0.923} & \multicolumn{1}{l}{\scriptsize 0.038} & \multicolumn{1}{l}{\scriptsize 24.226} \\ 
\multicolumn{1}{r}{\scriptsize \textbf{19}}& \multicolumn{1}{l}{\scriptsize \{00110\}+\{00000,01100,00111,01111\}+\{01101\}} &\multicolumn{1}{l}{\scriptsize red/red}& \multicolumn{1}{l}{\scriptsize 0.002} & \multicolumn{1}{l}{\scriptsize 0.012} & \multicolumn{1}{l}{\scriptsize 0.201} & \multicolumn{1}{l}{\scriptsize 0.827} & \multicolumn{1}{l}{\scriptsize 0.303} & \multicolumn{1}{l}{\scriptsize 2.729} \\ 
\multicolumn{1}{r}{\scriptsize \textbf{18}}& \multicolumn{1}{l}{\scriptsize \{00110\}+\{00000,01100,00101,00111\}+\{01101\}} &\multicolumn{1}{l}{\scriptsize red/red}& \multicolumn{1}{l}{\scriptsize 0.003} & \multicolumn{1}{l}{\scriptsize 0.014} & \multicolumn{1}{l}{\scriptsize 0.201} & \multicolumn{1}{l}{\scriptsize 0.827} & \multicolumn{1}{l}{\scriptsize 0.303} & \multicolumn{1}{l}{\scriptsize 2.729} \\ 
\multicolumn{1}{r}{\scriptsize \textbf{17}}& \multicolumn{1}{l}{\scriptsize \{01110\}+\{00000,01000,01100,01111\}+\{01101\}} &\multicolumn{1}{l}{\scriptsize red/red}& \multicolumn{1}{l}{\scriptsize 0.003} & \multicolumn{1}{l}{\scriptsize 0.013} & \multicolumn{1}{l}{\scriptsize 0.264} & \multicolumn{1}{l}{\scriptsize 0.742} & \multicolumn{1}{l}{\scriptsize 0.251} & \multicolumn{1}{l}{\scriptsize 2.956} \\ 
\multicolumn{1}{r}{\scriptsize \textbf{16}}& \multicolumn{1}{l}{\scriptsize \{00110\}+\{00000,00010,01011,00111\}+\{00011\}} &\multicolumn{1}{l}{\scriptsize red/red}& \multicolumn{1}{l}{\scriptsize 0.004} & \multicolumn{1}{l}{\scriptsize 0.003} & \multicolumn{1}{l}{\scriptsize 1.568} & \multicolumn{1}{l}{\scriptsize 0.755} & \multicolumn{1}{l}{\scriptsize 0.843} & \multicolumn{1}{l}{\scriptsize 0.896} \\ 
\multicolumn{1}{r}{\scriptsize \textbf{15}}& \multicolumn{1}{l}{\scriptsize \{00010\}+\{00000,01010,00110,01011\}+\{01110\}} &\multicolumn{1}{l}{\scriptsize red/red}& \multicolumn{1}{l}{\scriptsize 0.007} & \multicolumn{1}{l}{\scriptsize 0.010} & \multicolumn{1}{l}{\scriptsize 0.748} & \multicolumn{1}{l}{\scriptsize 0.606} & \multicolumn{1}{l}{\scriptsize 0.488} & \multicolumn{1}{l}{\scriptsize 1.242} \\ 
\multicolumn{1}{r}{\scriptsize \textbf{14}}& \multicolumn{1}{l}{\scriptsize \{01010\}+\{00000,00010,00110,01011\}+\{00111\}} &\multicolumn{1}{l}{\scriptsize red/red}& \multicolumn{1}{l}{\scriptsize 0.008} & \multicolumn{1}{l}{\scriptsize 0.005} & \multicolumn{1}{l}{\scriptsize 1.583} & \multicolumn{1}{l}{\scriptsize 0.443} & \multicolumn{1}{l}{\scriptsize 0.639} & \multicolumn{1}{l}{\scriptsize 0.693} \\ 
\multicolumn{1}{r}{\scriptsize \textbf{13}}& \multicolumn{1}{l}{\scriptsize \{01010\}+\{00000,00110,01110,01011\}+\{01111\}} &\multicolumn{1}{l}{\scriptsize red/red}& \multicolumn{1}{l}{\scriptsize 0.009} & \multicolumn{1}{l}{\scriptsize 0.024} & \multicolumn{1}{l}{\scriptsize 0.359} & \multicolumn{1}{l}{\scriptsize 0.475} & \multicolumn{1}{l}{\scriptsize 0.062} & \multicolumn{1}{l}{\scriptsize 7.686} \\ 
\multicolumn{1}{r}{\scriptsize \textbf{12}}& \multicolumn{1}{l}{\scriptsize \{01010\}+\{00000,01000,01110,01011\}+\{01111\}} &\multicolumn{1}{l}{\scriptsize red/red}& \multicolumn{1}{l}{\scriptsize 0.009} & \multicolumn{1}{l}{\scriptsize 0.024} & \multicolumn{1}{l}{\scriptsize 0.359} & \multicolumn{1}{l}{\scriptsize 0.475} & \multicolumn{1}{l}{\scriptsize 0.062} & \multicolumn{1}{l}{\scriptsize 7.686} \\ 
\multicolumn{1}{r}{\scriptsize \textbf{11}}& \multicolumn{1}{l}{\scriptsize \{00100\}+\{00000,01100,00110,00101\}+\{00111\}} &\multicolumn{1}{l}{\scriptsize red/red}& \multicolumn{1}{l}{\scriptsize 0.009} & \multicolumn{1}{l}{\scriptsize 0.018} & \multicolumn{1}{l}{\scriptsize 0.498} & \multicolumn{1}{l}{\scriptsize 0.533} & \multicolumn{1}{l}{\scriptsize 0.269} & \multicolumn{1}{l}{\scriptsize 1.981} \\ 
\multicolumn{1}{r}{\scriptsize \textbf{10}}& \multicolumn{1}{l}{\scriptsize \{01001\}+\{00000,00001,01101,00111\}+\{00101\}} &\multicolumn{1}{l}{\scriptsize red/red}& \multicolumn{1}{l}{\scriptsize 0.014} & \multicolumn{1}{l}{\scriptsize 0.014} & \multicolumn{1}{l}{\scriptsize 1.018} & \multicolumn{1}{l}{\scriptsize 0.288} & \multicolumn{1}{l}{\scriptsize 0.313} & \multicolumn{1}{l}{\scriptsize 0.920} \\ 
\multicolumn{1}{r}{\scriptsize \textbf{9}}& \multicolumn{1}{l}{\scriptsize \{00101\}+\{00000,01100,01101,00111\}+\{01111\}} &\multicolumn{1}{l}{\scriptsize red/red}& \multicolumn{1}{l}{\scriptsize 0.015} & \multicolumn{1}{l}{\scriptsize 0.026} & \multicolumn{1}{l}{\scriptsize 0.584} & \multicolumn{1}{l}{\scriptsize 0.228} & \multicolumn{1}{l}{\scriptsize 0.062} & \multicolumn{1}{l}{\scriptsize 3.695} \\ 
& \multicolumn{1}{l}{\scriptsize \{00101\}+\{00000,01100,00110,00111\}+\{01111\}} &\multicolumn{1}{l}{\scriptsize red/blue}& \multicolumn{1}{l}{\scriptsize 0.018} & \multicolumn{1}{l}{\scriptsize 0.040} & \multicolumn{1}{l}{\scriptsize 0.446} & \multicolumn{1}{l}{\scriptsize 0.321} & \multicolumn{1}{l}{\scriptsize 0.035} & \multicolumn{1}{l}{\scriptsize 9.119} \\ 
\multicolumn{1}{r}{\scriptsize \textbf{8}}& \multicolumn{1}{l}{\scriptsize \{01101\}+\{00000,01001,00111,01111\}+\{01011\}} &\multicolumn{1}{l}{\scriptsize red/red}& \multicolumn{1}{l}{\scriptsize 0.019} & \multicolumn{1}{l}{\scriptsize 0.003} & \multicolumn{1}{l}{\scriptsize 6.623} & \multicolumn{1}{l}{\scriptsize 0.068} & \multicolumn{1}{l}{\scriptsize 0.800} & \multicolumn{1}{l}{\scriptsize 0.085} \\ 
\multicolumn{1}{r}{\scriptsize \textbf{7}}& \multicolumn{1}{l}{\scriptsize \{01101\}+\{00000,01000,01001,01111\}+\{01011\}} &\multicolumn{1}{l}{\scriptsize red/red}& \multicolumn{1}{l}{\scriptsize 0.019} & \multicolumn{1}{l}{\scriptsize 0.003} & \multicolumn{1}{l}{\scriptsize 6.623} & \multicolumn{1}{l}{\scriptsize 0.068} & \multicolumn{1}{l}{\scriptsize 0.800} & \multicolumn{1}{l}{\scriptsize 0.085} \\ 
\multicolumn{1}{r}{\scriptsize \textbf{6}}& \multicolumn{1}{l}{\scriptsize \{01001\}+\{00000,00001,01011,00111\}+\{00011\}} &\multicolumn{1}{l}{\scriptsize red/red}& \multicolumn{1}{l}{\scriptsize 0.019} & \multicolumn{1}{l}{\scriptsize 0.005} & \multicolumn{1}{l}{\scriptsize 3.571} & \multicolumn{1}{l}{\scriptsize 0.153} & \multicolumn{1}{l}{\scriptsize 0.689} & \multicolumn{1}{l}{\scriptsize 0.222} \\ 
\multicolumn{1}{r}{\scriptsize \textbf{5}}& \multicolumn{1}{l}{\scriptsize \{01000\}+\{00000,01010,01110,01011\}+\{00110\}} &\multicolumn{1}{l}{\scriptsize red/red}& \multicolumn{1}{l}{\scriptsize 0.020} & \multicolumn{1}{l}{\scriptsize 0.011} & \multicolumn{1}{l}{\scriptsize 1.750} & \multicolumn{1}{l}{\scriptsize 0.169} & \multicolumn{1}{l}{\scriptsize 0.443} & \multicolumn{1}{l}{\scriptsize 0.381} \\ 
\multicolumn{1}{r}{\scriptsize \textbf{4}}& \multicolumn{1}{l}{\scriptsize \{01000\}+\{00000,01100,01110,01111\}+\{00110\}} &\multicolumn{1}{l}{\scriptsize red/red}& \multicolumn{1}{l}{\scriptsize 0.020} & \multicolumn{1}{l}{\scriptsize 0.011} & \multicolumn{1}{l}{\scriptsize 1.750} & \multicolumn{1}{l}{\scriptsize 0.169} & \multicolumn{1}{l}{\scriptsize 0.443} & \multicolumn{1}{l}{\scriptsize 0.381} \\ 
\multicolumn{1}{r}{\scriptsize \textbf{3}}& \multicolumn{1}{l}{\scriptsize \{01000\}+\{00000,01001,01011,01111\}+\{00111\}} &\multicolumn{1}{l}{\scriptsize red/red}& \multicolumn{1}{l}{\scriptsize 0.021} & \multicolumn{1}{l}{\scriptsize 0.013} & \multicolumn{1}{l}{\scriptsize 1.535} & \multicolumn{1}{l}{\scriptsize 0.140} & \multicolumn{1}{l}{\scriptsize 0.339} & \multicolumn{1}{l}{\scriptsize 0.413} \\ 
\multicolumn{1}{r}{\scriptsize \textbf{2}}& \multicolumn{1}{l}{\scriptsize \{01100\}+\{00000,01000,01101,01111\}+\{01001\}} &\multicolumn{1}{l}{\scriptsize blue/blue}& \multicolumn{1}{l}{\scriptsize 0.045} & \multicolumn{1}{l}{\scriptsize 0.037} & \multicolumn{1}{l}{\scriptsize 1.215} & \multicolumn{1}{l}{\scriptsize 0.000} & \multicolumn{1}{l}{\scriptsize 0.003} & \multicolumn{1}{l}{\scriptsize 0.176} \\ 
\multicolumn{1}{r}{\scriptsize \textbf{1}}& \multicolumn{1}{l}{\scriptsize \{01001\}+\{00000,01000,01011,01111\}+\{01110\}} &\multicolumn{1}{l}{\scriptsize blue/red}& \multicolumn{1}{l}{\scriptsize 0.048} & \multicolumn{1}{l}{\scriptsize 0.024} & \multicolumn{1}{l}{\scriptsize 1.993} & \multicolumn{1}{l}{\scriptsize 0.000} & \multicolumn{1}{l}{\scriptsize 0.056} & \multicolumn{1}{l}{\scriptsize 0.002} \\ 
& \multicolumn{1}{l}{\scriptsize \{01100\}+\{00000,01000,01110,01111\}+\{01011\}} &\multicolumn{1}{l}{\scriptsize blue/blue}& \multicolumn{1}{l}{\scriptsize 0.064} & \multicolumn{1}{l}{\scriptsize 0.034} & \multicolumn{1}{l}{\scriptsize 1.855} & \multicolumn{1}{l}{\scriptsize 0.000} & \multicolumn{1}{l}{\scriptsize 0.005} & \multicolumn{1}{l}{\scriptsize 0.000} \\ 
& \multicolumn{1}{l}{\scriptsize \{00010\}+\{00000,00011,01011,00111\}+\{00001\}} &\multicolumn{1}{l}{\scriptsize blue/blue}& \multicolumn{1}{l}{\scriptsize 0.065} & \multicolumn{1}{l}{\scriptsize 0.043} & \multicolumn{1}{l}{\scriptsize 1.518} & \multicolumn{1}{l}{\scriptsize 0.000} & \multicolumn{1}{l}{\scriptsize 0.001} & \multicolumn{1}{l}{\scriptsize 0.001} \\ 
& \multicolumn{1}{l}{\scriptsize \{01100\}+\{00000,01101,00111,01111\}+\{01001\}} &\multicolumn{1}{l}{\scriptsize blue/red}& \multicolumn{1}{l}{\scriptsize 0.066} & \multicolumn{1}{l}{\scriptsize 0.024} & \multicolumn{1}{l}{\scriptsize 2.775} & \multicolumn{1}{l}{\scriptsize 0.000} & \multicolumn{1}{l}{\scriptsize 0.105} & \multicolumn{1}{l}{\scriptsize 0.000} \\ 
& \multicolumn{1}{l}{\scriptsize \{00001\}+\{00000,00101,01101,00111\}+\{01100\}} &\multicolumn{1}{l}{\scriptsize blue/blue}& \multicolumn{1}{l}{\scriptsize 0.066} & \multicolumn{1}{l}{\scriptsize 0.033} & \multicolumn{1}{l}{\scriptsize 1.989} & \multicolumn{1}{l}{\scriptsize 0.000} & \multicolumn{1}{l}{\scriptsize 0.009} & \multicolumn{1}{l}{\scriptsize 0.000} \\ 
& \multicolumn{1}{l}{\scriptsize \{01001\}+\{00000,01011,00111,01111\}+\{00110\}} &\multicolumn{1}{l}{\scriptsize blue/blue}& \multicolumn{1}{l}{\scriptsize 0.068} & \multicolumn{1}{l}{\scriptsize 0.036} & \multicolumn{1}{l}{\scriptsize 1.917} & \multicolumn{1}{l}{\scriptsize 0.000} & \multicolumn{1}{l}{\scriptsize 0.007} & \multicolumn{1}{l}{\scriptsize 0.000} \\ 
& \multicolumn{1}{l}{\scriptsize \{01100\}+\{00000,00110,01110,01111\}+\{01011\}} &\multicolumn{1}{l}{\scriptsize blue/blue}& \multicolumn{1}{l}{\scriptsize 0.083} & \multicolumn{1}{l}{\scriptsize 0.045} & \multicolumn{1}{l}{\scriptsize 1.829} & \multicolumn{1}{l}{\scriptsize 0.000} & \multicolumn{1}{l}{\scriptsize 0.002} & \multicolumn{1}{l}{\scriptsize 0.000} \\ 
& \multicolumn{1}{l}{\scriptsize \{01100\}+\{00000,00110,00111,01111\}+\{01011\}} &\multicolumn{1}{l}{\scriptsize blue/red}& \multicolumn{1}{l}{\scriptsize 0.084} & \multicolumn{1}{l}{\scriptsize 0.021} & \multicolumn{1}{l}{\scriptsize 4.035} & \multicolumn{1}{l}{\scriptsize 0.000} & \multicolumn{1}{l}{\scriptsize 0.210} & \multicolumn{1}{l}{\scriptsize 0.000} \\ 
\bottomrule \end{tabular*}   \end{table}
                  
%\newpage
%\input{Figures/paralleltransports/singles/analyze_Lud2018SurvDataNormalized1starstarstarstarTOLud2018SurvDataNormalized0starstarstarstar_180tables.tex}
%\newpage
%\input{Figures/paralleltransports/singles/analyze_Lud2018SurvDataNormalizedstar0starstarstarTOLud2018SurvDataNormalizedstar1starstarstar_180tables.tex}
%\newpage
%\input{Figures/paralleltransports/singles/analyze_Lud2018SurvDataNormalizedstar1starstarstarTOLud2018SurvDataNormalizedstar0starstarstar_180tables.tex}
%

% Tab S9

\begin{table}[b]
\caption[Regressions for Eble ${0}{\ast}{\ast}{\ast}{\ast}$ and Eble ${1}{\ast}{\ast}{\ast}{\ast}$.]{Regressions over {\{0001\}+\{0000,1001,1011,0111\}+\{1111\}} for normalized lifespan data
for Eble ${0}{\ast}{\ast}{\ast}{\ast}$ and Eble ${1}{\ast}{\ast}{\ast}{\ast}$.}\label{reg:bipy1}                
\begin{tabular*}{\linewidth}{@{\extracolsep{\fill}} r  r  r r  r @{}}
  \toprule
         	&Coefficient	&Std. error&   $t$-statistic 	&	$p$-value 	\\
\midrule
$\beta_0$ & $0$ & $0$ &	nan & nan \\
$x_1$ & $-0.0270$ &	$0.009$ & $-2.987$ & $0.003$\\	
$x_2$ & $-0.0149$ &	$0.012$ & $-1.246$ & $0.213$\\	
$x_3$ & $-0.0156$ &	$0.012$ & $-1.306$ & $0.192$\\	
$x_4$ & $0.2039$ & $0.008$ & $26.022$	& $0.000$\\
 \midrule
$\beta_0$ & $0.2320$ & $0.005$ & $44.642$ & $0.000$	\\
$x_1$ & $0.0310$ & $0.005$ & $5.957$ & $0.000$	\\
$x_2$ & $0.0610$ & $0.007$ & $8.874$ & $0.000$	\\
$x_3$ & $-0.0185$ & $0.007$ & $-2.692$ & $0.007$	\\
$x_4$ & $-0.0861$ & $0.007$ & $-12.518$ & $0.000$	\\
\bottomrule

\end{tabular*}
\end{table}

\end{document}

\begin{figure}[H]
            \newcommand\maxepi{1}
            \newcommand\maxclusters{98}
            \newcommand\clusterlength{3pt}
            \newcommand\clusterspace{0.1}
            \newcommand\lw{3pt}

\begin{tikzpicture}[yscale=0.5,xscale=1.7, scale = 0.36]
        \tikzset{
        cluster/.style = {red, line width=\lw, join=round},
        significant/.style = {blue, line width=\lw, join=round},
        semisignificant/.style = {purple, line width=\lw, join=round},
        tick/.style = {black, thick},
        tick_thin/.style = {black, thin}}

\coordinate (offset) at (0, 101);
\foreach \nclusters\ngaps/\size in {0/0/1, 1/1/1, 2/2/1, 3/3/1, 4/4/1, 5/5/1, 6/6/1, 7/7/1, 8/8/1, 9/9/1, 10/10/1, 11/11/1, 12/12/1, 13/13/1, 14/14/1, 15/15/1, 16/16/1, 17/17/1, 18/18/1, 19/19/1, 20/20/1, 21/21/1, 22/22/1, 23/23/1, 24/24/1, 25/25/1, 26/26/1, 27/27/1, 28/28/1, 29/29/1, 30/30/1, 31/31/1, 32/32/1, 33/33/1, 34/34/1, 35/35/1, 36/36/1, 37/37/1, 38/38/1, 39/39/1, 40/40/1, 41/41/1, 42/42/1, 43/43/1, 44/44/1, 45/45/1, 46/46/1, 47/47/1, 48/48/1, 49/49/1, 50/50/1, 51/51/1, 52/52/1, 53/53/1, 54/54/1, 55/55/1, 56/56/1, 57/57/1, 58/58/1, 59/59/1, 60/60/1, 61/61/1, 62/62/1, 63/63/1, 64/64/1, 65/65/1, 66/66/1, 67/67/1, 68/68/1, 69/69/1, 70/70/1, 71/71/1, 72/72/1, 73/73/1, 74/74/1, 75/75/1, 76/76/1, 77/77/1, 78/78/1, 79/79/1, 80/80/1, 81/81/1, 82/82/1, 83/83/1, 84/84/1, 85/85/1, 86/86/1, 87/87/1, 88/88/1, 89/89/1, 90/90/1, 91/91/1, 92/92/1, 93/93/1, 94/94/1, 95/95/1, 96/96/1, 97/97/1, 98/98/1, 99/99/1, 100/100/1}{

        \draw[cluster] ($ (offset) + (\nclusters*\clusterlength,0) + (\ngaps*\clusterspace,0) $) 
        -- ($ (offset) + (\nclusters*\clusterlength,0) + (\ngaps*\clusterspace,0) + (\size*\clusterlength,0) $);}
 \draw[tick] ($ (offset) + (0,\lw) $) -- ($ (offset) - (0,2*\lw) $);\coordinate (offset) at (0.00101641935781323, 1);
\foreach \nclusters\ngaps/\size in {0/0/1}{
 
        \draw[cluster] ($ (offset) + (\nclusters*\clusterlength,0) + (\ngaps*\clusterspace,0) $) 
        -- ($ (offset) + (\nclusters*\clusterlength,0) + (\ngaps*\clusterspace,0) + (\size*\clusterlength,0) $);}
 \draw[tick] ($ (offset) + (0,\lw) $) -- ($ (offset) - (0,2*\lw) $);\coordinate (offset) at (0.00101641935781323, 1);
\foreach \nclusters\ngaps/\size in {0/0/1}{
 
        \draw[cluster] ($ (offset) + (\nclusters*\clusterlength,0) + (\ngaps*\clusterspace,0) $) 
        -- ($ (offset) + (\nclusters*\clusterlength,0) + (\ngaps*\clusterspace,0) + (\size*\clusterlength,0) $);}
 \draw[tick] ($ (offset) + (0,\lw) $) -- ($ (offset) - (0,2*\lw) $);\coordinate (offset) at (0.00108659799949197, 1);
\foreach \nclusters\ngaps/\size in {0/0/1}{
 
        \draw[cluster] ($ (offset) + (\nclusters*\clusterlength,0) + (\ngaps*\clusterspace,0) $) 
        -- ($ (offset) + (\nclusters*\clusterlength,0) + (\ngaps*\clusterspace,0) + (\size*\clusterlength,0) $);}
 \draw[tick] ($ (offset) + (0,\lw) $) -- ($ (offset) - (0,2*\lw) $);\coordinate (offset) at (0.00108659799949197, 1);
\foreach \nclusters\ngaps/\size in {0/0/1}{
 
        \draw[cluster] ($ (offset) + (\nclusters*\clusterlength,0) + (\ngaps*\clusterspace,0) $) 
        -- ($ (offset) + (\nclusters*\clusterlength,0) + (\ngaps*\clusterspace,0) + (\size*\clusterlength,0) $);}
 \draw[tick] ($ (offset) + (0,\lw) $) -- ($ (offset) - (0,2*\lw) $);\coordinate (offset) at (0.00108659799949197, 1);
\foreach \nclusters\ngaps/\size in {0/0/1}{
 
        \draw[cluster] ($ (offset) + (\nclusters*\clusterlength,0) + (\ngaps*\clusterspace,0) $) 
        -- ($ (offset) + (\nclusters*\clusterlength,0) + (\ngaps*\clusterspace,0) + (\size*\clusterlength,0) $);}
 \draw[tick] ($ (offset) + (0,\lw) $) -- ($ (offset) - (0,2*\lw) $);\coordinate (offset) at (0.00108659799949197, 1);
\foreach \nclusters\ngaps/\size in {0/0/1}{
 
        \draw[cluster] ($ (offset) + (\nclusters*\clusterlength,0) + (\ngaps*\clusterspace,0) $) 
        -- ($ (offset) + (\nclusters*\clusterlength,0) + (\ngaps*\clusterspace,0) + (\size*\clusterlength,0) $);}
 \draw[tick] ($ (offset) + (0,\lw) $) -- ($ (offset) - (0,2*\lw) $);\coordinate (offset) at (0.00108659799949197, 1);
\foreach \nclusters\ngaps/\size in {0/0/1}{
 
        \draw[cluster] ($ (offset) + (\nclusters*\clusterlength,0) + (\ngaps*\clusterspace,0) $) 
        -- ($ (offset) + (\nclusters*\clusterlength,0) + (\ngaps*\clusterspace,0) + (\size*\clusterlength,0) $);}
 \draw[tick] ($ (offset) + (0,\lw) $) -- ($ (offset) - (0,2*\lw) $);\coordinate (offset) at (0.00108659799949197, 1);
\foreach \nclusters\ngaps/\size in {0/0/1}{
 
        \draw[cluster] ($ (offset) + (\nclusters*\clusterlength,0) + (\ngaps*\clusterspace,0) $) 
        -- ($ (offset) + (\nclusters*\clusterlength,0) + (\ngaps*\clusterspace,0) + (\size*\clusterlength,0) $);}
 \draw[tick] ($ (offset) + (0,\lw) $) -- ($ (offset) - (0,2*\lw) $);\coordinate (offset) at (0.00152960152495589, 1);
\foreach \nclusters\ngaps/\size in {0/0/1}{
 
        \draw[cluster] ($ (offset) + (\nclusters*\clusterlength,0) + (\ngaps*\clusterspace,0) $) 
        -- ($ (offset) + (\nclusters*\clusterlength,0) + (\ngaps*\clusterspace,0) + (\size*\clusterlength,0) $);}
 \draw[tick] ($ (offset) + (0,\lw) $) -- ($ (offset) - (0,2*\lw) $);\coordinate (offset) at (0.00153668162772902, 1);
\foreach \nclusters\ngaps/\size in {0/0/1}{
 
        \draw[cluster] ($ (offset) + (\nclusters*\clusterlength,0) + (\ngaps*\clusterspace,0) $) 
        -- ($ (offset) + (\nclusters*\clusterlength,0) + (\ngaps*\clusterspace,0) + (\size*\clusterlength,0) $);}
 \draw[tick] ($ (offset) + (0,\lw) $) -- ($ (offset) - (0,2*\lw) $);\coordinate (offset) at (0.00485621396277331, 1);
\foreach \nclusters\ngaps/\size in {0/0/1}{
 
        \draw[cluster] ($ (offset) + (\nclusters*\clusterlength,0) + (\ngaps*\clusterspace,0) $) 
        -- ($ (offset) + (\nclusters*\clusterlength,0) + (\ngaps*\clusterspace,0) + (\size*\clusterlength,0) $);}
 \draw[tick] ($ (offset) + (0,\lw) $) -- ($ (offset) - (0,2*\lw) $);\coordinate (offset) at (0.00537055353984969, 1);
\foreach \nclusters\ngaps/\size in {0/0/1}{
 
        \draw[cluster] ($ (offset) + (\nclusters*\clusterlength,0) + (\ngaps*\clusterspace,0) $) 
        -- ($ (offset) + (\nclusters*\clusterlength,0) + (\ngaps*\clusterspace,0) + (\size*\clusterlength,0) $);}
 \draw[tick] ($ (offset) + (0,\lw) $) -- ($ (offset) - (0,2*\lw) $);\coordinate (offset) at (0.00537055353984969, 1);
\foreach \nclusters\ngaps/\size in {0/0/1}{
 
        \draw[cluster] ($ (offset) + (\nclusters*\clusterlength,0) + (\ngaps*\clusterspace,0) $) 
        -- ($ (offset) + (\nclusters*\clusterlength,0) + (\ngaps*\clusterspace,0) + (\size*\clusterlength,0) $);}
 \draw[tick] ($ (offset) + (0,\lw) $) -- ($ (offset) - (0,2*\lw) $);\coordinate (offset) at (0.00620138106385903, 1);
\foreach \nclusters\ngaps/\size in {0/0/1}{
 
        \draw[cluster] ($ (offset) + (\nclusters*\clusterlength,0) + (\ngaps*\clusterspace,0) $) 
        -- ($ (offset) + (\nclusters*\clusterlength,0) + (\ngaps*\clusterspace,0) + (\size*\clusterlength,0) $);}
 \draw[tick] ($ (offset) + (0,\lw) $) -- ($ (offset) - (0,2*\lw) $);\coordinate (offset) at (0.00654811757158768, 1);
\foreach \nclusters\ngaps/\size in {0/0/1}{
 
        \draw[cluster] ($ (offset) + (\nclusters*\clusterlength,0) + (\ngaps*\clusterspace,0) $) 
        -- ($ (offset) + (\nclusters*\clusterlength,0) + (\ngaps*\clusterspace,0) + (\size*\clusterlength,0) $);}
 \draw[tick] ($ (offset) + (0,\lw) $) -- ($ (offset) - (0,2*\lw) $);\coordinate (offset) at (0.00760822509113018, 1);
\foreach \nclusters\ngaps/\size in {0/0/1}{
 
        \draw[cluster] ($ (offset) + (\nclusters*\clusterlength,0) + (\ngaps*\clusterspace,0) $) 
        -- ($ (offset) + (\nclusters*\clusterlength,0) + (\ngaps*\clusterspace,0) + (\size*\clusterlength,0) $);}
 \draw[tick] ($ (offset) + (0,\lw) $) -- ($ (offset) - (0,2*\lw) $);\coordinate (offset) at (0.00792689341272064, 1);
\foreach \nclusters\ngaps/\size in {0/0/1}{
 
        \draw[cluster] ($ (offset) + (\nclusters*\clusterlength,0) + (\ngaps*\clusterspace,0) $) 
        -- ($ (offset) + (\nclusters*\clusterlength,0) + (\ngaps*\clusterspace,0) + (\size*\clusterlength,0) $);}
 \draw[tick] ($ (offset) + (0,\lw) $) -- ($ (offset) - (0,2*\lw) $);\coordinate (offset) at (0.00971242792554662, 1);
\foreach \nclusters\ngaps/\size in {0/0/1}{
 
        \draw[cluster] ($ (offset) + (\nclusters*\clusterlength,0) + (\ngaps*\clusterspace,0) $) 
        -- ($ (offset) + (\nclusters*\clusterlength,0) + (\ngaps*\clusterspace,0) + (\size*\clusterlength,0) $);}
 \draw[tick] ($ (offset) + (0,\lw) $) -- ($ (offset) - (0,2*\lw) $);\coordinate (offset) at (0.0101425637287007, 1);
\foreach \nclusters\ngaps/\size in {0/0/1}{
 
        \draw[cluster] ($ (offset) + (\nclusters*\clusterlength,0) + (\ngaps*\clusterspace,0) $) 
        -- ($ (offset) + (\nclusters*\clusterlength,0) + (\ngaps*\clusterspace,0) + (\size*\clusterlength,0) $);}
 \draw[tick] ($ (offset) + (0,\lw) $) -- ($ (offset) - (0,2*\lw) $);\coordinate (offset) at (0.0101425637287007, 1);
\foreach \nclusters\ngaps/\size in {0/0/1}{
 
        \draw[cluster] ($ (offset) + (\nclusters*\clusterlength,0) + (\ngaps*\clusterspace,0) $) 
        -- ($ (offset) + (\nclusters*\clusterlength,0) + (\ngaps*\clusterspace,0) + (\size*\clusterlength,0) $);}
 \draw[tick] ($ (offset) + (0,\lw) $) -- ($ (offset) - (0,2*\lw) $);\coordinate (offset) at (0.0101425637287007, 1);
\foreach \nclusters\ngaps/\size in {0/0/1}{
 
        \draw[cluster] ($ (offset) + (\nclusters*\clusterlength,0) + (\ngaps*\clusterspace,0) $) 
        -- ($ (offset) + (\nclusters*\clusterlength,0) + (\ngaps*\clusterspace,0) + (\size*\clusterlength,0) $);}
 \draw[tick] ($ (offset) + (0,\lw) $) -- ($ (offset) - (0,2*\lw) $);\coordinate (offset) at (0.0106539655278787, 1);
\foreach \nclusters\ngaps/\size in {0/0/1}{
 
        \draw[cluster] ($ (offset) + (\nclusters*\clusterlength,0) + (\ngaps*\clusterspace,0) $) 
        -- ($ (offset) + (\nclusters*\clusterlength,0) + (\ngaps*\clusterspace,0) + (\size*\clusterlength,0) $);}
 \draw[tick] ($ (offset) + (0,\lw) $) -- ($ (offset) - (0,2*\lw) $);\coordinate (offset) at (0.0112561639703216, 1);
\foreach \nclusters\ngaps/\size in {0/0/1}{
 
        \draw[cluster] ($ (offset) + (\nclusters*\clusterlength,0) + (\ngaps*\clusterspace,0) $) 
        -- ($ (offset) + (\nclusters*\clusterlength,0) + (\ngaps*\clusterspace,0) + (\size*\clusterlength,0) $);}
 \draw[tick] ($ (offset) + (0,\lw) $) -- ($ (offset) - (0,2*\lw) $);\coordinate (offset) at (0.0112561639703216, 1);
\foreach \nclusters\ngaps/\size in {0/0/1}{
 
        \draw[cluster] ($ (offset) + (\nclusters*\clusterlength,0) + (\ngaps*\clusterspace,0) $) 
        -- ($ (offset) + (\nclusters*\clusterlength,0) + (\ngaps*\clusterspace,0) + (\size*\clusterlength,0) $);}
 \draw[tick] ($ (offset) + (0,\lw) $) -- ($ (offset) - (0,2*\lw) $);\coordinate (offset) at (0.0114608169688865, 1);
\foreach \nclusters\ngaps/\size in {0/0/1}{
 
        \draw[cluster] ($ (offset) + (\nclusters*\clusterlength,0) + (\ngaps*\clusterspace,0) $) 
        -- ($ (offset) + (\nclusters*\clusterlength,0) + (\ngaps*\clusterspace,0) + (\size*\clusterlength,0) $);}
 \draw[tick] ($ (offset) + (0,\lw) $) -- ($ (offset) - (0,2*\lw) $);\coordinate (offset) at (0.0114608169688865, 1);
\foreach \nclusters\ngaps/\size in {0/0/1}{
 
        \draw[cluster] ($ (offset) + (\nclusters*\clusterlength,0) + (\ngaps*\clusterspace,0) $) 
        -- ($ (offset) + (\nclusters*\clusterlength,0) + (\ngaps*\clusterspace,0) + (\size*\clusterlength,0) $);}
 \draw[tick] ($ (offset) + (0,\lw) $) -- ($ (offset) - (0,2*\lw) $);\coordinate (offset) at (0.0116217557947888, 1);
\foreach \nclusters\ngaps/\size in {0/0/1}{
 
        \draw[cluster] ($ (offset) + (\nclusters*\clusterlength,0) + (\ngaps*\clusterspace,0) $) 
        -- ($ (offset) + (\nclusters*\clusterlength,0) + (\ngaps*\clusterspace,0) + (\size*\clusterlength,0) $);}
 \draw[tick] ($ (offset) + (0,\lw) $) -- ($ (offset) - (0,2*\lw) $);\coordinate (offset) at (0.0121085296965464, 1);
\foreach \nclusters\ngaps/\size in {0/0/1}{
 
        \draw[cluster] ($ (offset) + (\nclusters*\clusterlength,0) + (\ngaps*\clusterspace,0) $) 
        -- ($ (offset) + (\nclusters*\clusterlength,0) + (\ngaps*\clusterspace,0) + (\size*\clusterlength,0) $);}
 \draw[tick] ($ (offset) + (0,\lw) $) -- ($ (offset) - (0,2*\lw) $);\coordinate (offset) at (0.0121085296965464, 1);
\foreach \nclusters\ngaps/\size in {0/0/1}{
 
        \draw[cluster] ($ (offset) + (\nclusters*\clusterlength,0) + (\ngaps*\clusterspace,0) $) 
        -- ($ (offset) + (\nclusters*\clusterlength,0) + (\ngaps*\clusterspace,0) + (\size*\clusterlength,0) $);}
 \draw[tick] ($ (offset) + (0,\lw) $) -- ($ (offset) - (0,2*\lw) $);\coordinate (offset) at (0.0124241795476727, 1);
\foreach \nclusters\ngaps/\size in {0/0/1}{
 
        \draw[cluster] ($ (offset) + (\nclusters*\clusterlength,0) + (\ngaps*\clusterspace,0) $) 
        -- ($ (offset) + (\nclusters*\clusterlength,0) + (\ngaps*\clusterspace,0) + (\size*\clusterlength,0) $);}
 \draw[tick] ($ (offset) + (0,\lw) $) -- ($ (offset) - (0,2*\lw) $);\coordinate (offset) at (0.0129445627377298, 1);
\foreach \nclusters\ngaps/\size in {0/0/1}{
 
        \draw[cluster] ($ (offset) + (\nclusters*\clusterlength,0) + (\ngaps*\clusterspace,0) $) 
        -- ($ (offset) + (\nclusters*\clusterlength,0) + (\ngaps*\clusterspace,0) + (\size*\clusterlength,0) $);}
 \draw[tick] ($ (offset) + (0,\lw) $) -- ($ (offset) - (0,2*\lw) $);\coordinate (offset) at (0.0132046020919773, 1);
\foreach \nclusters\ngaps/\size in {0/0/1}{
 
        \draw[cluster] ($ (offset) + (\nclusters*\clusterlength,0) + (\ngaps*\clusterspace,0) $) 
        -- ($ (offset) + (\nclusters*\clusterlength,0) + (\ngaps*\clusterspace,0) + (\size*\clusterlength,0) $);}
 \draw[tick] ($ (offset) + (0,\lw) $) -- ($ (offset) - (0,2*\lw) $);\coordinate (offset) at (0.0132046020919773, 1);
\foreach \nclusters\ngaps/\size in {0/0/1}{
 
        \draw[cluster] ($ (offset) + (\nclusters*\clusterlength,0) + (\ngaps*\clusterspace,0) $) 
        -- ($ (offset) + (\nclusters*\clusterlength,0) + (\ngaps*\clusterspace,0) + (\size*\clusterlength,0) $);}
 \draw[tick] ($ (offset) + (0,\lw) $) -- ($ (offset) - (0,2*\lw) $);\coordinate (offset) at (0.0138700047555484, 1);
\foreach \nclusters\ngaps/\size in {0/0/1}{
 
        \draw[cluster] ($ (offset) + (\nclusters*\clusterlength,0) + (\ngaps*\clusterspace,0) $) 
        -- ($ (offset) + (\nclusters*\clusterlength,0) + (\ngaps*\clusterspace,0) + (\size*\clusterlength,0) $);}
 \draw[tick] ($ (offset) + (0,\lw) $) -- ($ (offset) - (0,2*\lw) $);\coordinate (offset) at (0.0138700047555484, 1);
\foreach \nclusters\ngaps/\size in {0/0/1}{
 
        \draw[cluster] ($ (offset) + (\nclusters*\clusterlength,0) + (\ngaps*\clusterspace,0) $) 
        -- ($ (offset) + (\nclusters*\clusterlength,0) + (\ngaps*\clusterspace,0) + (\size*\clusterlength,0) $);}
 \draw[tick] ($ (offset) + (0,\lw) $) -- ($ (offset) - (0,2*\lw) $);\coordinate (offset) at (0.0154027091699658, 1);
\foreach \nclusters\ngaps/\size in {0/0/1}{
 
        \draw[cluster] ($ (offset) + (\nclusters*\clusterlength,0) + (\ngaps*\clusterspace,0) $) 
        -- ($ (offset) + (\nclusters*\clusterlength,0) + (\ngaps*\clusterspace,0) + (\size*\clusterlength,0) $);}
 \draw[tick] ($ (offset) + (0,\lw) $) -- ($ (offset) - (0,2*\lw) $);\coordinate (offset) at (0.0158537868254413, 1);
\foreach \nclusters\ngaps/\size in {0/0/1}{
 
        \draw[cluster] ($ (offset) + (\nclusters*\clusterlength,0) + (\ngaps*\clusterspace,0) $) 
        -- ($ (offset) + (\nclusters*\clusterlength,0) + (\ngaps*\clusterspace,0) + (\size*\clusterlength,0) $);}
 \draw[tick] ($ (offset) + (0,\lw) $) -- ($ (offset) - (0,2*\lw) $);\coordinate (offset) at (0.0158537868254413, 1);
\foreach \nclusters\ngaps/\size in {0/0/1}{
 
        \draw[cluster] ($ (offset) + (\nclusters*\clusterlength,0) + (\ngaps*\clusterspace,0) $) 
        -- ($ (offset) + (\nclusters*\clusterlength,0) + (\ngaps*\clusterspace,0) + (\size*\clusterlength,0) $);}
 \draw[tick] ($ (offset) + (0,\lw) $) -- ($ (offset) - (0,2*\lw) $);\coordinate (offset) at (0.0161776551184645, 1);
\foreach \nclusters\ngaps/\size in {0/0/1}{
 
        \draw[cluster] ($ (offset) + (\nclusters*\clusterlength,0) + (\ngaps*\clusterspace,0) $) 
        -- ($ (offset) + (\nclusters*\clusterlength,0) + (\ngaps*\clusterspace,0) + (\size*\clusterlength,0) $);}
 \draw[tick] ($ (offset) + (0,\lw) $) -- ($ (offset) - (0,2*\lw) $);\coordinate (offset) at (0.0161776551184645, 1);
\foreach \nclusters\ngaps/\size in {0/0/1}{
 
        \draw[cluster] ($ (offset) + (\nclusters*\clusterlength,0) + (\ngaps*\clusterspace,0) $) 
        -- ($ (offset) + (\nclusters*\clusterlength,0) + (\ngaps*\clusterspace,0) + (\size*\clusterlength,0) $);}
 \draw[tick] ($ (offset) + (0,\lw) $) -- ($ (offset) - (0,2*\lw) $);\coordinate (offset) at (0.0162742011609848, 1);
\foreach \nclusters\ngaps/\size in {0/0/1}{
 
        \draw[cluster] ($ (offset) + (\nclusters*\clusterlength,0) + (\ngaps*\clusterspace,0) $) 
        -- ($ (offset) + (\nclusters*\clusterlength,0) + (\ngaps*\clusterspace,0) + (\size*\clusterlength,0) $);}
 \draw[tick] ($ (offset) + (0,\lw) $) -- ($ (offset) - (0,2*\lw) $);\coordinate (offset) at (0.0173978528536697, 1);
\foreach \nclusters\ngaps/\size in {0/0/1}{
 
        \draw[cluster] ($ (offset) + (\nclusters*\clusterlength,0) + (\ngaps*\clusterspace,0) $) 
        -- ($ (offset) + (\nclusters*\clusterlength,0) + (\ngaps*\clusterspace,0) + (\size*\clusterlength,0) $);}
 \draw[tick] ($ (offset) + (0,\lw) $) -- ($ (offset) - (0,2*\lw) $);\coordinate (offset) at (0.0173978528536697, 1);
\foreach \nclusters\ngaps/\size in {0/0/1}{
 
        \draw[cluster] ($ (offset) + (\nclusters*\clusterlength,0) + (\ngaps*\clusterspace,0) $) 
        -- ($ (offset) + (\nclusters*\clusterlength,0) + (\ngaps*\clusterspace,0) + (\size*\clusterlength,0) $);}
 \draw[tick] ($ (offset) + (0,\lw) $) -- ($ (offset) - (0,2*\lw) $);\coordinate (offset) at (0.0173978528536697, 1);
\foreach \nclusters\ngaps/\size in {0/0/1}{
 
        \draw[cluster] ($ (offset) + (\nclusters*\clusterlength,0) + (\ngaps*\clusterspace,0) $) 
        -- ($ (offset) + (\nclusters*\clusterlength,0) + (\ngaps*\clusterspace,0) + (\size*\clusterlength,0) $);}
 \draw[tick] ($ (offset) + (0,\lw) $) -- ($ (offset) - (0,2*\lw) $);\coordinate (offset) at (0.0175335837343096, 1);
\foreach \nclusters\ngaps/\size in {0/0/1}{
 
        \draw[cluster] ($ (offset) + (\nclusters*\clusterlength,0) + (\ngaps*\clusterspace,0) $) 
        -- ($ (offset) + (\nclusters*\clusterlength,0) + (\ngaps*\clusterspace,0) + (\size*\clusterlength,0) $);}
 \draw[tick] ($ (offset) + (0,\lw) $) -- ($ (offset) - (0,2*\lw) $);\coordinate (offset) at (0.0175335837343096, 1);
\foreach \nclusters\ngaps/\size in {0/0/1}{
 
        \draw[cluster] ($ (offset) + (\nclusters*\clusterlength,0) + (\ngaps*\clusterspace,0) $) 
        -- ($ (offset) + (\nclusters*\clusterlength,0) + (\ngaps*\clusterspace,0) + (\size*\clusterlength,0) $);}
 \draw[tick] ($ (offset) + (0,\lw) $) -- ($ (offset) - (0,2*\lw) $);\coordinate (offset) at (0.0175335837343096, 1);
\foreach \nclusters\ngaps/\size in {0/0/1}{
 
        \draw[cluster] ($ (offset) + (\nclusters*\clusterlength,0) + (\ngaps*\clusterspace,0) $) 
        -- ($ (offset) + (\nclusters*\clusterlength,0) + (\ngaps*\clusterspace,0) + (\size*\clusterlength,0) $);}
 \draw[tick] ($ (offset) + (0,\lw) $) -- ($ (offset) - (0,2*\lw) $);\coordinate (offset) at (0.0175335837343096, 1);
\foreach \nclusters\ngaps/\size in {0/0/1}{
 
        \draw[cluster] ($ (offset) + (\nclusters*\clusterlength,0) + (\ngaps*\clusterspace,0) $) 
        -- ($ (offset) + (\nclusters*\clusterlength,0) + (\ngaps*\clusterspace,0) + (\size*\clusterlength,0) $);}
 \draw[tick] ($ (offset) + (0,\lw) $) -- ($ (offset) - (0,2*\lw) $);\coordinate (offset) at (0.0178895458229643, 1);
\foreach \nclusters\ngaps/\size in {0/0/1}{
 
        \draw[cluster] ($ (offset) + (\nclusters*\clusterlength,0) + (\ngaps*\clusterspace,0) $) 
        -- ($ (offset) + (\nclusters*\clusterlength,0) + (\ngaps*\clusterspace,0) + (\size*\clusterlength,0) $);}
 \draw[tick] ($ (offset) + (0,\lw) $) -- ($ (offset) - (0,2*\lw) $);\coordinate (offset) at (0.0178895458229643, 1);
\foreach \nclusters\ngaps/\size in {0/0/1}{
 
        \draw[cluster] ($ (offset) + (\nclusters*\clusterlength,0) + (\ngaps*\clusterspace,0) $) 
        -- ($ (offset) + (\nclusters*\clusterlength,0) + (\ngaps*\clusterspace,0) + (\size*\clusterlength,0) $);}
 \draw[tick] ($ (offset) + (0,\lw) $) -- ($ (offset) - (0,2*\lw) $);\coordinate (offset) at (0.0178895458229643, 1);
\foreach \nclusters\ngaps/\size in {0/0/1}{
 
        \draw[cluster] ($ (offset) + (\nclusters*\clusterlength,0) + (\ngaps*\clusterspace,0) $) 
        -- ($ (offset) + (\nclusters*\clusterlength,0) + (\ngaps*\clusterspace,0) + (\size*\clusterlength,0) $);}
 \draw[tick] ($ (offset) + (0,\lw) $) -- ($ (offset) - (0,2*\lw) $);\coordinate (offset) at (0.0188643890614518, 1);
\foreach \nclusters\ngaps/\size in {0/0/1}{
 
        \draw[cluster] ($ (offset) + (\nclusters*\clusterlength,0) + (\ngaps*\clusterspace,0) $) 
        -- ($ (offset) + (\nclusters*\clusterlength,0) + (\ngaps*\clusterspace,0) + (\size*\clusterlength,0) $);}
 \draw[tick] ($ (offset) + (0,\lw) $) -- ($ (offset) - (0,2*\lw) $);\coordinate (offset) at (0.019428832856994, 1);
\foreach \nclusters\ngaps/\size in {0/0/1}{
 
        \draw[cluster] ($ (offset) + (\nclusters*\clusterlength,0) + (\ngaps*\clusterspace,0) $) 
        -- ($ (offset) + (\nclusters*\clusterlength,0) + (\ngaps*\clusterspace,0) + (\size*\clusterlength,0) $);}
 \draw[tick] ($ (offset) + (0,\lw) $) -- ($ (offset) - (0,2*\lw) $);\coordinate (offset) at (0.019428832856994, 1);
\foreach \nclusters\ngaps/\size in {0/0/1}{
 
        \draw[cluster] ($ (offset) + (\nclusters*\clusterlength,0) + (\ngaps*\clusterspace,0) $) 
        -- ($ (offset) + (\nclusters*\clusterlength,0) + (\ngaps*\clusterspace,0) + (\size*\clusterlength,0) $);}
 \draw[tick] ($ (offset) + (0,\lw) $) -- ($ (offset) - (0,2*\lw) $);\coordinate (offset) at (0.019428832856994, 1);
\foreach \nclusters\ngaps/\size in {0/0/1}{
 
        \draw[cluster] ($ (offset) + (\nclusters*\clusterlength,0) + (\ngaps*\clusterspace,0) $) 
        -- ($ (offset) + (\nclusters*\clusterlength,0) + (\ngaps*\clusterspace,0) + (\size*\clusterlength,0) $);}
 \draw[tick] ($ (offset) + (0,\lw) $) -- ($ (offset) - (0,2*\lw) $);\coordinate (offset) at (0.019428832856994, 1);
\foreach \nclusters\ngaps/\size in {0/0/1}{
 
        \draw[cluster] ($ (offset) + (\nclusters*\clusterlength,0) + (\ngaps*\clusterspace,0) $) 
        -- ($ (offset) + (\nclusters*\clusterlength,0) + (\ngaps*\clusterspace,0) + (\size*\clusterlength,0) $);}
 \draw[tick] ($ (offset) + (0,\lw) $) -- ($ (offset) - (0,2*\lw) $);\coordinate (offset) at (0.019428832856994, 1);
\foreach \nclusters\ngaps/\size in {0/0/1}{
 
        \draw[cluster] ($ (offset) + (\nclusters*\clusterlength,0) + (\ngaps*\clusterspace,0) $) 
        -- ($ (offset) + (\nclusters*\clusterlength,0) + (\ngaps*\clusterspace,0) + (\size*\clusterlength,0) $);}
 \draw[tick] ($ (offset) + (0,\lw) $) -- ($ (offset) - (0,2*\lw) $);\coordinate (offset) at (0.0203494359758325, 1);
\foreach \nclusters\ngaps/\size in {0/0/1}{
 
        \draw[cluster] ($ (offset) + (\nclusters*\clusterlength,0) + (\ngaps*\clusterspace,0) $) 
        -- ($ (offset) + (\nclusters*\clusterlength,0) + (\ngaps*\clusterspace,0) + (\size*\clusterlength,0) $);}
 \draw[tick] ($ (offset) + (0,\lw) $) -- ($ (offset) - (0,2*\lw) $);\coordinate (offset) at (0.0204007908029853, 1);
\foreach \nclusters\ngaps/\size in {0/0/1}{
 
        \draw[cluster] ($ (offset) + (\nclusters*\clusterlength,0) + (\ngaps*\clusterspace,0) $) 
        -- ($ (offset) + (\nclusters*\clusterlength,0) + (\ngaps*\clusterspace,0) + (\size*\clusterlength,0) $);}
 \draw[tick] ($ (offset) + (0,\lw) $) -- ($ (offset) - (0,2*\lw) $);\coordinate (offset) at (0.0205793241710582, 1);
\foreach \nclusters\ngaps/\size in {0/0/1}{
 
        \draw[cluster] ($ (offset) + (\nclusters*\clusterlength,0) + (\ngaps*\clusterspace,0) $) 
        -- ($ (offset) + (\nclusters*\clusterlength,0) + (\ngaps*\clusterspace,0) + (\size*\clusterlength,0) $);}
 \draw[tick] ($ (offset) + (0,\lw) $) -- ($ (offset) - (0,2*\lw) $);\coordinate (offset) at (0.0212055781030452, 1);
\foreach \nclusters\ngaps/\size in {0/0/1}{
 
        \draw[cluster] ($ (offset) + (\nclusters*\clusterlength,0) + (\ngaps*\clusterspace,0) $) 
        -- ($ (offset) + (\nclusters*\clusterlength,0) + (\ngaps*\clusterspace,0) + (\size*\clusterlength,0) $);}
 \draw[tick] ($ (offset) + (0,\lw) $) -- ($ (offset) - (0,2*\lw) $);\coordinate (offset) at (0.0221601394591951, 1);
\foreach \nclusters\ngaps/\size in {0/0/1}{
 
        \draw[cluster] ($ (offset) + (\nclusters*\clusterlength,0) + (\ngaps*\clusterspace,0) $) 
        -- ($ (offset) + (\nclusters*\clusterlength,0) + (\ngaps*\clusterspace,0) + (\size*\clusterlength,0) $);}
 \draw[tick] ($ (offset) + (0,\lw) $) -- ($ (offset) - (0,2*\lw) $);\coordinate (offset) at (0.0221601394591951, 1);
\foreach \nclusters\ngaps/\size in {0/0/1}{
 
        \draw[cluster] ($ (offset) + (\nclusters*\clusterlength,0) + (\ngaps*\clusterspace,0) $) 
        -- ($ (offset) + (\nclusters*\clusterlength,0) + (\ngaps*\clusterspace,0) + (\size*\clusterlength,0) $);}
 \draw[tick] ($ (offset) + (0,\lw) $) -- ($ (offset) - (0,2*\lw) $);\coordinate (offset) at (0.023762956699802, 1);
\foreach \nclusters\ngaps/\size in {0/0/1}{
 
        \draw[cluster] ($ (offset) + (\nclusters*\clusterlength,0) + (\ngaps*\clusterspace,0) $) 
        -- ($ (offset) + (\nclusters*\clusterlength,0) + (\ngaps*\clusterspace,0) + (\size*\clusterlength,0) $);}
 \draw[tick] ($ (offset) + (0,\lw) $) -- ($ (offset) - (0,2*\lw) $);\coordinate (offset) at (0.0237953633987278, 1);
\foreach \nclusters\ngaps/\size in {0/0/1}{
 
        \draw[cluster] ($ (offset) + (\nclusters*\clusterlength,0) + (\ngaps*\clusterspace,0) $) 
        -- ($ (offset) + (\nclusters*\clusterlength,0) + (\ngaps*\clusterspace,0) + (\size*\clusterlength,0) $);}
 \draw[tick] ($ (offset) + (0,\lw) $) -- ($ (offset) - (0,2*\lw) $);\coordinate (offset) at (0.0237953633987278, 1);
\foreach \nclusters\ngaps/\size in {0/0/1}{
 
        \draw[cluster] ($ (offset) + (\nclusters*\clusterlength,0) + (\ngaps*\clusterspace,0) $) 
        -- ($ (offset) + (\nclusters*\clusterlength,0) + (\ngaps*\clusterspace,0) + (\size*\clusterlength,0) $);}
 \draw[tick] ($ (offset) + (0,\lw) $) -- ($ (offset) - (0,2*\lw) $);\coordinate (offset) at (0.0237953633987278, 1);
\foreach \nclusters\ngaps/\size in {0/0/1}{
 
        \draw[cluster] ($ (offset) + (\nclusters*\clusterlength,0) + (\ngaps*\clusterspace,0) $) 
        -- ($ (offset) + (\nclusters*\clusterlength,0) + (\ngaps*\clusterspace,0) + (\size*\clusterlength,0) $);}
 \draw[tick] ($ (offset) + (0,\lw) $) -- ($ (offset) - (0,2*\lw) $);\coordinate (offset) at (0.0259498777109078, 1);
\foreach \nclusters\ngaps/\size in {0/0/1}{
 
        \draw[cluster] ($ (offset) + (\nclusters*\clusterlength,0) + (\ngaps*\clusterspace,0) $) 
        -- ($ (offset) + (\nclusters*\clusterlength,0) + (\ngaps*\clusterspace,0) + (\size*\clusterlength,0) $);}
 \draw[tick] ($ (offset) + (0,\lw) $) -- ($ (offset) - (0,2*\lw) $);\coordinate (offset) at (0.0274589423774891, 1);
\foreach \nclusters\ngaps/\size in {0/0/1}{
 
        \draw[cluster] ($ (offset) + (\nclusters*\clusterlength,0) + (\ngaps*\clusterspace,0) $) 
        -- ($ (offset) + (\nclusters*\clusterlength,0) + (\ngaps*\clusterspace,0) + (\size*\clusterlength,0) $);}
 \draw[tick] ($ (offset) + (0,\lw) $) -- ($ (offset) - (0,2*\lw) $);\coordinate (offset) at (0.0286322223409476, 1);
\foreach \nclusters\ngaps/\size in {0/0/1}{
 
        \draw[cluster] ($ (offset) + (\nclusters*\clusterlength,0) + (\ngaps*\clusterspace,0) $) 
        -- ($ (offset) + (\nclusters*\clusterlength,0) + (\ngaps*\clusterspace,0) + (\size*\clusterlength,0) $);}
 \draw[tick] ($ (offset) + (0,\lw) $) -- ($ (offset) - (0,2*\lw) $);\coordinate (offset) at (0.0286322223409476, 1);
\foreach \nclusters\ngaps/\size in {0/0/1}{
 
        \draw[cluster] ($ (offset) + (\nclusters*\clusterlength,0) + (\ngaps*\clusterspace,0) $) 
        -- ($ (offset) + (\nclusters*\clusterlength,0) + (\ngaps*\clusterspace,0) + (\size*\clusterlength,0) $);}
 \draw[tick] ($ (offset) + (0,\lw) $) -- ($ (offset) - (0,2*\lw) $);\coordinate (offset) at (0.029964337763661, 1);
\foreach \nclusters\ngaps/\size in {0/0/1}{
 
        \draw[cluster] ($ (offset) + (\nclusters*\clusterlength,0) + (\ngaps*\clusterspace,0) $) 
        -- ($ (offset) + (\nclusters*\clusterlength,0) + (\ngaps*\clusterspace,0) + (\size*\clusterlength,0) $);}
 \draw[tick] ($ (offset) + (0,\lw) $) -- ($ (offset) - (0,2*\lw) $);\coordinate (offset) at (0.029964337763661, 1);
\foreach \nclusters\ngaps/\size in {0/0/1}{
 
        \draw[cluster] ($ (offset) + (\nclusters*\clusterlength,0) + (\ngaps*\clusterspace,0) $) 
        -- ($ (offset) + (\nclusters*\clusterlength,0) + (\ngaps*\clusterspace,0) + (\size*\clusterlength,0) $);}
 \draw[tick] ($ (offset) + (0,\lw) $) -- ($ (offset) - (0,2*\lw) $);\coordinate (offset) at (0.0303424155913995, 1);
\foreach \nclusters\ngaps/\size in {0/0/1}{
 
        \draw[cluster] ($ (offset) + (\nclusters*\clusterlength,0) + (\ngaps*\clusterspace,0) $) 
        -- ($ (offset) + (\nclusters*\clusterlength,0) + (\ngaps*\clusterspace,0) + (\size*\clusterlength,0) $);}
 \draw[tick] ($ (offset) + (0,\lw) $) -- ($ (offset) - (0,2*\lw) $);\coordinate (offset) at (0.0314354369216361, 1);
\foreach \nclusters\ngaps/\size in {0/0/1}{
 
        \draw[cluster] ($ (offset) + (\nclusters*\clusterlength,0) + (\ngaps*\clusterspace,0) $) 
        -- ($ (offset) + (\nclusters*\clusterlength,0) + (\ngaps*\clusterspace,0) + (\size*\clusterlength,0) $);}
 \draw[tick] ($ (offset) + (0,\lw) $) -- ($ (offset) - (0,2*\lw) $);\coordinate (offset) at (0.0322772254839305, 1);
\foreach \nclusters\ngaps/\size in {0/0/1}{
 
        \draw[cluster] ($ (offset) + (\nclusters*\clusterlength,0) + (\ngaps*\clusterspace,0) $) 
        -- ($ (offset) + (\nclusters*\clusterlength,0) + (\ngaps*\clusterspace,0) + (\size*\clusterlength,0) $);}
 \draw[tick] ($ (offset) + (0,\lw) $) -- ($ (offset) - (0,2*\lw) $);\coordinate (offset) at (0.0336058956469446, 1);
\foreach \nclusters\ngaps/\size in {0/0/1}{
 
        \draw[cluster] ($ (offset) + (\nclusters*\clusterlength,0) + (\ngaps*\clusterspace,0) $) 
        -- ($ (offset) + (\nclusters*\clusterlength,0) + (\ngaps*\clusterspace,0) + (\size*\clusterlength,0) $);}
 \draw[tick] ($ (offset) + (0,\lw) $) -- ($ (offset) - (0,2*\lw) $);\coordinate (offset) at (0.0341051469860245, 1);
\foreach \nclusters\ngaps/\size in {0/0/1}{
 
        \draw[cluster] ($ (offset) + (\nclusters*\clusterlength,0) + (\ngaps*\clusterspace,0) $) 
        -- ($ (offset) + (\nclusters*\clusterlength,0) + (\ngaps*\clusterspace,0) + (\size*\clusterlength,0) $);}
 \draw[tick] ($ (offset) + (0,\lw) $) -- ($ (offset) - (0,2*\lw) $);\coordinate (offset) at (0.0371617178811987, 1);
\foreach \nclusters\ngaps/\size in {0/0/1}{
 
        \draw[cluster] ($ (offset) + (\nclusters*\clusterlength,0) + (\ngaps*\clusterspace,0) $) 
        -- ($ (offset) + (\nclusters*\clusterlength,0) + (\ngaps*\clusterspace,0) + (\size*\clusterlength,0) $);}
 \draw[tick] ($ (offset) + (0,\lw) $) -- ($ (offset) - (0,2*\lw) $);\coordinate (offset) at (0.0371617178811987, 1);
\foreach \nclusters\ngaps/\size in {0/0/1}{
 
        \draw[cluster] ($ (offset) + (\nclusters*\clusterlength,0) + (\ngaps*\clusterspace,0) $) 
        -- ($ (offset) + (\nclusters*\clusterlength,0) + (\ngaps*\clusterspace,0) + (\size*\clusterlength,0) $);}
 \draw[tick] ($ (offset) + (0,\lw) $) -- ($ (offset) - (0,2*\lw) $);\coordinate (offset) at (0.0371617178811987, 1);
\foreach \nclusters\ngaps/\size in {0/0/1}{
 
        \draw[cluster] ($ (offset) + (\nclusters*\clusterlength,0) + (\ngaps*\clusterspace,0) $) 
        -- ($ (offset) + (\nclusters*\clusterlength,0) + (\ngaps*\clusterspace,0) + (\size*\clusterlength,0) $);}
 \draw[tick] ($ (offset) + (0,\lw) $) -- ($ (offset) - (0,2*\lw) $);\coordinate (offset) at (0.0371617178811987, 1);
\foreach \nclusters\ngaps/\size in {0/0/1}{
 
        \draw[cluster] ($ (offset) + (\nclusters*\clusterlength,0) + (\ngaps*\clusterspace,0) $) 
        -- ($ (offset) + (\nclusters*\clusterlength,0) + (\ngaps*\clusterspace,0) + (\size*\clusterlength,0) $);}
 \draw[tick] ($ (offset) + (0,\lw) $) -- ($ (offset) - (0,2*\lw) $);\coordinate (offset) at (0.0372423794510973, 1);
\foreach \nclusters\ngaps/\size in {0/0/1}{
 
        \draw[cluster] ($ (offset) + (\nclusters*\clusterlength,0) + (\ngaps*\clusterspace,0) $) 
        -- ($ (offset) + (\nclusters*\clusterlength,0) + (\ngaps*\clusterspace,0) + (\size*\clusterlength,0) $);}
 \draw[tick] ($ (offset) + (0,\lw) $) -- ($ (offset) - (0,2*\lw) $);\coordinate (offset) at (0.0372423794510973, 1);
\foreach \nclusters\ngaps/\size in {0/0/1}{
 
        \draw[cluster] ($ (offset) + (\nclusters*\clusterlength,0) + (\ngaps*\clusterspace,0) $) 
        -- ($ (offset) + (\nclusters*\clusterlength,0) + (\ngaps*\clusterspace,0) + (\size*\clusterlength,0) $);}
 \draw[tick] ($ (offset) + (0,\lw) $) -- ($ (offset) - (0,2*\lw) $);\coordinate (offset) at (0.0410245285206469, 1);
\foreach \nclusters\ngaps/\size in {0/0/1}{
 
        \draw[cluster] ($ (offset) + (\nclusters*\clusterlength,0) + (\ngaps*\clusterspace,0) $) 
        -- ($ (offset) + (\nclusters*\clusterlength,0) + (\ngaps*\clusterspace,0) + (\size*\clusterlength,0) $);}
 \draw[tick] ($ (offset) + (0,\lw) $) -- ($ (offset) - (0,2*\lw) $);\coordinate (offset) at (0.0413289471330374, 1);
\foreach \nclusters\ngaps/\size in {0/0/1}{
 
        \draw[cluster] ($ (offset) + (\nclusters*\clusterlength,0) + (\ngaps*\clusterspace,0) $) 
        -- ($ (offset) + (\nclusters*\clusterlength,0) + (\ngaps*\clusterspace,0) + (\size*\clusterlength,0) $);}
 \draw[tick] ($ (offset) + (0,\lw) $) -- ($ (offset) - (0,2*\lw) $);\coordinate (offset) at (0.0413289471330374, 1);
\foreach \nclusters\ngaps/\size in {0/0/1}{
 
        \draw[cluster] ($ (offset) + (\nclusters*\clusterlength,0) + (\ngaps*\clusterspace,0) $) 
        -- ($ (offset) + (\nclusters*\clusterlength,0) + (\ngaps*\clusterspace,0) + (\size*\clusterlength,0) $);}
 \draw[tick] ($ (offset) + (0,\lw) $) -- ($ (offset) - (0,2*\lw) $);\coordinate (offset) at (0.0414992459239586, 1);
\foreach \nclusters\ngaps/\size in {0/0/1}{
 
        \draw[cluster] ($ (offset) + (\nclusters*\clusterlength,0) + (\ngaps*\clusterspace,0) $) 
        -- ($ (offset) + (\nclusters*\clusterlength,0) + (\ngaps*\clusterspace,0) + (\size*\clusterlength,0) $);}
 \draw[tick] ($ (offset) + (0,\lw) $) -- ($ (offset) - (0,2*\lw) $);\coordinate (offset) at (0.0438875077561685, 1);
\foreach \nclusters\ngaps/\size in {0/0/1}{
 
        \draw[cluster] ($ (offset) + (\nclusters*\clusterlength,0) + (\ngaps*\clusterspace,0) $) 
        -- ($ (offset) + (\nclusters*\clusterlength,0) + (\ngaps*\clusterspace,0) + (\size*\clusterlength,0) $);}
 \draw[tick] ($ (offset) + (0,\lw) $) -- ($ (offset) - (0,2*\lw) $);\coordinate (offset) at (0.0452735932454064, 1);
\foreach \nclusters\ngaps/\size in {0/0/1}{
 
        \draw[cluster] ($ (offset) + (\nclusters*\clusterlength,0) + (\ngaps*\clusterspace,0) $) 
        -- ($ (offset) + (\nclusters*\clusterlength,0) + (\ngaps*\clusterspace,0) + (\size*\clusterlength,0) $);}
 \draw[tick] ($ (offset) + (0,\lw) $) -- ($ (offset) - (0,2*\lw) $);\coordinate (offset) at (0.0452735932454064, 1);
\foreach \nclusters\ngaps/\size in {0/0/1}{
 
        \draw[cluster] ($ (offset) + (\nclusters*\clusterlength,0) + (\ngaps*\clusterspace,0) $) 
        -- ($ (offset) + (\nclusters*\clusterlength,0) + (\ngaps*\clusterspace,0) + (\size*\clusterlength,0) $);}
 \draw[tick] ($ (offset) + (0,\lw) $) -- ($ (offset) - (0,2*\lw) $);\coordinate (offset) at (0.0455780118577969, 1);
\foreach \nclusters\ngaps/\size in {0/0/1}{
 
        \draw[cluster] ($ (offset) + (\nclusters*\clusterlength,0) + (\ngaps*\clusterspace,0) $) 
        -- ($ (offset) + (\nclusters*\clusterlength,0) + (\ngaps*\clusterspace,0) + (\size*\clusterlength,0) $);}
 \draw[tick] ($ (offset) + (0,\lw) $) -- ($ (offset) - (0,2*\lw) $);\coordinate (offset) at (0.0455780118577969, 1);
\foreach \nclusters\ngaps/\size in {0/0/1}{
 
        \draw[cluster] ($ (offset) + (\nclusters*\clusterlength,0) + (\ngaps*\clusterspace,0) $) 
        -- ($ (offset) + (\nclusters*\clusterlength,0) + (\ngaps*\clusterspace,0) + (\size*\clusterlength,0) $);}
 \draw[tick] ($ (offset) + (0,\lw) $) -- ($ (offset) - (0,2*\lw) $);\coordinate (offset) at (0.0455780118577969, 1);
\foreach \nclusters\ngaps/\size in {0/0/1}{
 
        \draw[cluster] ($ (offset) + (\nclusters*\clusterlength,0) + (\ngaps*\clusterspace,0) $) 
        -- ($ (offset) + (\nclusters*\clusterlength,0) + (\ngaps*\clusterspace,0) + (\size*\clusterlength,0) $);}
 \draw[tick] ($ (offset) + (0,\lw) $) -- ($ (offset) - (0,2*\lw) $);\coordinate (offset) at (0.0455780118577969, 1);
\foreach \nclusters\ngaps/\size in {0/0/1}{
 
        \draw[cluster] ($ (offset) + (\nclusters*\clusterlength,0) + (\ngaps*\clusterspace,0) $) 
        -- ($ (offset) + (\nclusters*\clusterlength,0) + (\ngaps*\clusterspace,0) + (\size*\clusterlength,0) $);}
 \draw[tick] ($ (offset) + (0,\lw) $) -- ($ (offset) - (0,2*\lw) $);\coordinate (offset) at (0.0471082978198876, 1);
\foreach \nclusters\ngaps/\size in {0/0/1}{
 
        \draw[cluster] ($ (offset) + (\nclusters*\clusterlength,0) + (\ngaps*\clusterspace,0) $) 
        -- ($ (offset) + (\nclusters*\clusterlength,0) + (\ngaps*\clusterspace,0) + (\size*\clusterlength,0) $);}
 \draw[tick] ($ (offset) + (0,\lw) $) -- ($ (offset) - (0,2*\lw) $);\coordinate (offset) at (0.047722557505166, 1);
\foreach \nclusters\ngaps/\size in {0/0/1}{
 
        \draw[cluster] ($ (offset) + (\nclusters*\clusterlength,0) + (\ngaps*\clusterspace,0) $) 
        -- ($ (offset) + (\nclusters*\clusterlength,0) + (\ngaps*\clusterspace,0) + (\size*\clusterlength,0) $);}
 \draw[tick] ($ (offset) + (0,\lw) $) -- ($ (offset) - (0,2*\lw) $);\coordinate (offset) at (0.0495051730150695, 1);
\foreach \nclusters\ngaps/\size in {0/0/1}{
 
        \draw[cluster] ($ (offset) + (\nclusters*\clusterlength,0) + (\ngaps*\clusterspace,0) $) 
        -- ($ (offset) + (\nclusters*\clusterlength,0) + (\ngaps*\clusterspace,0) + (\size*\clusterlength,0) $);}
 \draw[tick] ($ (offset) + (0,\lw) $) -- ($ (offset) - (0,2*\lw) $);\coordinate (offset) at (0.0508322834452551, 1);
\foreach \nclusters\ngaps/\size in {0/0/1}{
 
        \draw[cluster] ($ (offset) + (\nclusters*\clusterlength,0) + (\ngaps*\clusterspace,0) $) 
        -- ($ (offset) + (\nclusters*\clusterlength,0) + (\ngaps*\clusterspace,0) + (\size*\clusterlength,0) $);}
 \draw[tick] ($ (offset) + (0,\lw) $) -- ($ (offset) - (0,2*\lw) $);\coordinate (offset) at (0.0518451067214478, 1);
\foreach \nclusters\ngaps/\size in {0/0/1}{
 
        \draw[cluster] ($ (offset) + (\nclusters*\clusterlength,0) + (\ngaps*\clusterspace,0) $) 
        -- ($ (offset) + (\nclusters*\clusterlength,0) + (\ngaps*\clusterspace,0) + (\size*\clusterlength,0) $);}
 \draw[tick] ($ (offset) + (0,\lw) $) -- ($ (offset) - (0,2*\lw) $); \draw[thin, gray, dotted ,step=1] (0,0) grid ($ (\maxepi+2,\maxclusters) $);
        \foreach \x in {5,10,...,\maxepi} {
            \ifthenelse{\x<\maxepi}{\draw[tick,black,thin] (\x+0.5,-0.4) -- (\x+0.5,-0.2); }{}
            \draw[tick,black,thin] (\x,-0.4) -- (\x,-0.1);     
            \node at (\x,-0.4) [below,black] { $\x$ };}\foreach \y in {1,5,10,...,\maxclusters} {
            \draw[tick,black,thin] (-0.2,\y) -- (-0.05,\y);
            \node at (-0.2,\y) [left,black] { $\y$ };}
        \draw[tick,black,thin]  ($ (\maxepi+2,\maxclusters+1) $) -- ($ (-0.2,\maxclusters+1) $) -- (-0.2,-0.4) -- ($ (\maxepi+2,-0.4) $) ;
\node at ($(\maxepi*0.5+2.5,-2.5-1)$) [left,black] { epistatic weight };\end{tikzpicture}

\subcaption{productmodelWeinreich\_from\_TanSupp0starstarstarstarTOWeinreich\_from\_TanSupp1starstarstarstar}
\end{figure}

 \vspace{1cm}